\documentclass[10pt,twocolumn,letterpaper]{article}

\usepackage[pagenumbers]{cvpr} 
\usepackage{multicol}
\usepackage{multirow}

\usepackage{float}

\definecolor{cvprblue}{rgb}{0.21,0.49,0.74}
\usepackage[pagebackref,breaklinks,colorlinks,allcolors=cvprblue]{hyperref}

\def\paperID{7} 
\def\confName{CVPR}
\def\confYear{2026}

\title{Smaller and Faster 3DGS via Post-Training Dictionary Learning}

\author{
Jiarong Gong, Jonas Unger, Ehsan Miandji \\
Linköping University \\
Department of Science and technology \\
{\tt\small jiarong.gong@liu.se, jonas.unger@liu.se, ehsan.miandji@liu.se}
}

\begin{document}
\twocolumn[{
\maketitle
\begin{center}
\includegraphics[width=0.95\textwidth]{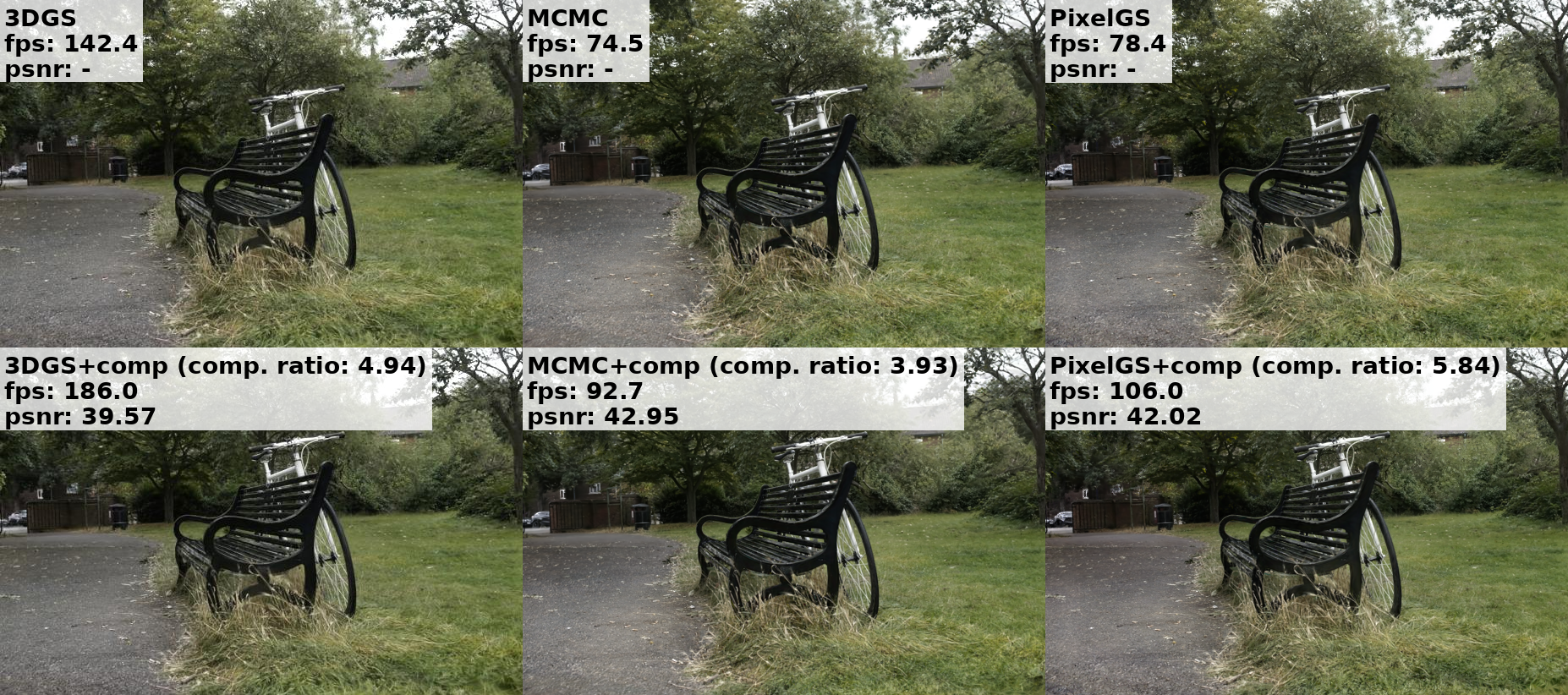}
\captionof{figure}{Visual comparison of baseline methods (\textit{3DGS}, \textit{MCMC}, and \textit{PixelGS}, top row) and their compressed counterparts (bottom row). Each panel in bottom row reports the corresponding compression ratio, rendering speed (FPS), and PSNR, highlighting that our dictionary-learning-based compression substantially reduces model size and improves rendering efficiency while preserving visual fidelity. Note that the reported PSNR values here are computed with respect to the renderings of original methods, not the Ground Truth.}
\label{fig:teaser}
\end{center}
}]
\begin{abstract}
3D Gaussian Splatting (3DGS) is a promising neural scene representation for real-time rendering, but trained models often suffer from large memory footprints, limiting deployment on less powerful devices. 
Existing compression techniques often lead to architectures with several additional trainable parameters. 
While achieving outstanding compression ratios, they introduce noticeable drops in image quality.
In this work, we introduce the first dictionary-learning-based compression framework for 3DGS. The proposed post-training compression pipeline can be deployed in virtually any 3DGS model without the need for re-training or modifications to existing 3DGS models. Our compression framework is straightforward to implement, yet provides significant compression capabilities, preserves image quality, and improves real-time rendering performance. Across 13 benchmark scenes, our approach achieves an average compression ratio of \textbf{3.95$\times$}, \textbf{3.10$\times$}, and \textbf{4.55$\times$} when applied to \textit{3DGS}, \textit{3DGS-MCMC}, and \textit{PixelGS}, respectively. This yields consistent rendering speedups of \textbf{23.3\%}, \textbf{24.3\%}, and \textbf{25.3\%}, while maintaining image quality.
\end{abstract}    
\section{Introduction}

3D Gaussian Splatting (3DGS) \cite{kerbl20233d} is a radiance field model that enables real-time rendering speed with comparable quality compared to state-of-the-art (SOTA) NeRF methods \cite{barron2023zip}, that has led to a large body of research \cite{zhang2024pixel, kheradmand20243d}. 
3DGS represents radiance fields using anisotropic Gaussians. While achieving a significant reduction in storage costs when compared to raw radiance field data, the excessive memory costs often hinders its adoption for complex scenes. 
As such, 3DGS compression has become an active research field \cite{girish2024eagles, navaneet2023compact3d, lee2024compact, morgenstern2024compact}. For instance, Navaneet et al.\cite{navaneet2023compact3d} quantize Gaussian parameters with K-means codebooks and reduce the number of Gaussians of a 3D model via opacity penalization and pruning; Similarly, Girish et al.\cite{girish2024eagles} propose to quantize latent embeddings and prune low-influence Gaussians. However, existing pipelines give rise to a drop in image quality. 
Additionally, some works concentrate on the reduction of the number of primitives.
Fridovich et al.\cite{fridovich2022plenoxels} explicitly uses voxel grid to store opacity and spherical harmonics which are used to compute color and makes the voxel grid sparse via voxel pruning. \cite{girish2024eagles, navaneet2023compact3d} reduces the number of Gaussians by eliminating those with low contributions through opacity, importance score thresholds, etc. For a detailed survey of related work, we refer readers to \cite{bagdasarian20253dgs}.

In this paper, we propose a fundamentally different approach via a dictionary-learning-based compression pipeline that applies compression as a post-training process, hence enabling compression of existing SOTA 3DGS models with virtually no noticeable degradation in image quality.
Since each 3D Gaussian is described by a set of attributes, namely position, covariance, Spherical Harmonics (SH) coefficients for color, and opacity. SH coefficients account for over 80\% of the storage cost, we, thus, train a dictionary based on SH coefficients, followed by Orthogonal Matching Pursuit (OMP) \cite{pati1993orthogonal} to obtain sparse codes. We show that the radiance field reconstruction, e.g. during rendering, can be directly evaluated using the sparse coefficients to obtain directional radiance values. Given the inherent sparsity introduced by our dictionary, our method achieves a higher rendering speed, while reducing the memory footprint and preserving the image quality of a given 3DGS model. 

The proposed compression pipeline not only ensures controllable rendering quality via representation sparsity, but also provides flexibility for integration into virtually any 3DGS method.
We evaluate this plug-and-play strategy on three SOTA 3DGS methods, and the results consistently demonstrate the effectiveness of our method. To summarize, this paper introduces a  3DGS compression framework with the following contributions:
\begin{itemize}
    \item We introduce a dictionary learning technology for 3DGS compression that can be applied to existing methods as a light-weight post training process. Our concise representation drastically reduces the storage cost and, equally importantly, maintains image quality.
    \item With a dictionary and a set of sparse coefficients replacing the SH coefficients, our new sparse rendering algorithm improves the rendering speed by more than 23.3\%.
\end{itemize}
\section{Method}

\subsection{Dictionary Learning and Sparse Representation}  \label{subsec:DL}

We briefly recall the dictionary learning framework, which underpins our compression method. 
Given a data matrix 
$X = [x_1, \dots, x_N] \in \mathbb{R}^{d \times N}$, 
the goal is to find an overcomplete dictionary 
$D=[a_1,\dots,a_m] \in \mathbb{R}^{d \times m}$, where $m > d$, and sparse codes $\alpha_i \in \mathbb{R}^m$, $i\in\{1,\dots,N\}$, such that
\begin{equation}
\min_{D, \{\alpha_i\}} \; \sum_{i=1}^{N} \|x_i - D \alpha_i\|_2^2,
\quad \text{s.t. } \|\alpha_i\|_0 \leq k,\ \|a_j\|_2 = 1,\ \forall j .
\label{eq:dl_objective}
\end{equation}

The dictionary learning problem is typically solved approximately via an alternating minimization strategy, whereby the sparse codes $\{\alpha_i\}$ and the dictionary $D$ are updated in turn while keeping the other fixed.
Having the dictionary $D$, and given a new test set, the task is then to obtain the most sparse coefficients while minimizing the error. In our framework, sparse coefficients are achieved via OMP, which efficiently selects the most relevant atoms from the dictionary until either/both a target sparsity $k$ or/and error threshold (tolerance) $\epsilon$ is reached. Since $k\ll m$, it has led to several applications in compression, see e.g. \cite{ehsan-tog}. We utilize a learned dictionary, and corresponding sparse coefficients, to compress a subset of 3DGS attributes that exhibit the highest storage/memory cost, as described in what follows.

\subsection{Post-training 3DGS compression}

Each Gaussian splat in 3DGS is parameterized by its position, covariance, opacity, and SH coefficients. Formally, each Gaussian is described by 59 parameters: 3 for position, 3 for scale, 4 for rotation, 1 for opacity, and 48 for the SH coefficients. This makes SH coefficients the dominant factor in storage overhead.

Our compression strategy targets this issue. Instead of storing all SH coefficients explicitly, we apply learned dictionary and sparse codes to represent SH coefficients in a compact form (with the DC term of SH excluded) after training.
Each Gaussian's SH coefficients are replaced by a shared dictionary and its corresponding sparse codes. For storage, we only need the non-zero elements and their indices for sparse codes. Since each sparse code might have a different number of non-zero elements, we store them in a \textit{Compressed-Sparse-Column} (CSC) format.

To accelerate rendering, we modify the radiance calculation based on the shared dictionary and the corresponding non-zero elements in the sparse codes. In the original computation, it needs to read the whole SH coefficients from global memory while our method reads the small shared dictionary (size: 45 × 90 FP32, 15.8 KB) and non-zeros of the sparse codes. From our tests (on RTX 4090 and RTX 4070ti), GPUs are memory-bound devices in terms of radiance computation which means memory-reading takes much more time than radiance computation. Since our method takes smaller information, it costs smaller time than the original. The motivation for acceleration is as follow.

Taking an RTX 4090 as an example, the peak FP32 compute throughput and memory bandwidth are 82.6 TFLOPS and 1008 GB/s, respectively. Under a simplified counting model, the original radiance computation costs about 96 FLOPs per Gaussian, while the compressed version costs about 1,152 FLOPs per Gaussian (about 1,149 FLOPs when the average sparsity is k = 11.7). 
In terms of data movement, the original SH path reads about 192~B per Gaussian, while the compressed path reads about 110~B per Gaussian ($k = 11.7$). 
The dictionary is a shared 15.8~KB table, accounted for as an amortized term $15.8~\text{KB} / N_{\text{vis}}$ per frame; due to high cache reuse, this term is negligible for large $N_{\text{vis}}$.


\begin{equation}
    \label{eq:t_orig}
    \begin{aligned}
    T_{\text{orig}} &= T_{\text{comp}} + T_{\text{mem}} \\
    &= \frac{96}{82.6 \times 10^{12}} + \frac{192}{1008 \times 10^9} \\
    &\approx 0.0012\ \text{ns} + 0.1905\ \text{ns} = 0.1917\ \text{ns}
    \end{aligned}
\end{equation}
\vspace{-4pt}
\begin{equation}
    \label{eq:t_opt}
    \begin{aligned}
    T_{\text{opt}} &= T_{\text{comp}} + T_{\text{mem}} \\
    &= \frac{1152}{82.6 \times 10^{12}} + \frac{110}{1008 \times 10^9} \\
    &\approx 0.0139\ \text{ns} + 0.1091\ \text{ns} = 0.1230\ \text{ns}
    \end{aligned}
\end{equation}

\section{Experiments}

\begin{table*}[t]
  \small
  \centering
  \caption{Mean per-scene PSNR (dB), SSIM, LPIPS, and FPS of compressive renderings relative to the original for tol = 0.10.
    Higher PSNR/SSIM and lower LPIPS indicate closer agreement with the original.
    "FPS(-Comp)" denotes the rendering speed of the original baselines.}
  \label{tab:pairwise_comparison1}
  \setlength{\tabcolsep}{3pt}
  \resizebox{\textwidth}{!}{
  \begin{tabular}{c c | ccccccccccccc | c}
    \toprule
    \textbf{Method}  & \textbf{Metric}
      & Bicycle & Bonsai & Counter & Garden & Kitchen & Room
      & Stump & Truck & Flowers & Playroom & Train & Treehill & Drjohnson
      & \textbf{Mean} \\
    \midrule

    \multirow{6}{*}{3DGS+comp}
      & PSNR  & 38.54 & 41.03 & 40.69 & 37.97 & 39.59 & 41.20 & 38.51 & 39.34 & 37.80 & 40.03 & 39.64 & 39.18 & 40.84 & \textbf{39.57} \\
      & SSIM  & 0.9830 & 0.9863 & 0.9818 & 0.9810 & 0.9851 & 0.9829 & 0.9819 & 0.9859 & 0.9870 & 0.9821 & 0.9873 & 0.9829 & 0.9871 & \textbf{0.9842} \\
      & LPIPS & 0.0301 & 0.0264 & 0.0263 & 0.0296 & 0.0217 & 0.0301 & 0.0392 & 0.0210 & 0.0257 & 0.0260 & 0.0210 & 0.0311 & 0.0311 & \textbf{0.0276} \\
      & Comp Ratio & 4.94 & 4.65 & 3.76 & 3.16 & 3.36 & 4.91 & 4.15 & 3.92 & 3.40 & 3.56 & 3.39 & 3.76 & 4.33 & \textbf{3.95} \\
      \cmidrule{2-16}
      & FPS & 186.0 & 268.9 & 186.7 & 223.7 & 158.8 & 186.9 & 197.1 & 225.9 & 305.7 & 203.2 & 193.2 & 203.7 & 173.4 & \textbf{208.7} \\
      & FPS(-Comp)   & 142.4 & 240.0 & 170.5 & 188.4 & 143.9 & 167.7 & 147.7 & 193.4 & 250.4 & 177.7 & 178.6 & 163.6 & 136.2 & \textbf{169.3} \\
    \midrule

    \multirow{6}{*}{MCMC+comp}
      & PSNR  & 43.20 & 43.09 & 43.56 & 42.85 & 42.64 & 43.64 & 44.21 & 42.51 & 43.41 & 42.17 & 42.29 & 43.22 & 41.57 & \textbf{42.95} \\
      & SSIM  & 0.9974 & 0.9951 & 0.9947 & 0.9969 & 0.9964 & 0.9944 & 0.9966 & 0.9969 & 0.9974 & 0.9946 & 0.9970 & 0.9967 & 0.9952 & \textbf{0.9961} \\
      & LPIPS & 0.0094 & 0.0067 & 0.0065 & 0.0080 & 0.0060 & 0.0056 & 0.0126 & 0.0051 & 0.0076 & 0.0045 & 0.0062 & 0.0096 & 0.0109 & \textbf{0.0076} \\
      &  Comp  Ratio     &  3.93 & 3.22 & 2.67 & 2.85 & 2.64 & 3.72 & 2.80 & 3.29 & 2.25 & 2.54 & 3.43 & 3.38 & 3.56 & \textbf{3.10} \\
      \cmidrule{2-16}
      & FPS            & 92.7 & 182.3 & 125.5 & 101.5 & 129.7 & 142.5 & 92.1 & 148.8 & 74.1 & 182.3 & 216.5 & 85.3 & 253.2 & \textbf{140.5} \\
      & FPS(-Comp)             & 74.5 & 167.1 & 118.5 & 82.9  & 118.0 & 132.2 & 78.4 & 127.7 & 67.6 & 157.1 & 196.5 & 72.4 & 177.6 & \textbf{113.0} \\
    \midrule

    \multirow{6}{*}{PixelGS+comp}
      & PSNR  & 42.41 & 42.18 & 42.73 & 41.97 & 41.85 & 42.11 & 42.78 & 41.65 & 42.24 & 41.20 & 41.50 & 42.00 & 41.61 & \textbf{42.02} \\
      & SSIM  & 0.9954 & 0.9932 & 0.9914 & 0.9950 & 0.9948 & 0.9913 & 0.9943 & 0.9952 & 0.9958 & 0.9915 & 0.9958 & 0.9945 & 0.9940 & \textbf{0.9940} \\
      & LPIPS & 0.0123 & 0.0115 & 0.0101 & 0.0104 & 0.0086 & 0.0152 & 0.0164 & 0.0078 & 0.0106 & 0.0114 & 0.0084 & 0.0140 & 0.0164 & \textbf{0.0118} \\
      & Comp Ratio  & 5.84 & 5.13 & 4.40 & 3.56 & 3.69 & 5.53 & 4.36 & 4.56 & 3.59 & 4.45 & 4.29 & 4.18 & 5.59 & \textbf{4.55} \\
      \cmidrule{2-16}
      & FPS       & 106.0 & 216.0 & 123.9 & 138.0 & 129.4 & 158.5 & 116.8 & 117.7 & 114.4 & 152.5 & 119.9 & 97.3 & 134.5 & \textbf{132.7} \\
      & FPS(-Comp)         & 78.4  & 183.5 & 106.9 & 105.6 & 107.5 & 133.0 & 88.1 & 93.1   & 89.6  & 122.9 & 99.0 & 73.8 & 95.3 & \textbf{105.9} \\
    \bottomrule
  \end{tabular}
  }
\end{table*}

\subsection{Experimental Setup}
Our compressive framework can be easily integrated into SOTA 3DGS methods. To verify the effectiveness of our methods, we make 3 sets of comparisons, \textit{3DGS} \cite{kerbl20233d} vs. \textit{3DGS+comp}, \textit{MCMC}\cite{kheradmand20243d} vs. \textit{MCMC+comp}, and\textit{ PixelGS}\cite{zhang2024pixel} vs. \textit{PixelGS+comp}.
All experiments are conducted on an NVIDIA GPU RTX 4090.
We report standard evaluation metrics: PSNR, SSIM, and LPIPS for reconstruction quality, compression ratio for storage efficiency, and FPS for rendering speed performance. The experimental scenes include Mip-NeRF-360, Tanks\&Temples, and Deep-Blending datasets, 13 scenes in total.

\subsection{Effect of Controllable OMP Tolerance Parameter on Rendering Quality and Efficiency}

As previously mentioned, the sparsity is controlled by a tolerance parameter when performing OMP algorithm.
It, thus, directly decides the quality of the reconstructed SH coefficients and serves as a key factor determining the trade-off between compression ratio and rendering quality. 
We now investigate how varying the tolerance parameter influences the compressive mothods. 
We study the impact of tolerance by varying it across {0.03, 0.05, 0.1, 0.2}, which we selected based on preliminary experiments showing that these values cover the typical trade-off range between high-fidelity reconstruction and aggressive compression. 
For better understanding, we use compression ratio as an indicator of tolerance in the following figures.

Figure. \ref{fig:psnr_vs_cr_3dgs} (more figures per scene for the 3 SOTA 3DGS methods can be found in \textbf{Supplementary Material} Section 1) plots PSNR against compression ratio, revealing the expected degradation in quality as compression increases. However the rate of degradation diminishes. Extrapolating this trend, we anticipate that quality will ultimately plateau: once compression is sufficiently strong, the sparse codes finally have a single non-zero entry whose value remains fixed; consequently, the SH-coefficient reconstruction error reaches its upper bound and ceases to grow.
Note that the reported PSNR values are computed with respect to original renderings, not the GT. Thus, it basically indicates the visual similarity between baselines and their compressive counterparts.

\begin{figure}[t]
    \centering
    \includegraphics[width=0.95\linewidth]{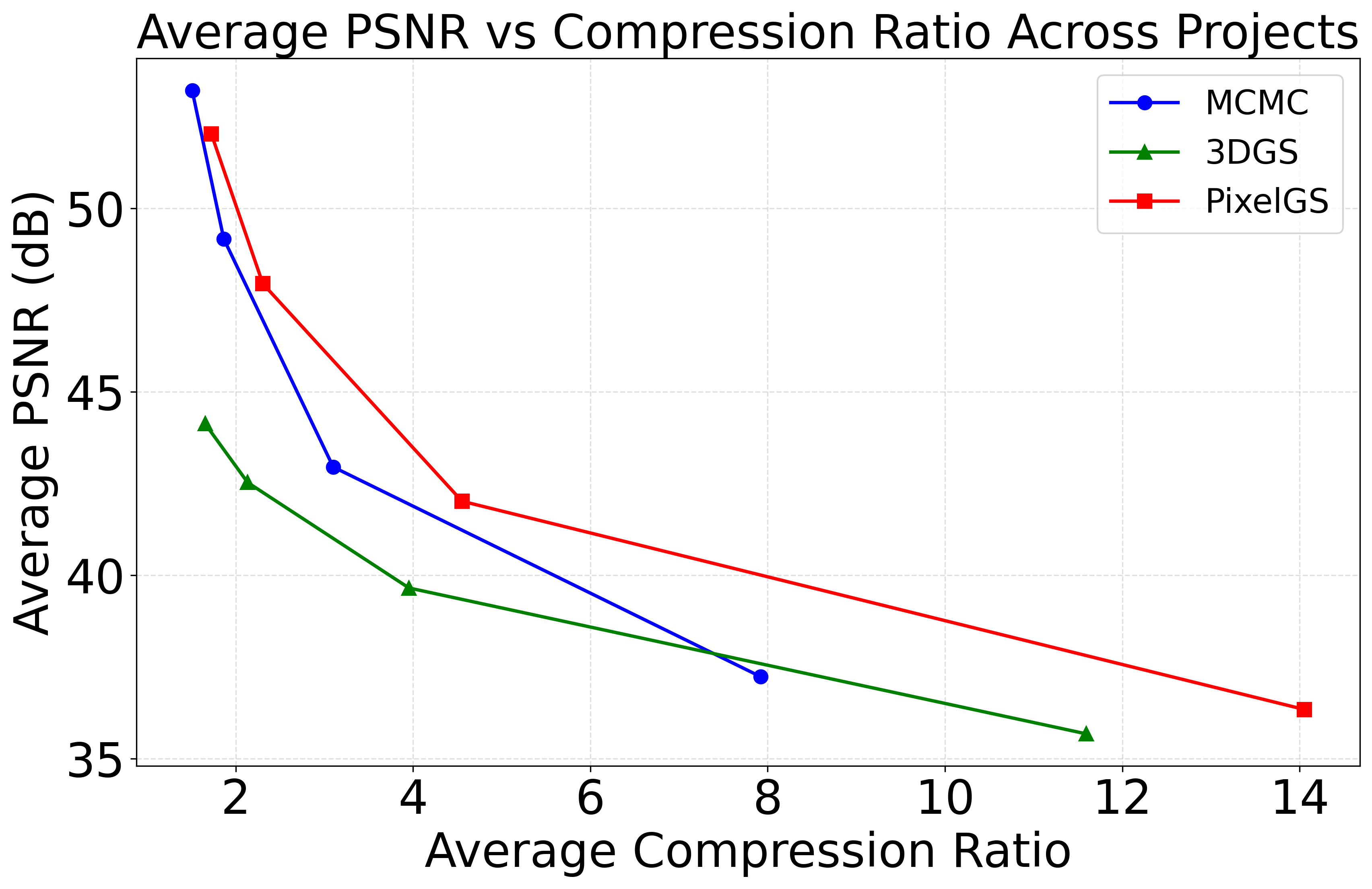}
    \caption{Averaged PSNR vs. Compression Ratio for three SOTA algorithms utilizing our novel post-training compression method.}
    \label{fig:psnr_vs_cr_3dgs}
\end{figure}

Tolerance also affects the rendering speed.
Figure. \ref{fig:fps_vs_cr_mcmc} (see more figures in \textbf{Supplementary Material} Section 2) quantify the trade-off between compression strength and rendering speed, plotting FPS against compression ratio. Figure. \ref{fig:fps_vs_cr_mcmc} shows the frame-rate rises monotonically with stronger compression, but the rendering efficiency gain diminishes and the curves eventually plateau. This saturation aligns with the quality analysis given previously.

To further quantify the rendering fidelity and efficiency of our compression method, Table~\ref{tab:pairwise_comparison1} reports PSNR, SSIM, Lpips, and FPS means over each scene for the compressed renderings of each pipeline \textit{3DGS+comp}, \textit{MCMC+comp}, and \textit{PixelGS+comp} at tolerance 0.1, against the corresponding original baseline renderings. A table records the same metric results over different tolerances is included in \textbf{Supplementary Material} Section 3. 
From Table~\ref{tab:pairwise_comparison1}, it is evident that our method preserves rendering quality since the minimum PSNR value is 37.80 dB and the maximum PSNR 44.21 dB; from \textbf{Metric} "Comp Ratio" and "FPS", we can also easily find that across all frameworks, our method consistently reduces memory usage and improves rendering efficiency.
More concretely, when our compressive framework is applied to baselines, our method improves the rendering speed substantially, with average FPS gains of 23.3\%, 24.3\%, and 25.3\%. Meanwhile, the SH coefficients achieve significant compression, with mean ratios of 3.95$\times$, 3.10$\times$, and 4.55$\times$, respectively.


To ensure a rigorous evaluation, we further benchmark our framework against Ground Truth (GT) images across three SOTA methods. As detailed in \textbf{Supplementary Material} Section 4, the performance gap between our compressed models and the original baselines is negligible, e.g., with an average PSNR decrease of $0.14$ dB for 3DGS.
In contrast, existing methods like EAGLES \cite{girish2024eagles} incur more pronounced quality degradation (e.g., a $0.21$ dB average PSNR drop) and significant performance volatility in complex scenes (e.g., up to $0.76$ dB and $0.62$ dB drops in Bonsai and Counter, respectively) due to Scalar Quantization. Notably, unlike our streamlined post-training approach, EAGLES requires a burdensome optimization process via Quantization-Aware Training (QAT). This necessitates the Straight-Through Estimator (STE) to bypass non-differentiable gradients, which significantly increases implementation complexity and computational overhead.

\begin{figure}[t]
    \centering
    \includegraphics[width=1\linewidth]{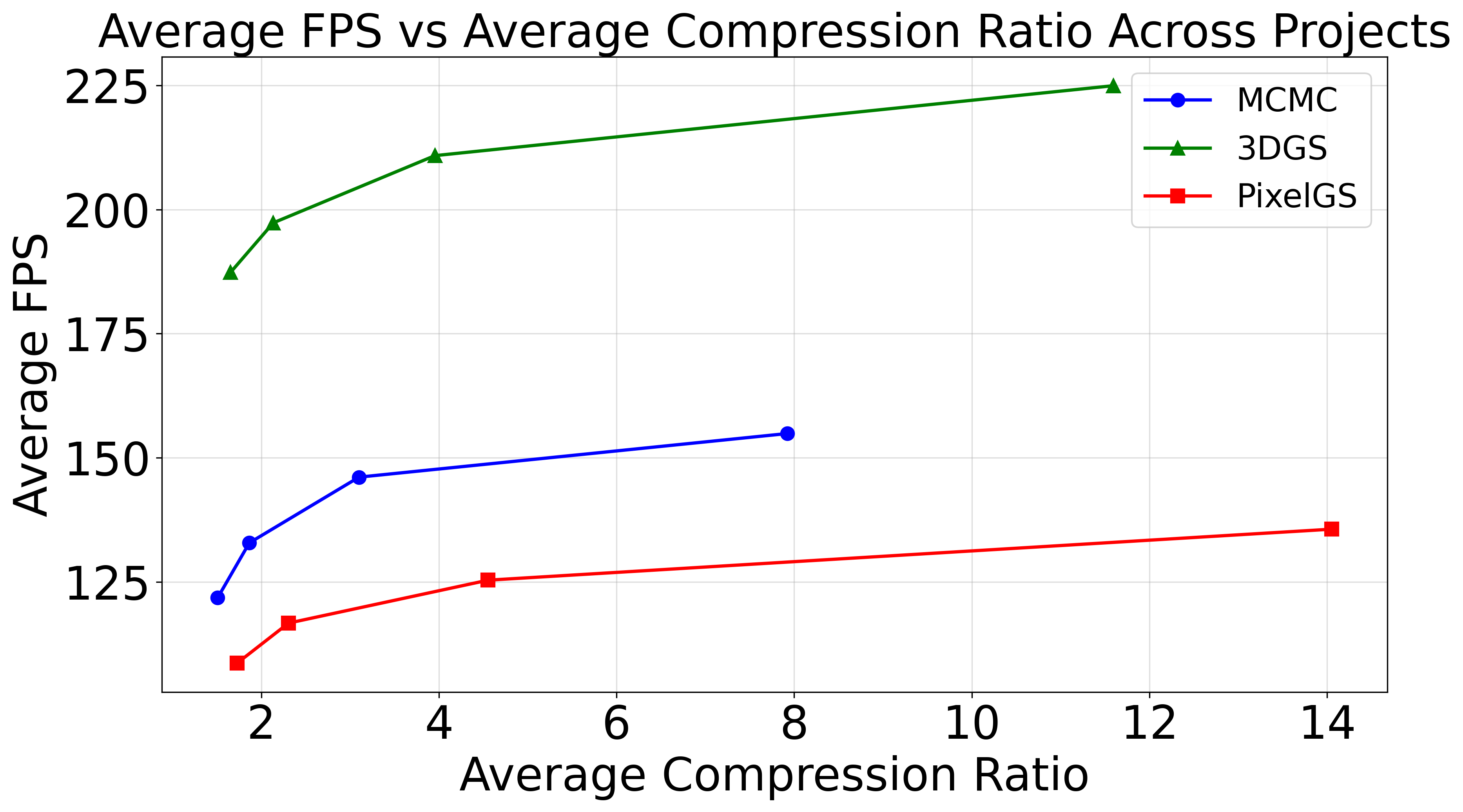}
    \caption{
    Averaged FPS vs. Compression Ratio for three SOTA algorithms. \textbf{\textit{PixelGS}} presents a stronger compression than the other two.}
    \label{fig:fps_vs_cr_mcmc}
\end{figure}



\section{Conclusion}

In this paper, we presented the first dictionary-learning-based post-training compression framework for 3D Gaussian Splatting. This approach brings benefits in two dimensions. First, dictionary-learning compression significantly reduces the number of Gaussian parameters, achieving compression ratios of 3.95$\times$, 3.10$\times$, and 4.55$\times$ for 3DGS, 3DGS-MCMC, and PixelGS, respectively, while preserving rendering fidelity. Second, leveraging the compact representation, we optimized the rasterization process, leading to consistent rendering speedups of more than 23\% across all frameworks. These improvements confirm that dictionary learning provides an effective and generalizable compression strategy for Gaussian splatting methods.


{
    \small
    \bibliographystyle{ieeenat_fullname}
    \bibliography{main}
}

\clearpage
\appendix
\renewcommand{\thesection}{\arabic{section}}
\onecolumn
\clearpage
\raggedbottom
\setlength{\parskip}{0pt}
\setlength{\floatsep}{4pt plus 1pt minus 1pt}
\setlength{\intextsep}{4pt plus 1pt minus 1pt}
\setlength{\textfloatsep}{6pt plus 1pt minus 1pt}

\begin{center}
  {\LARGE \textbf{Supplementary Material}}\\[6pt]
  {\Large \textbf{Smaller and Faster 3D Gaussian Splatting via Post-Training Dictionary Learning}}\\[6pt]

  {\normalsize This supplementary material contains additional qualitative and quantitative per-scene results referenced in the main paper. There are 13 scenes in total, but here we only present results on 8 scenes to make this supplementary file $\leq$ 10MB which is the upper limit size of a file uploaded through submission system}
\end{center}

\bigskip
\hrule
\bigskip

\section{The Effect of Compression Ratio on the Rendered Image Quality}
Here we provide additional figures illustrating the effect of compression ratio (controlled by the OMP tolerance) on rendered image quality, measured in PSNR, across 13 scenes. In total, our evaluation covers 13 scenes for each of the three baselines combined with our compression method (\textit{3DGS+comp}, \textit{MCMC+comp}, and \textit{PixelGS+comp}). ``Max PSNR'' and ``Min PSNR'' are obtained by evaluating the renderings from all test camera viewpoints. For instance, the \textbf{Bicycle} scene contains 25 test cameras, and the reported Max/Min PSNR corresponds to the highest and lowest PSNR among these views.

\subsection{3DGS+Comp}

\begin{figure}[H]
    \centering
    \begin{minipage}{0.44\textwidth}
        \centering
        \includegraphics[width=\linewidth]{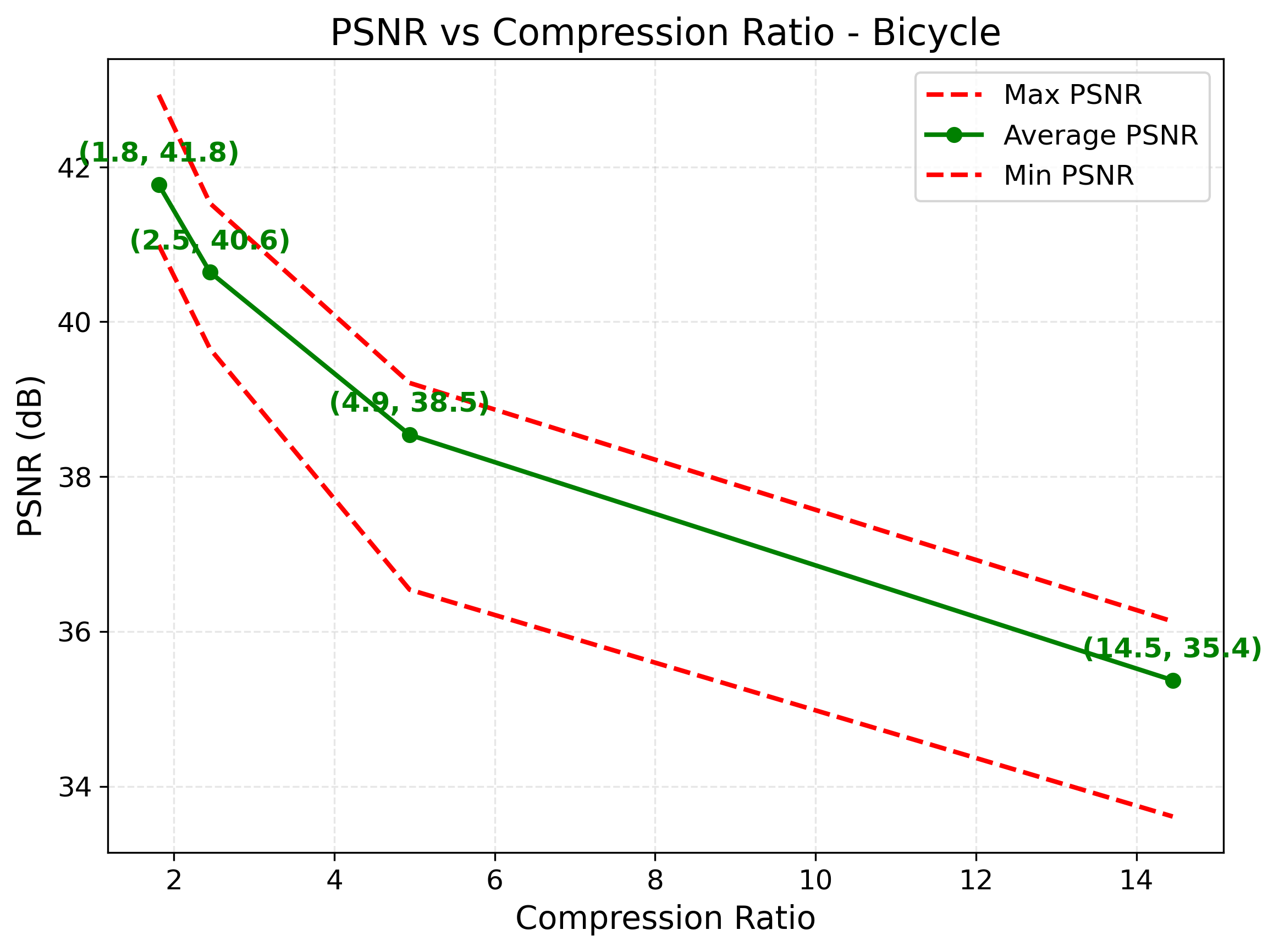}
        \caption{Effect of compression ratio on rendering quality for the \textit{Bicycle} scene.}
    \end{minipage}
    \hfill
    \begin{minipage}{0.44\textwidth}
        \centering
        \includegraphics[width=\linewidth]{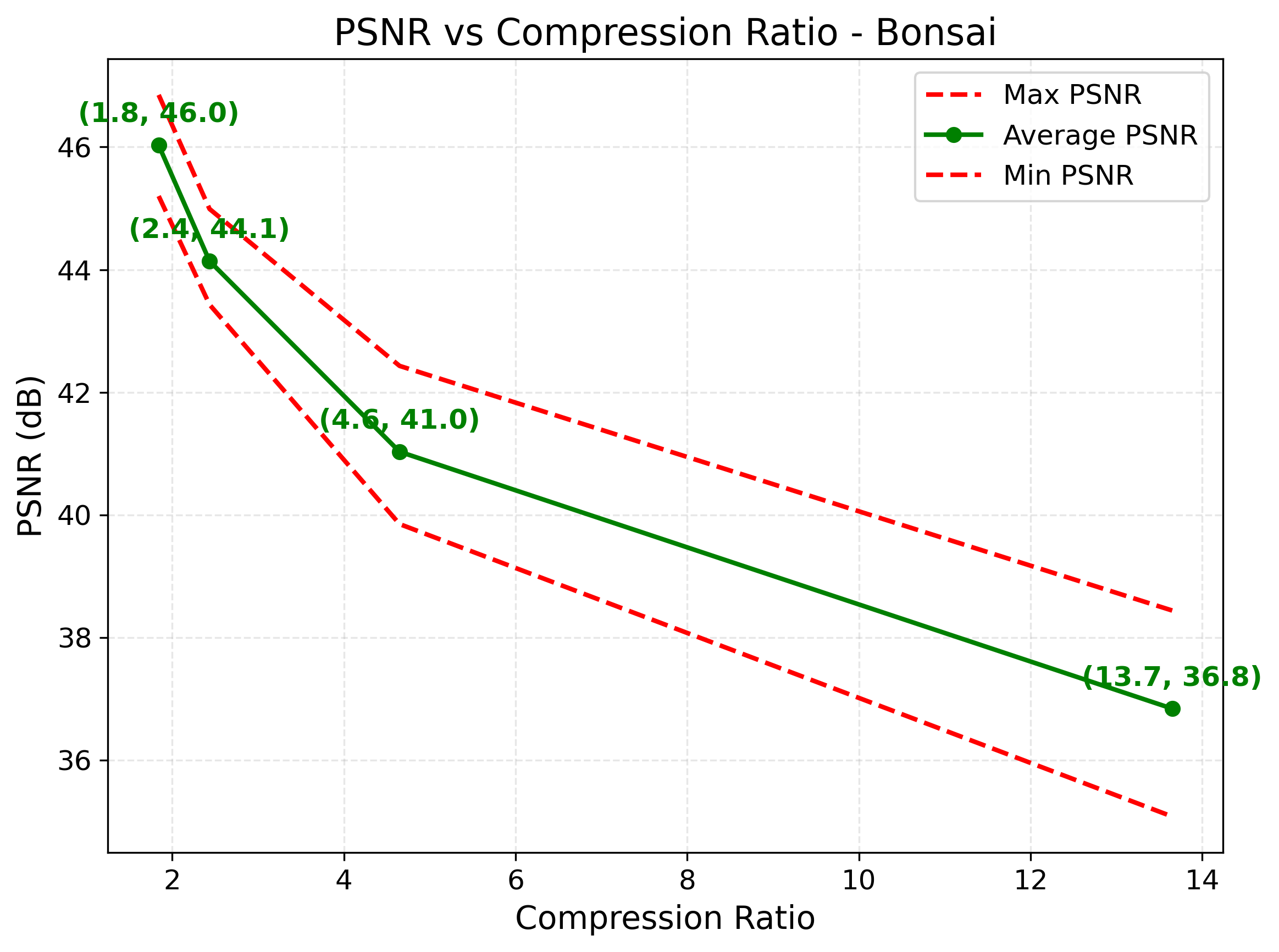}
        \caption{Effect of compression ratio on rendering quality for the \textit{Bonsai} scene.}
    \end{minipage}
\end{figure}

\begin{figure}[H]
    \centering
    \begin{minipage}{0.44\textwidth}
        \centering
        \includegraphics[width=\linewidth]{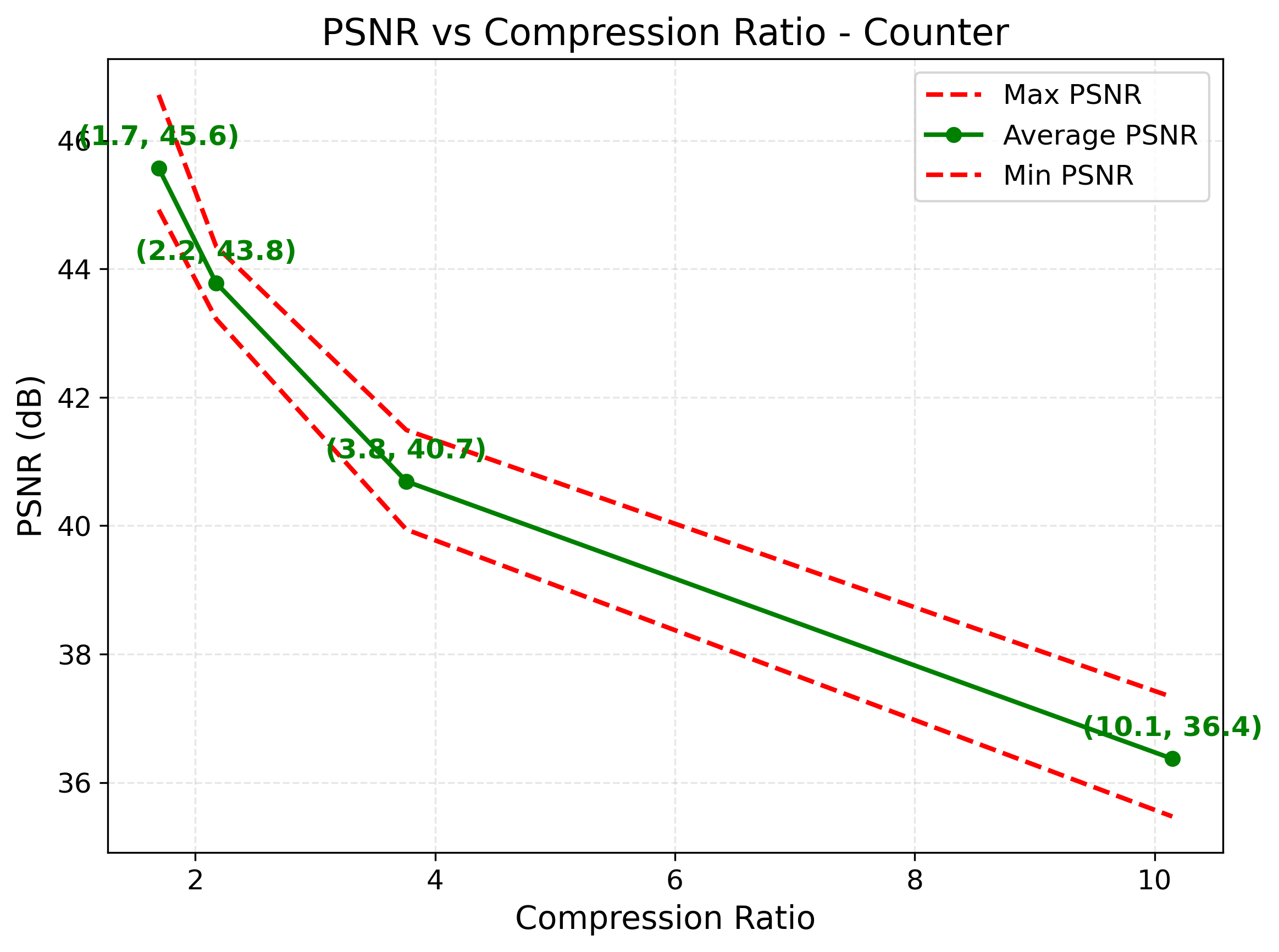}
        \caption{Effect of compression ratio on rendering quality for the \textit{Counter} scene.}
    \end{minipage}
    \hfill
    \begin{minipage}{0.44\textwidth}
        \centering
        \includegraphics[width=\linewidth]{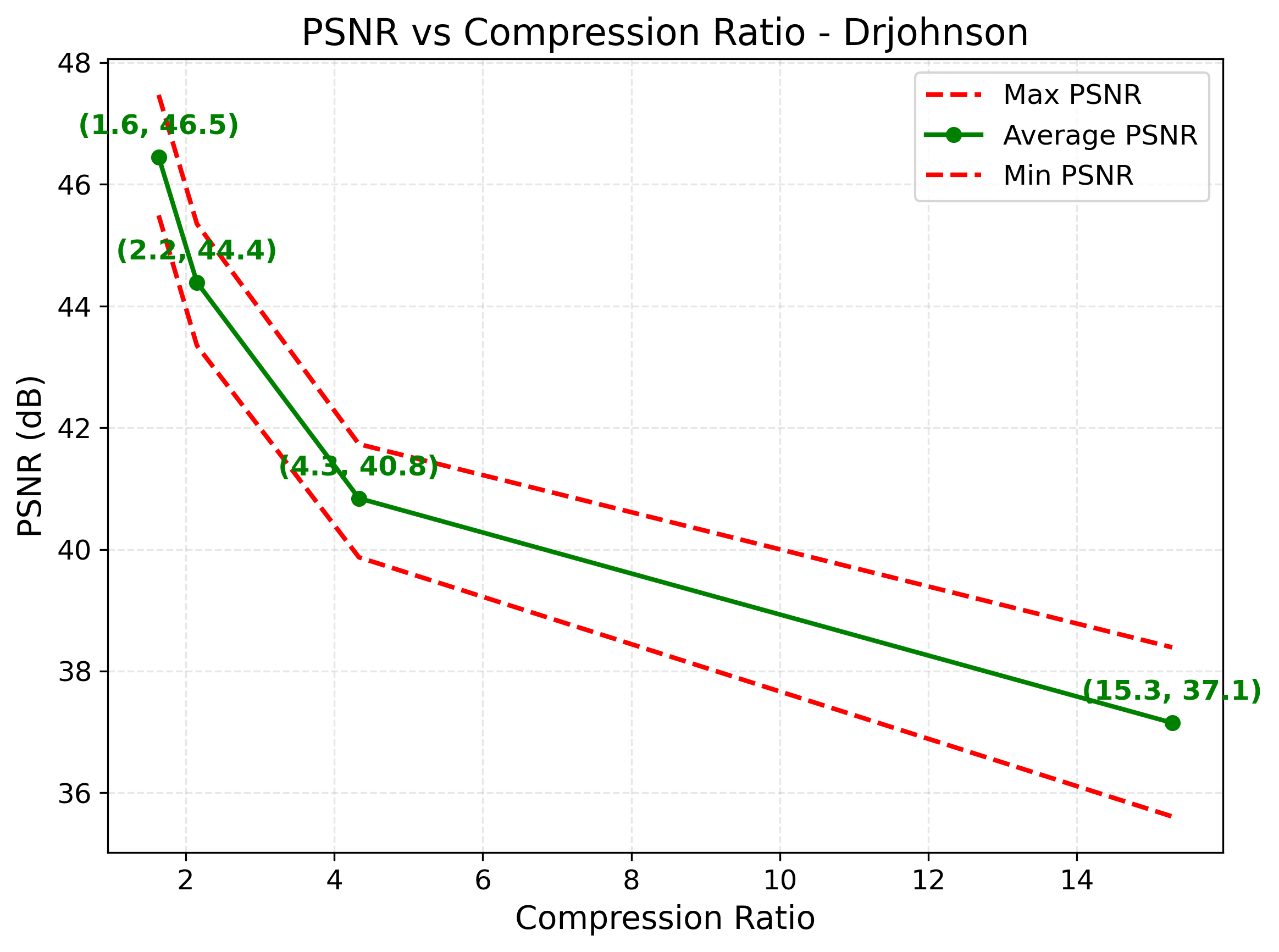}
        \caption{Effect of compression ratio on rendering quality for the \textit{Drjohnson} scene.}
    \end{minipage}
\end{figure}

\begin{figure}[H]
    \centering
    \begin{minipage}{0.44\textwidth}
        \centering
        \includegraphics[width=\linewidth]{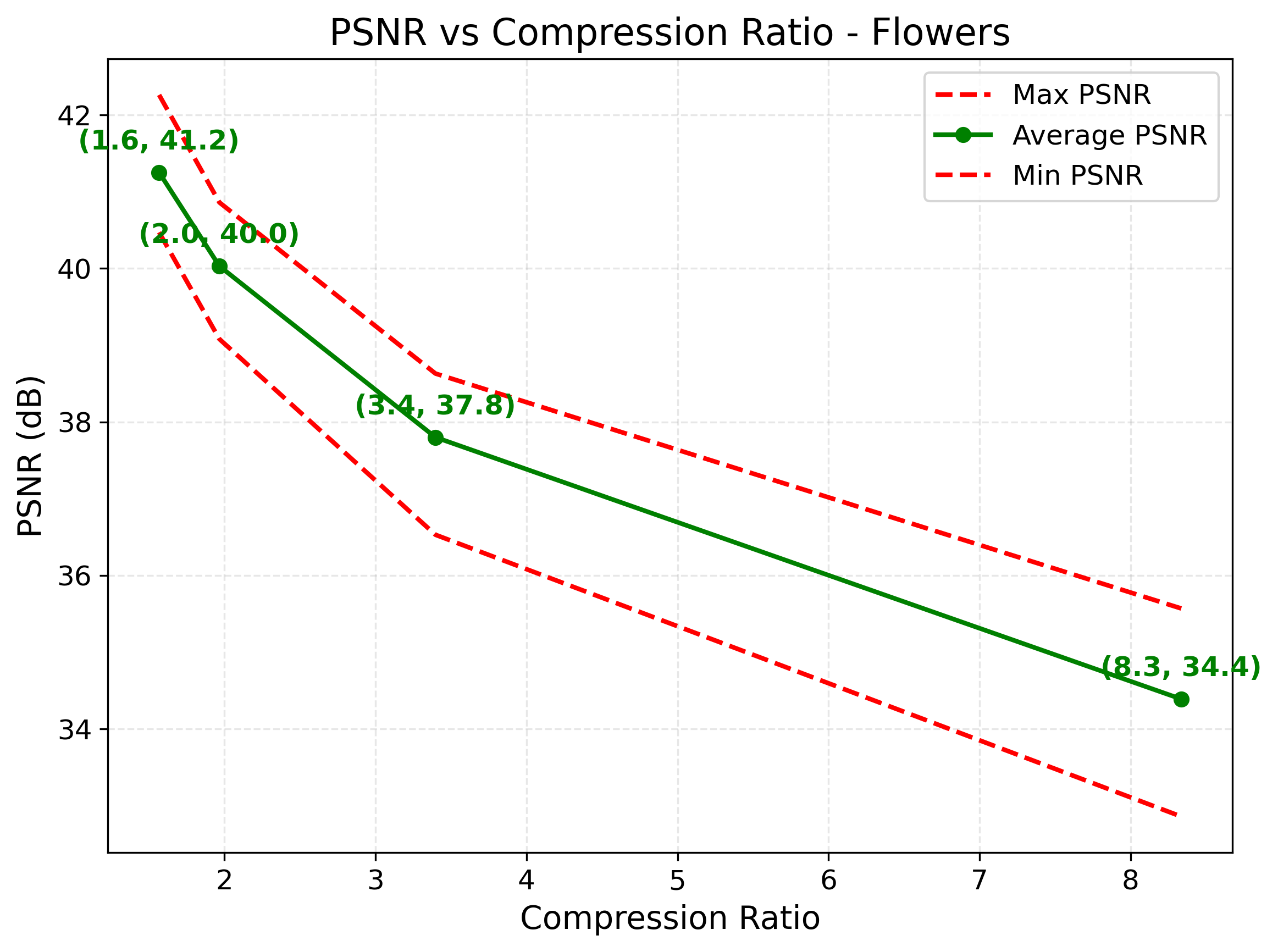}
        \caption{Effect of compression ratio on rendering quality for the \textit{Flowers} scene.}
    \end{minipage}
    \hfill
    \begin{minipage}{0.44\textwidth}
        \centering
        \includegraphics[width=\linewidth]{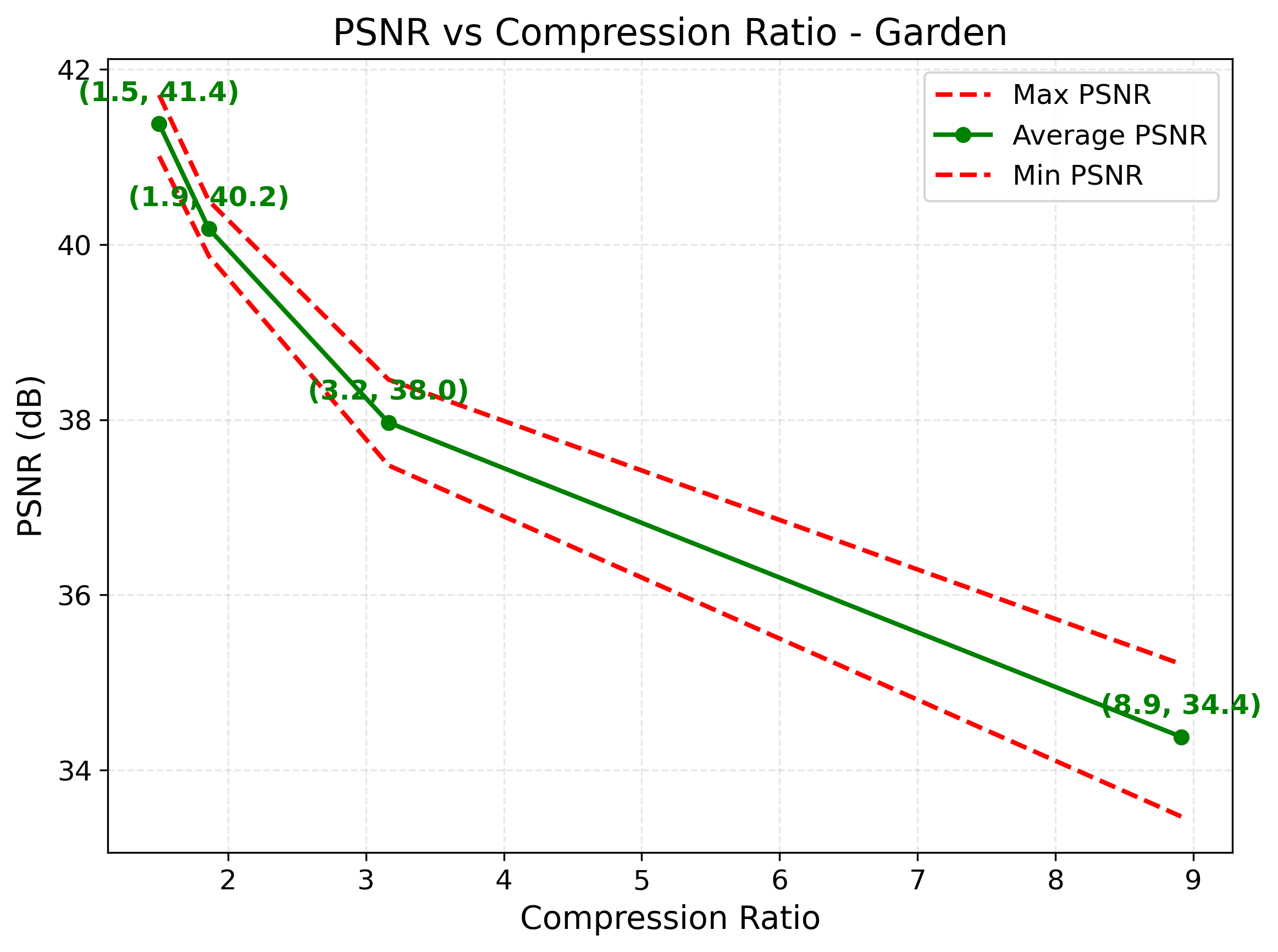}
        \caption{Effect of compression ratio on rendering quality for the \textit{Garden} scene.}
    \end{minipage}
\end{figure}

\begin{figure}[H]
    \centering
    \begin{minipage}{0.44\textwidth}
        \centering
        \includegraphics[width=\linewidth]{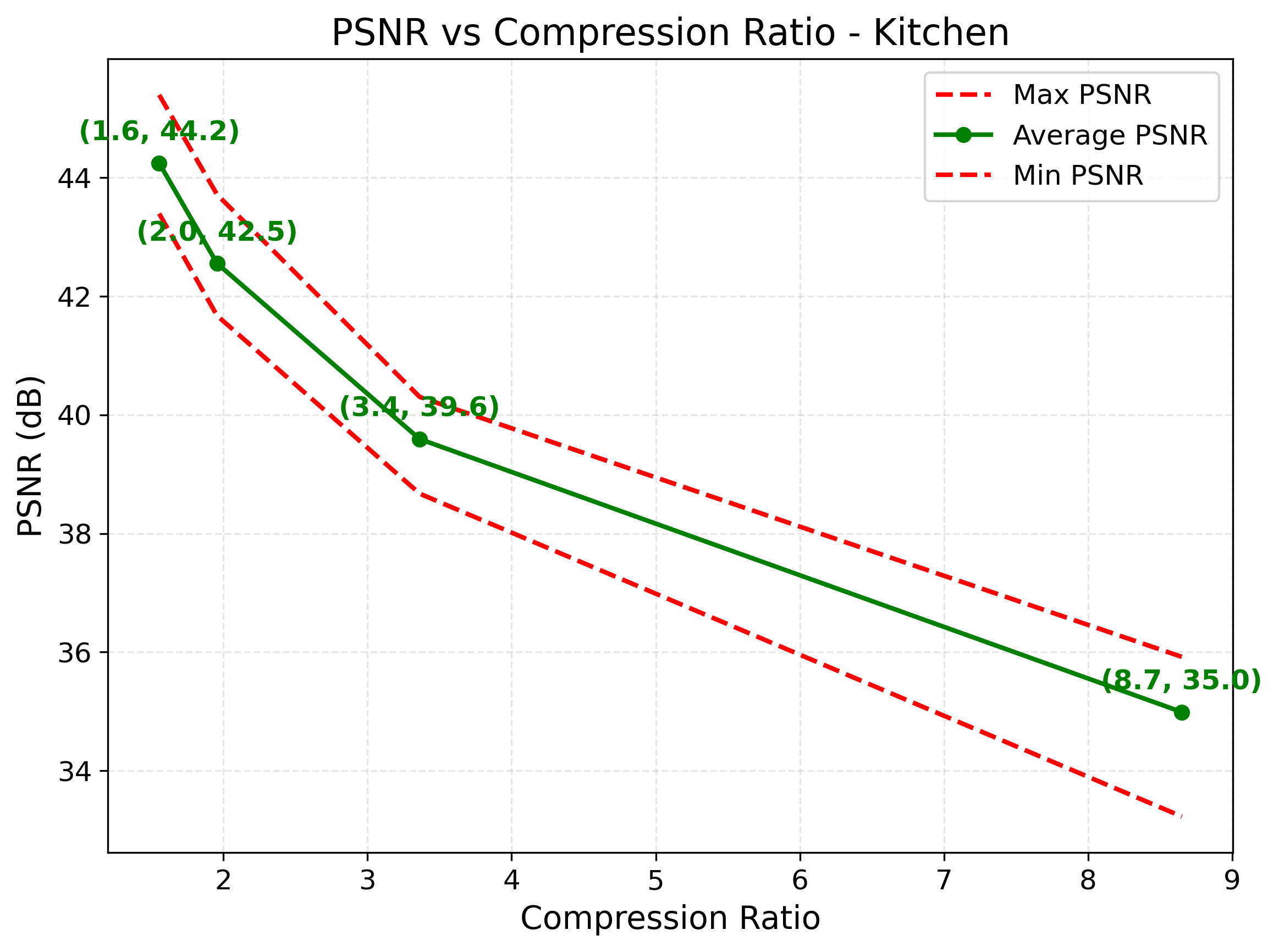}
        \caption{Effect of compression ratio on rendering quality for the \textit{Kitchen} scene.}
    \end{minipage}
    \hfill
    \begin{minipage}{0.44\textwidth}
        \centering
        \includegraphics[width=\linewidth]{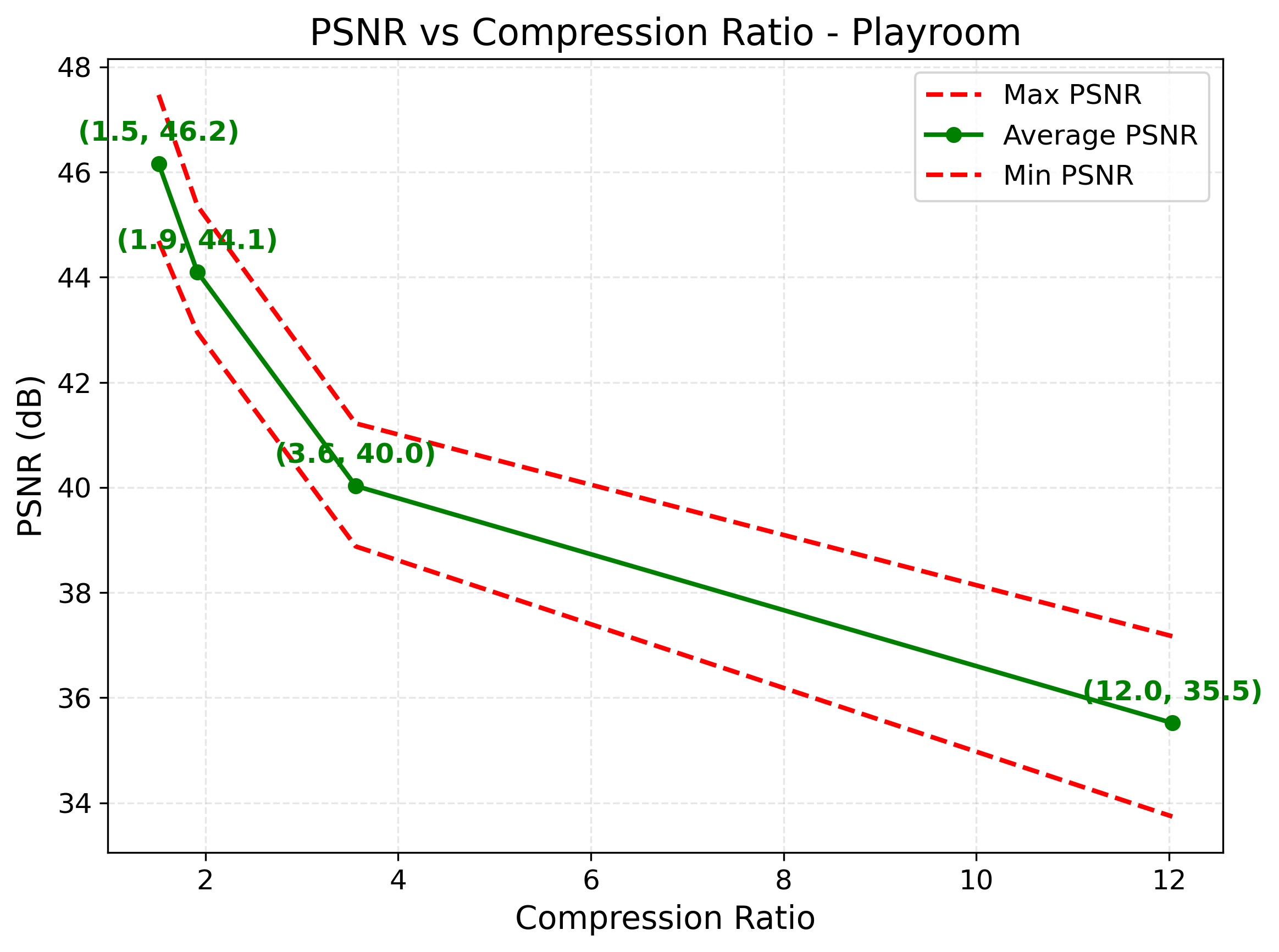}
        \caption{Effect of compression ratio on rendering quality for the \textit{Playroom} scene.}
    \end{minipage}
\end{figure}

\subsection{MCMC+Comp}

\begin{figure}[H]
    \centering
    \begin{minipage}{0.44\textwidth}
        \centering
        \includegraphics[width=\linewidth]{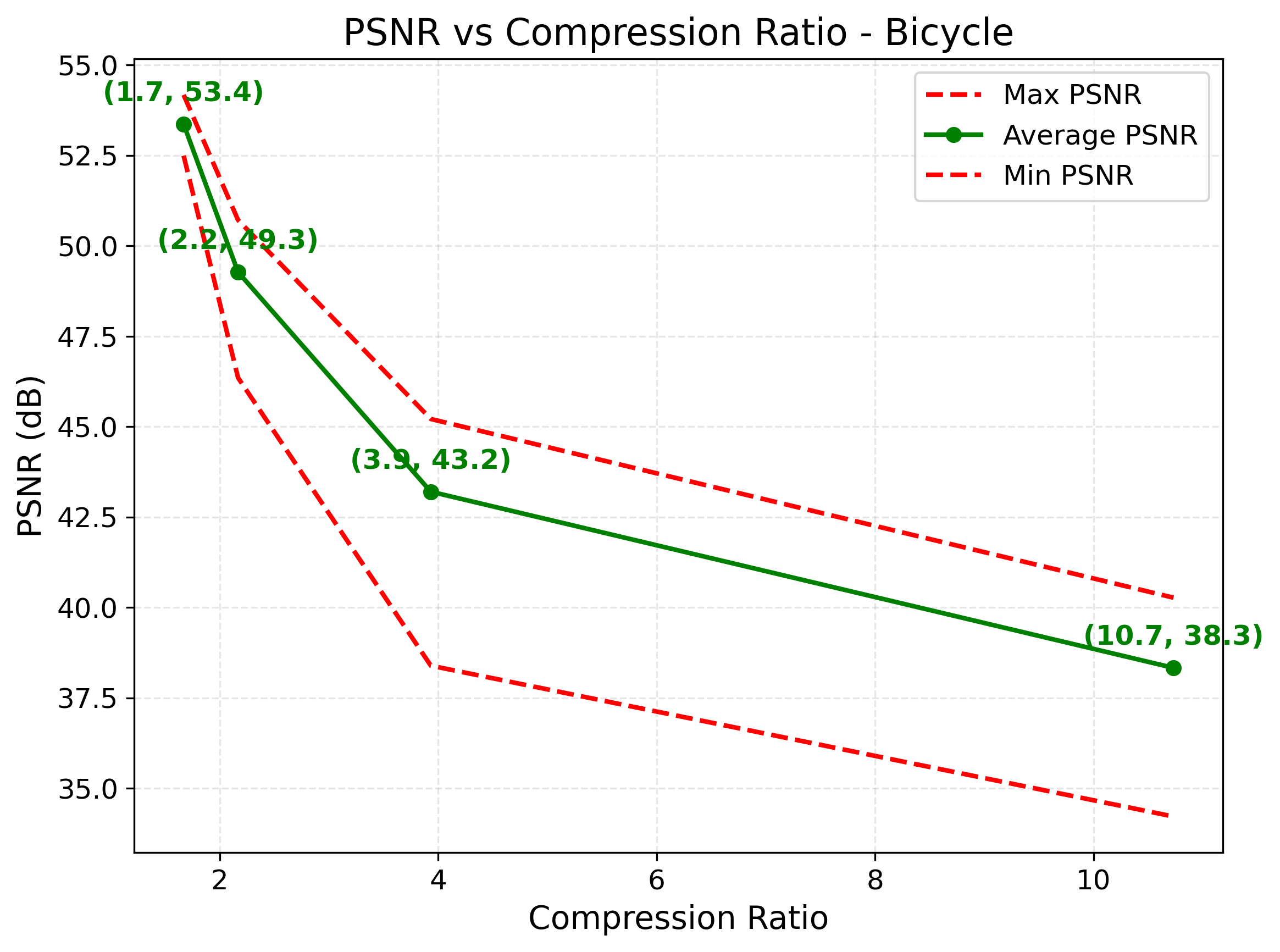}
        \caption{Effect of compression ratio on rendering quality for the \textit{Bicycle} scene.}
    \end{minipage}
    \hfill
    \begin{minipage}{0.44\textwidth}
        \centering
        \includegraphics[width=\linewidth]{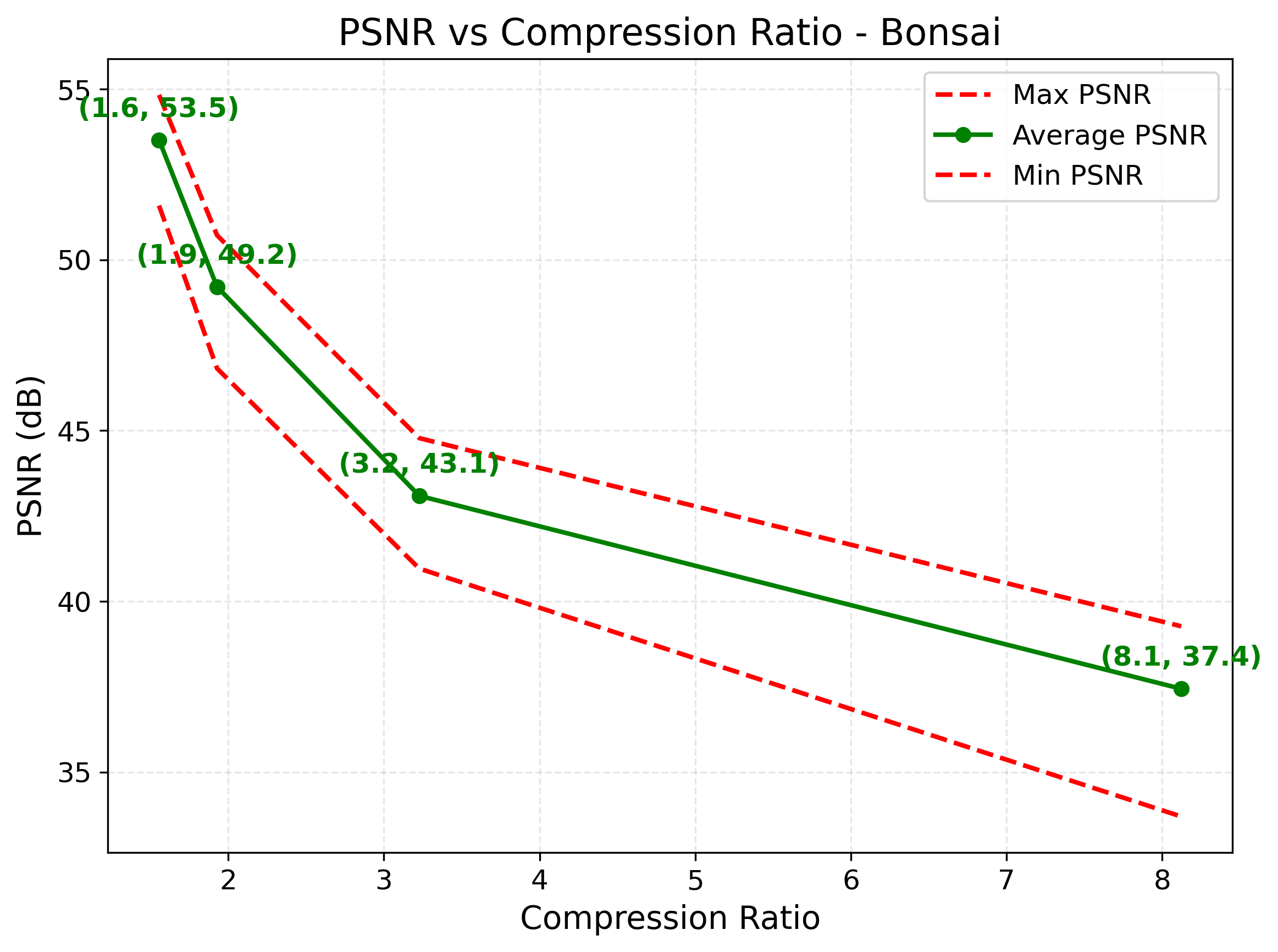}
        \caption{Effect of compression ratio on rendering quality for the \textit{Bonsai} scene.}
    \end{minipage}
\end{figure}

\begin{figure}[H]
    \centering
    \begin{minipage}{0.44\textwidth}
        \centering
        \includegraphics[width=\linewidth]{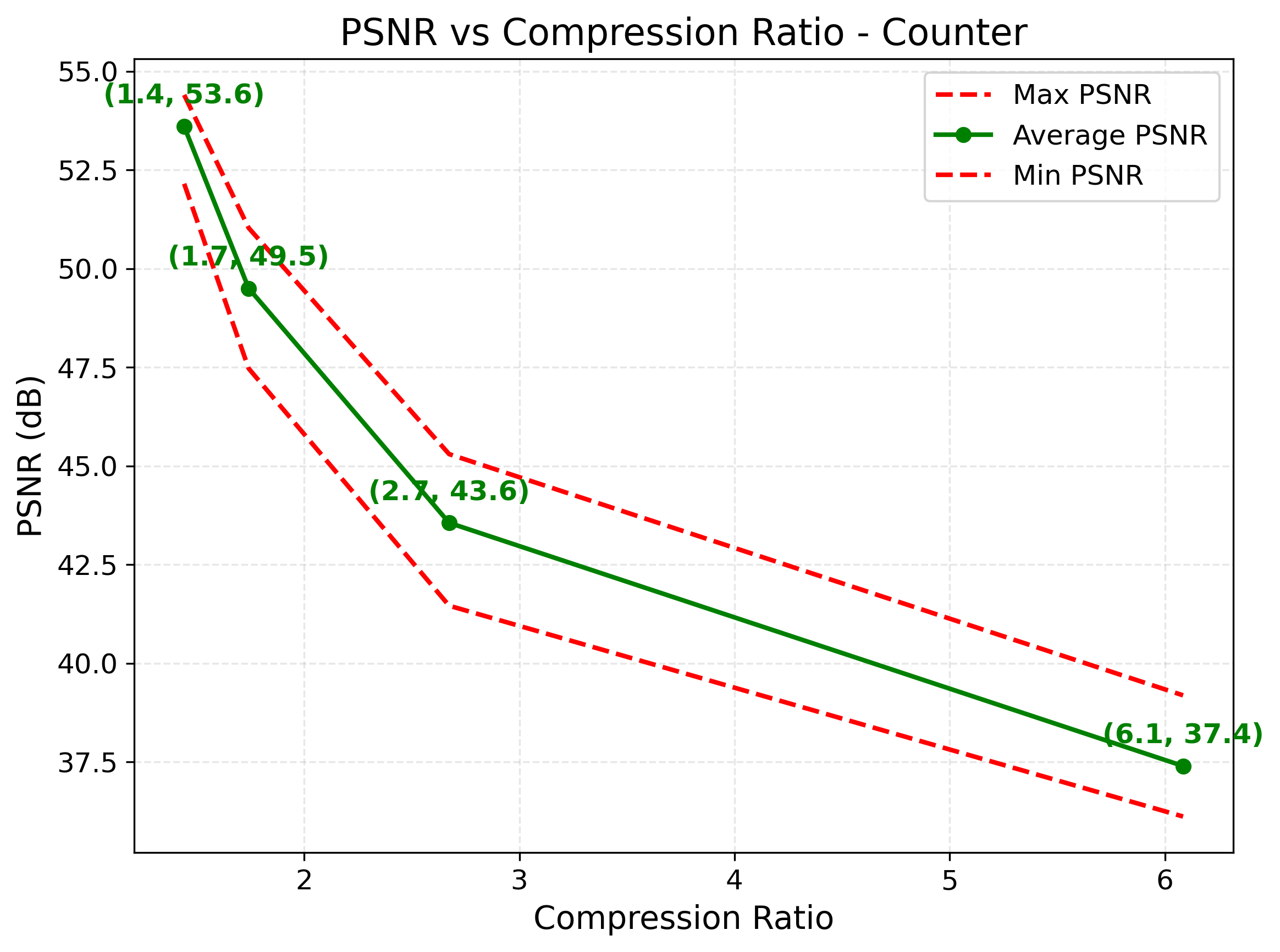}
        \caption{Effect of compression ratio on rendering quality for the \textit{Counter} scene.}
    \end{minipage}
    \hfill
    \begin{minipage}{0.44\textwidth}
        \centering
        \includegraphics[width=\linewidth]{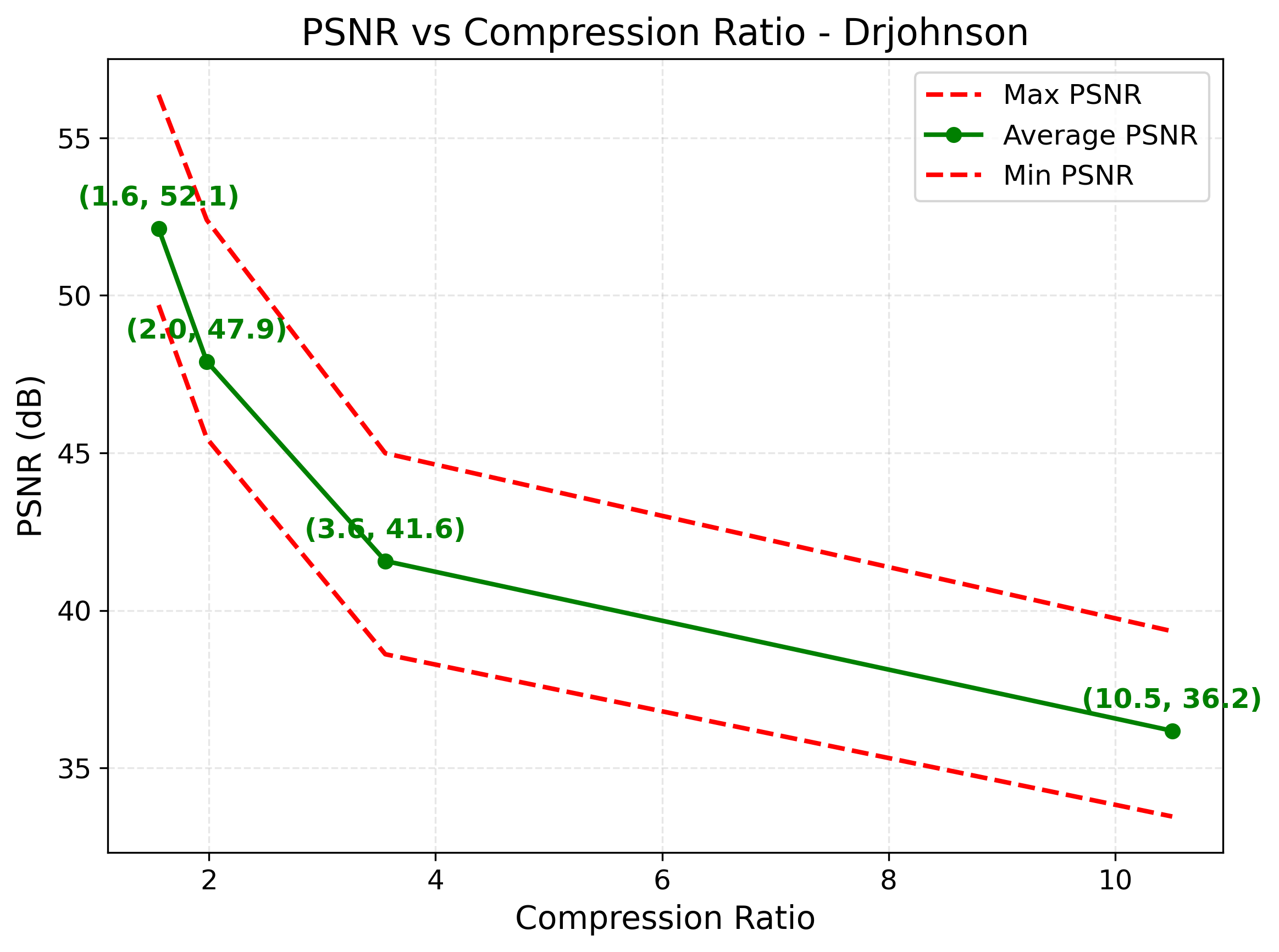}
        \caption{Effect of compression ratio on rendering quality for the \textit{Drjohnson} scene.}
    \end{minipage}
\end{figure}

\begin{figure}[H]
    \centering
    \begin{minipage}{0.44\textwidth}
        \centering
        \includegraphics[width=\linewidth]{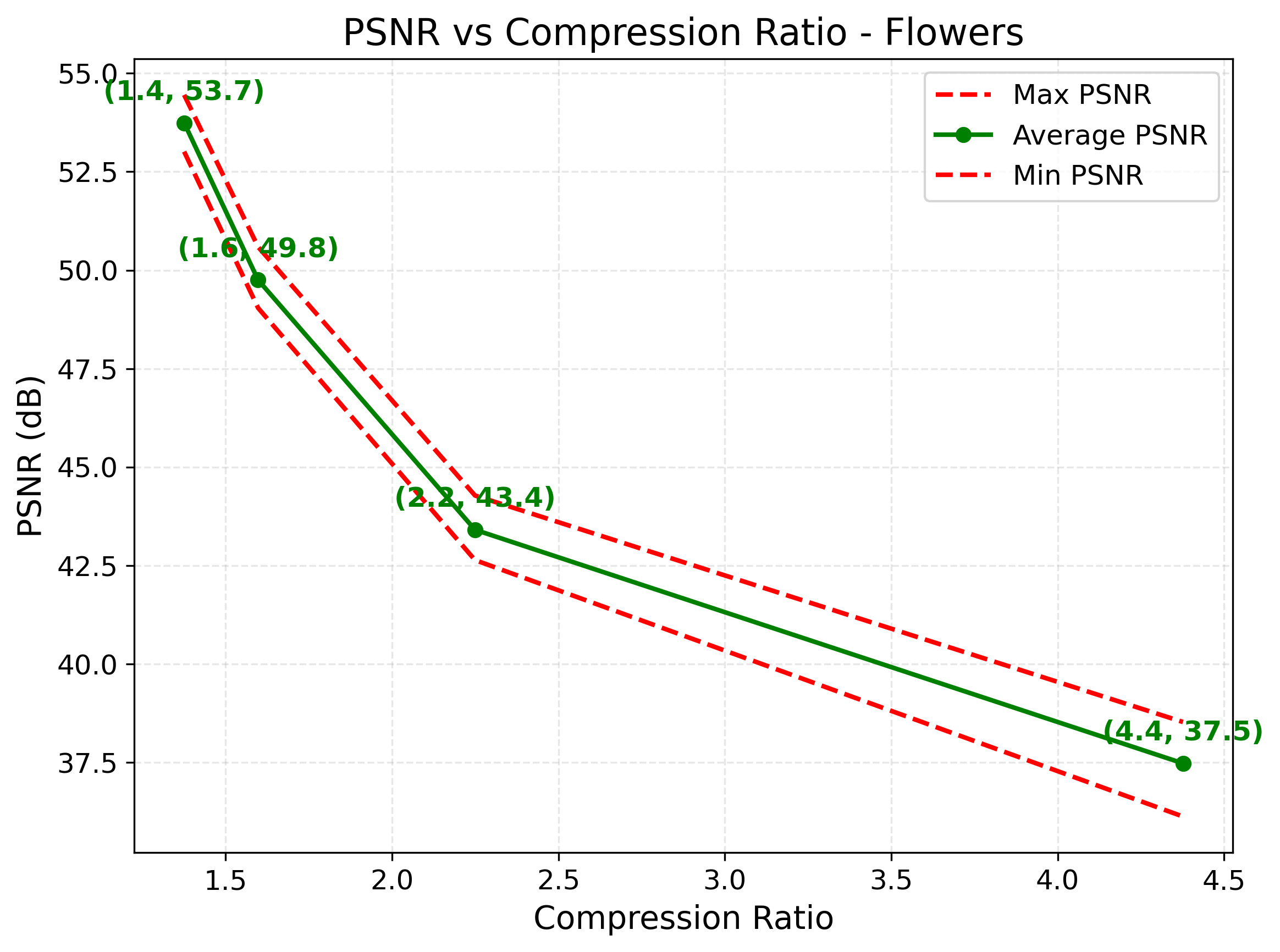}
        \caption{Effect of compression ratio on rendering quality for the \textit{Flowers} scene.}
    \end{minipage}
    \hfill
    \begin{minipage}{0.44\textwidth}
        \centering
        \includegraphics[width=\linewidth]{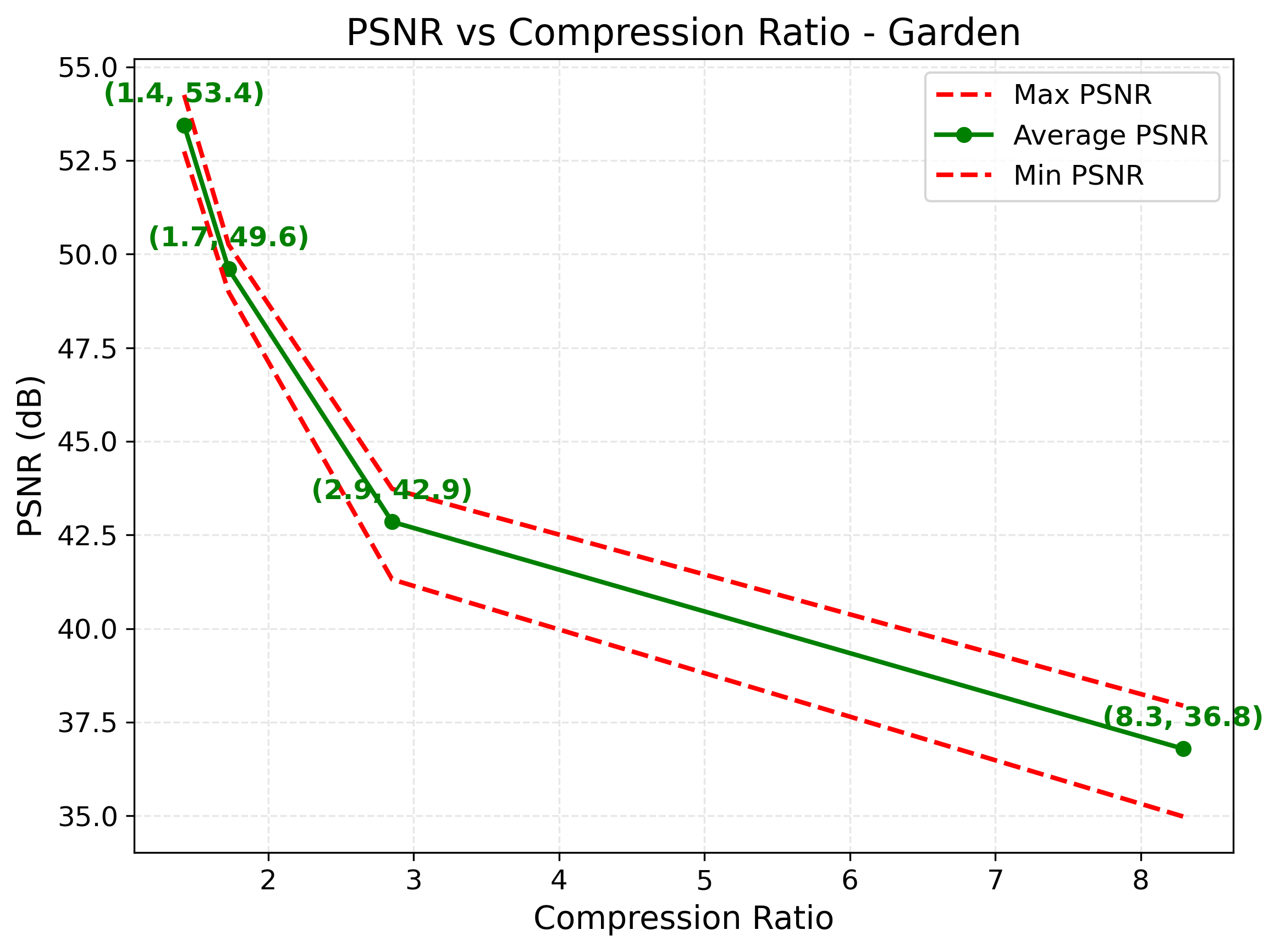}
        \caption{Effect of compression ratio on rendering quality for the \textit{Garden} scene.}
    \end{minipage}
\end{figure}

\begin{figure}[H]
    \centering
    \begin{minipage}{0.44\textwidth}
        \centering
        \includegraphics[width=\linewidth]{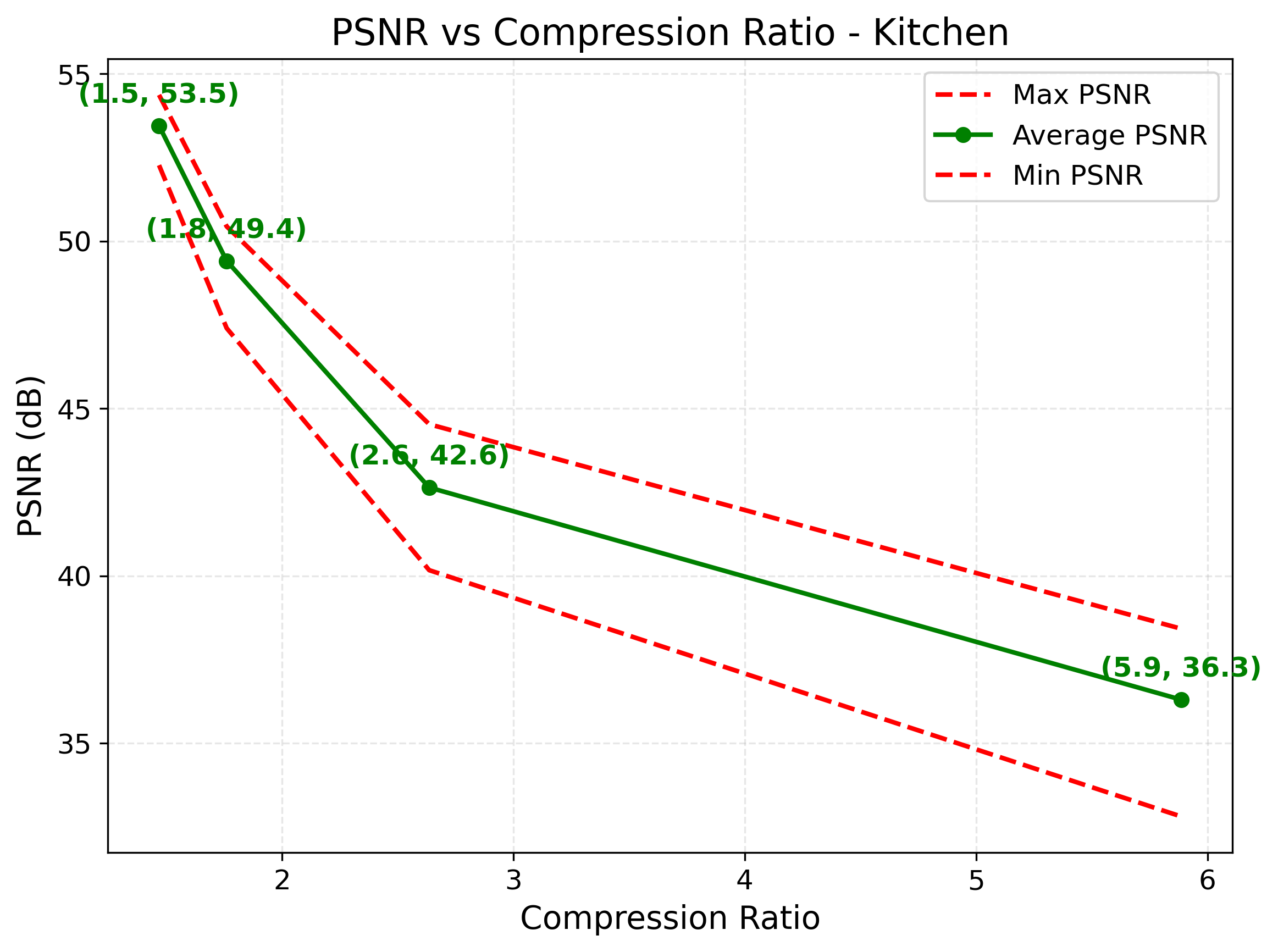}
        \caption{Effect of compression ratio on rendering quality for the \textit{Kitchen} scene.}
    \end{minipage}
    \hfill
    \begin{minipage}{0.44\textwidth}
        \centering
        \includegraphics[width=\linewidth]{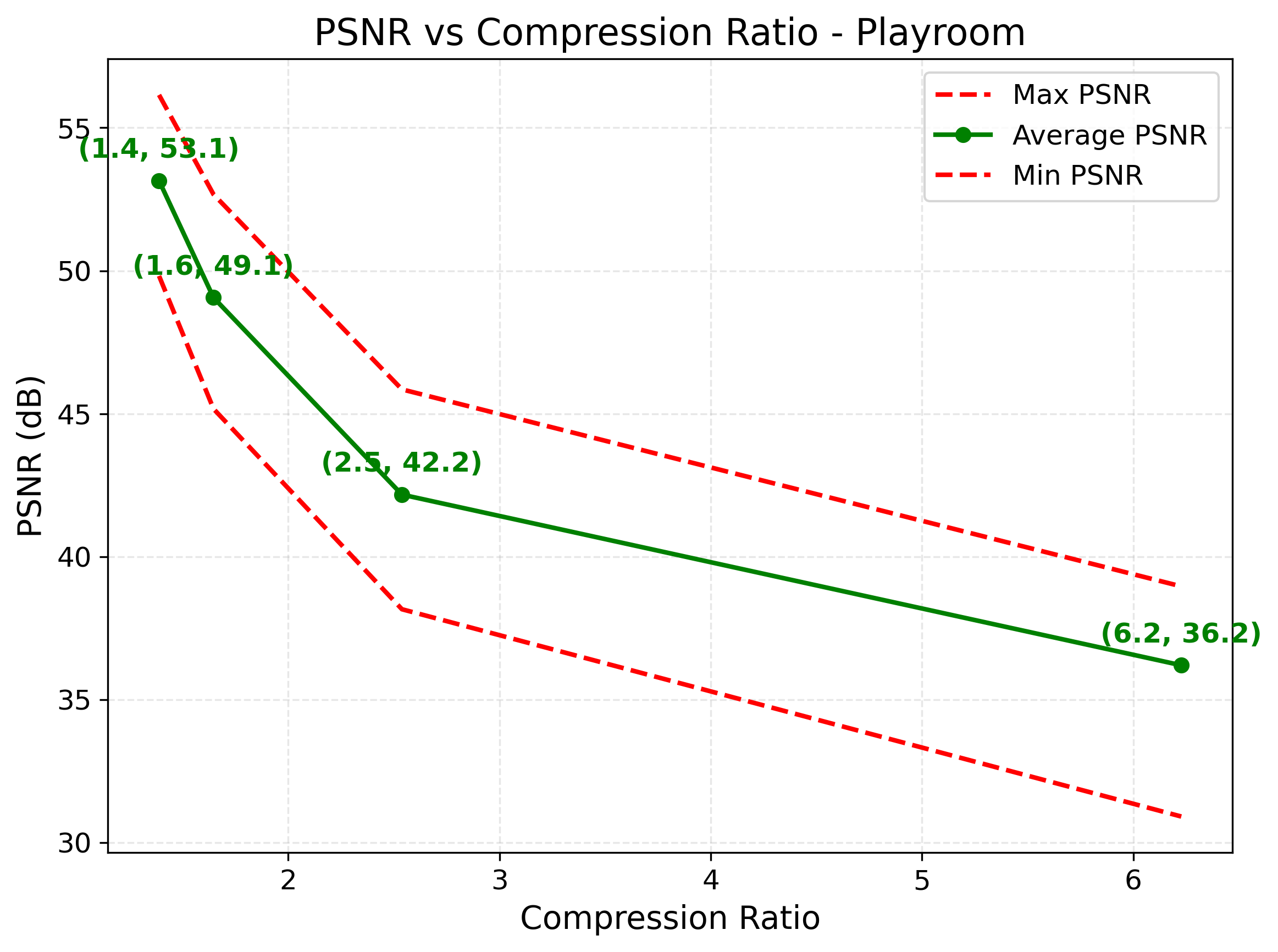}
        \caption{Effect of compression ratio on rendering quality for the \textit{Playroom} scene.}
    \end{minipage}
\end{figure}

\subsection{PixelGS+Comp}

\begin{figure}[H]
    \centering
    \begin{minipage}{0.44\textwidth}
        \centering
        \includegraphics[width=\linewidth]{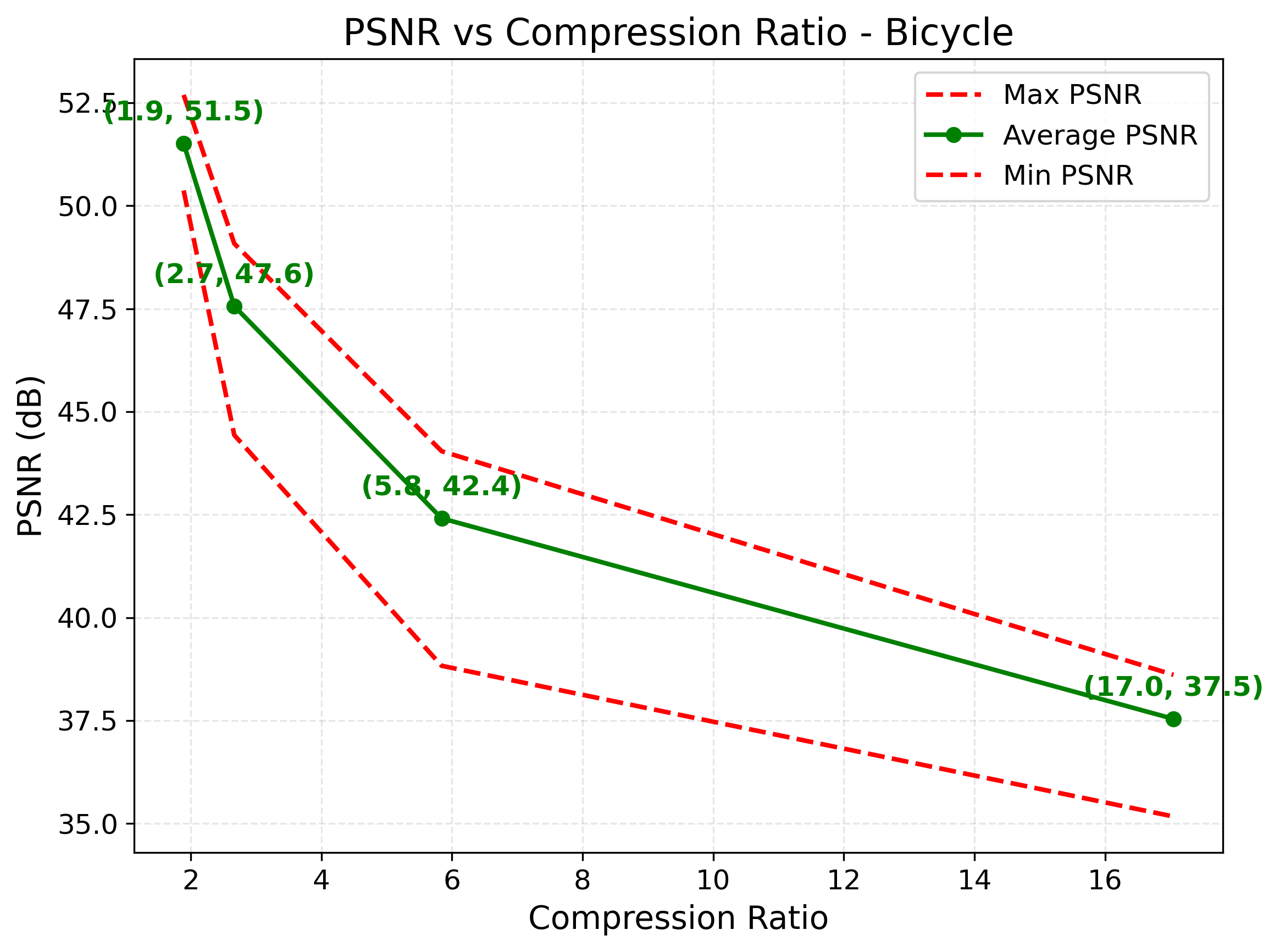}
        \caption{Effect of compression ratio on rendering quality for the \textit{Bicycle} scene.}
    \end{minipage}
    \hfill
    \begin{minipage}{0.44\textwidth}
        \centering
        \includegraphics[width=\linewidth]{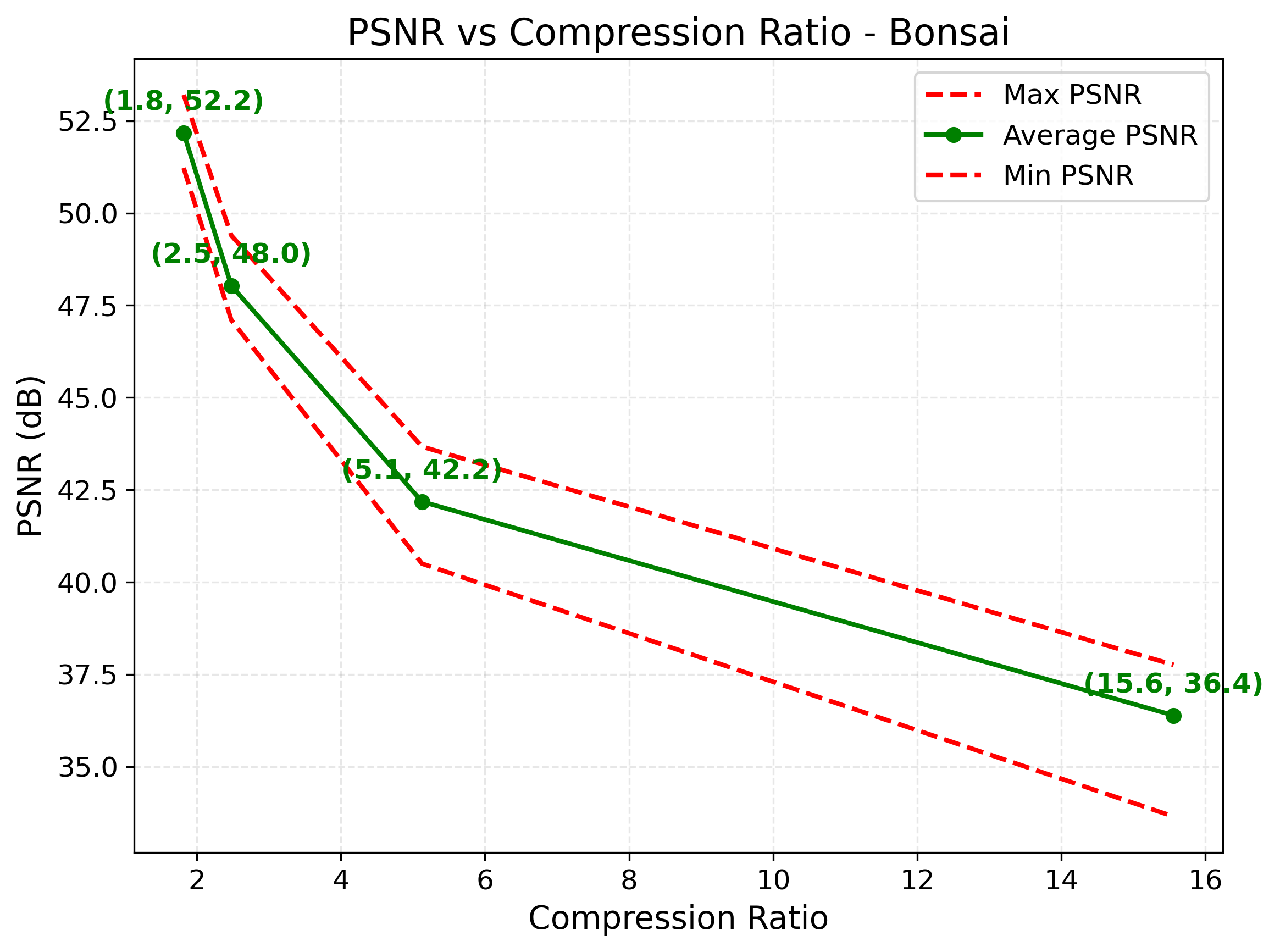}
        \caption{Effect of compression ratio on rendering quality for the \textit{Bonsai} scene.}
    \end{minipage}
\end{figure}

\begin{figure}[H]
    \centering
    \begin{minipage}{0.44\textwidth}
        \centering
        \includegraphics[width=\linewidth]{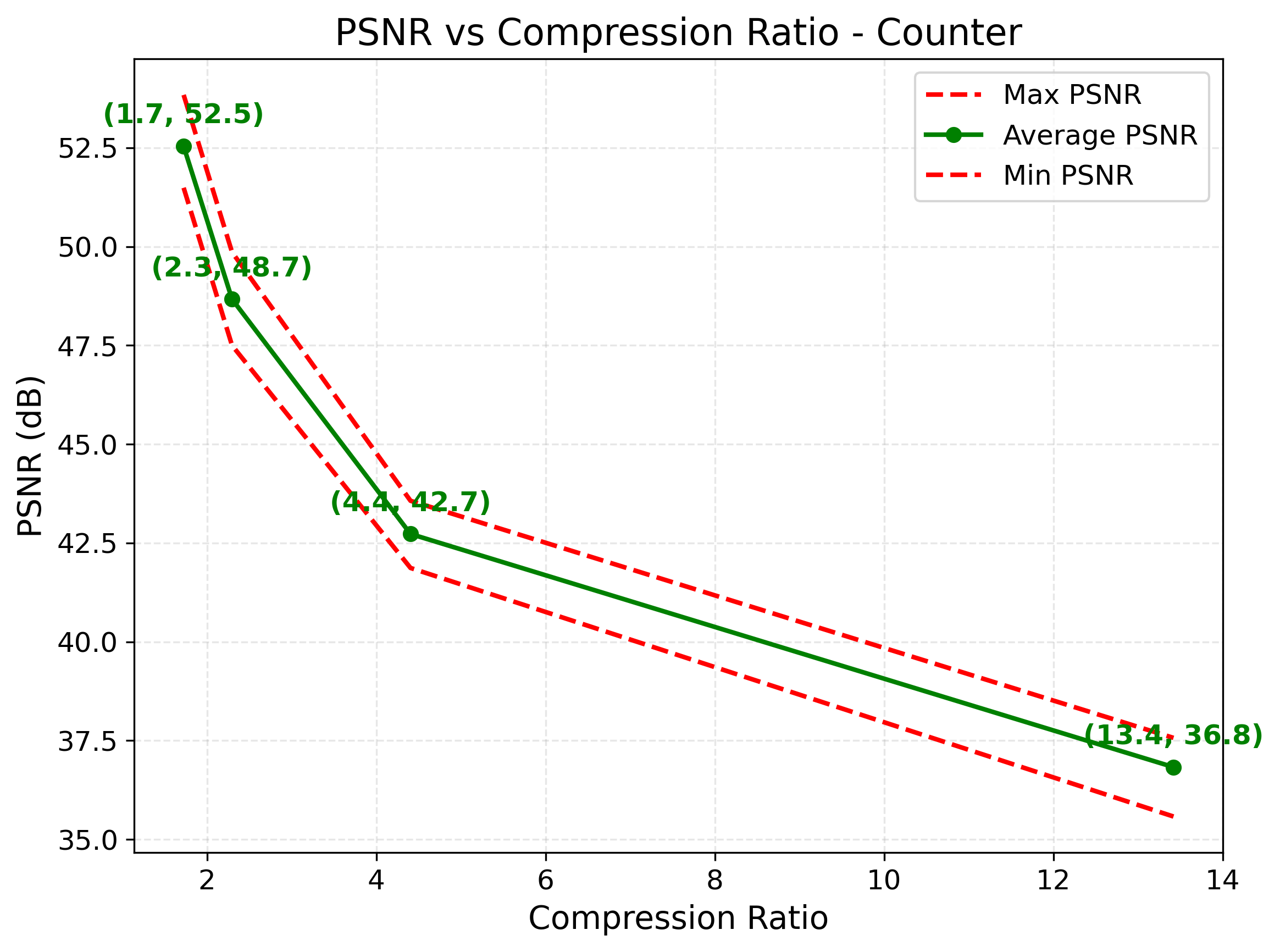}
        \caption{Effect of compression ratio on rendering quality for the \textit{Counter} scene.}
    \end{minipage}
    \hfill
    \begin{minipage}{0.44\textwidth}
        \centering
        \includegraphics[width=\linewidth]{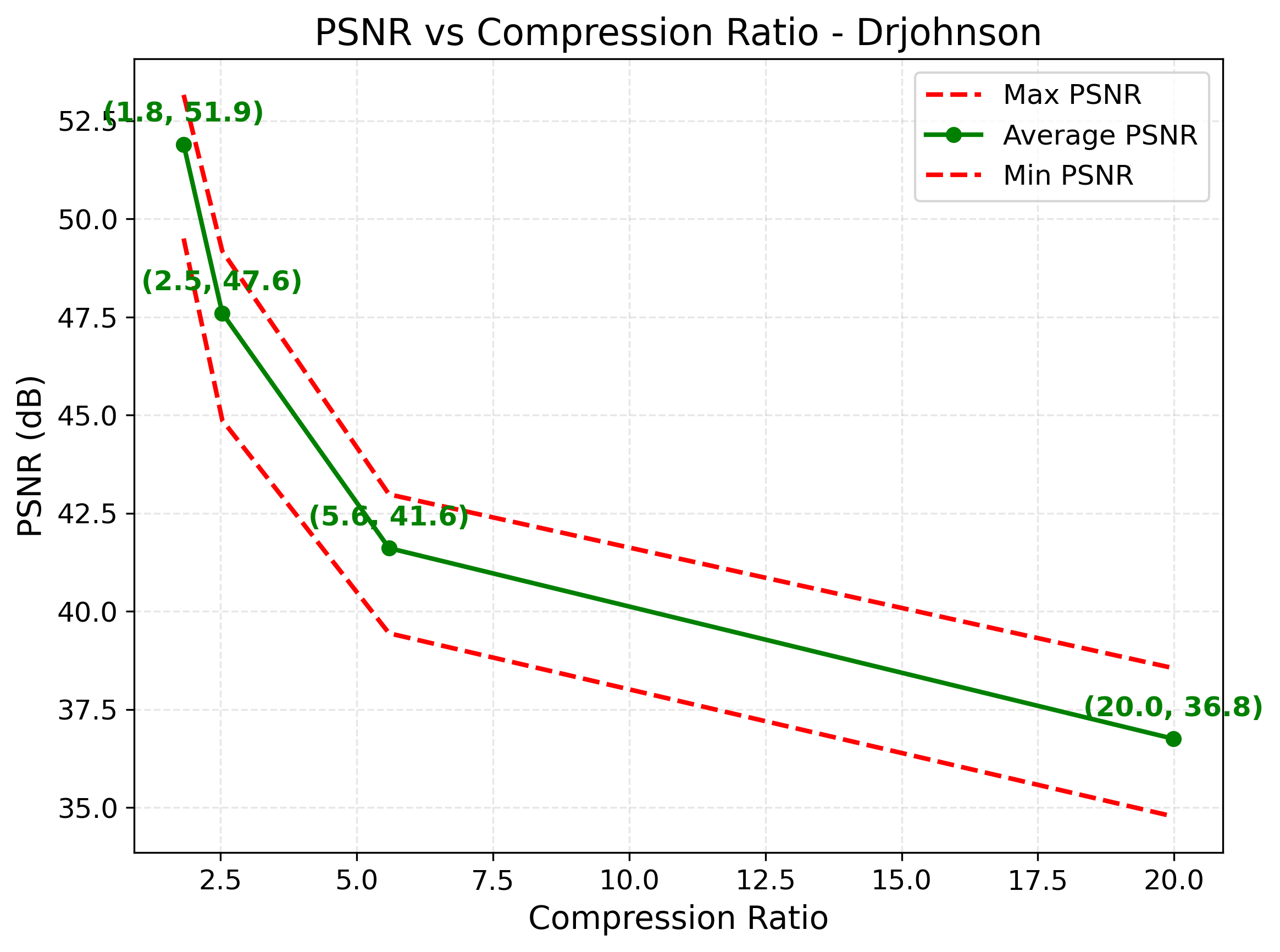}
        \caption{Effect of compression ratio on rendering quality for the \textit{Drjohnson} scene.}
    \end{minipage}
\end{figure}

\begin{figure}[H]
    \centering
    \begin{minipage}{0.44\textwidth}
        \centering
        \includegraphics[width=\linewidth]{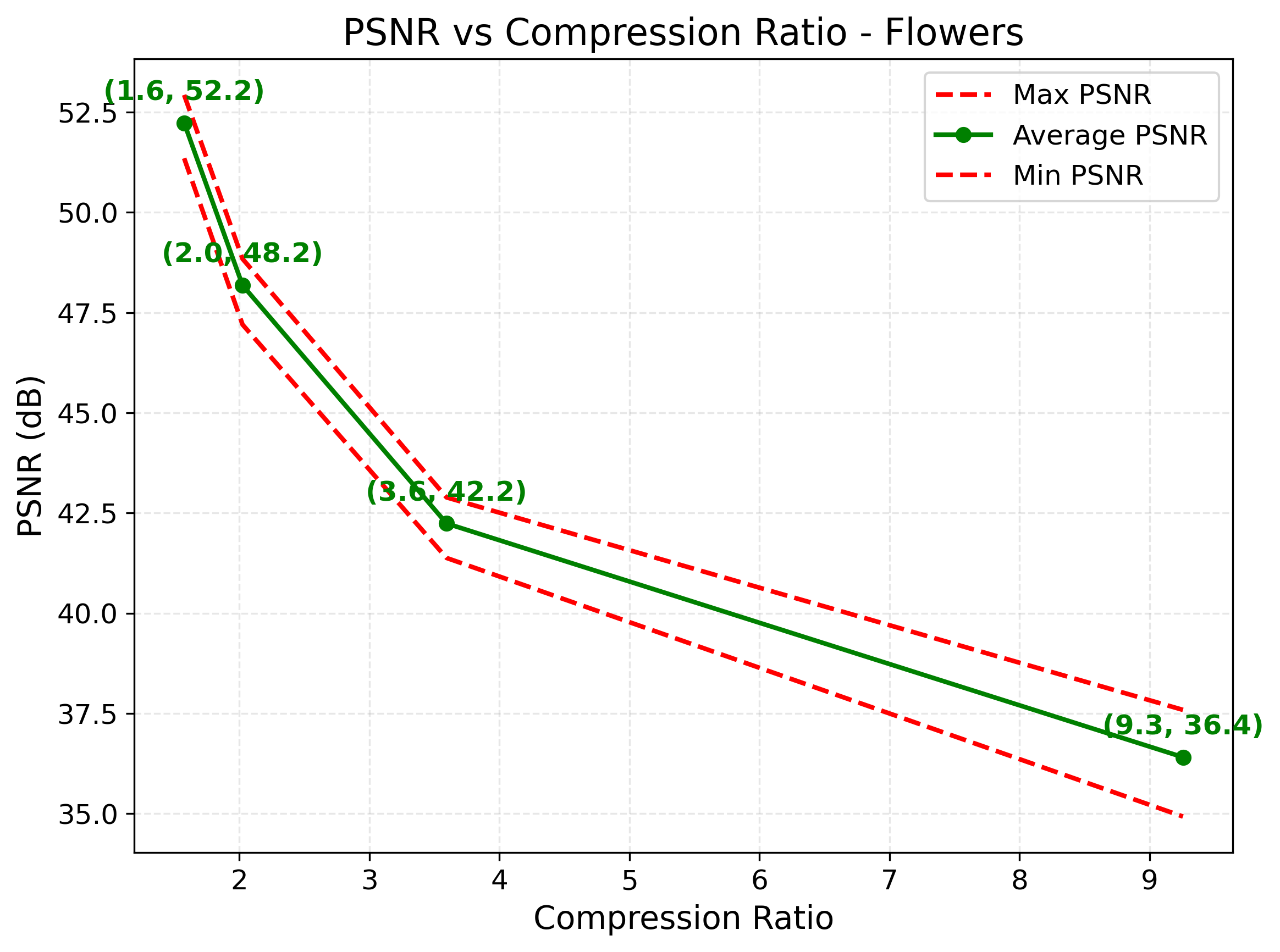}
        \caption{Effect of compression ratio on rendering quality for the \textit{Flowers} scene.}
    \end{minipage}
    \hfill
    \begin{minipage}{0.44\textwidth}
        \centering
        \includegraphics[width=\linewidth]{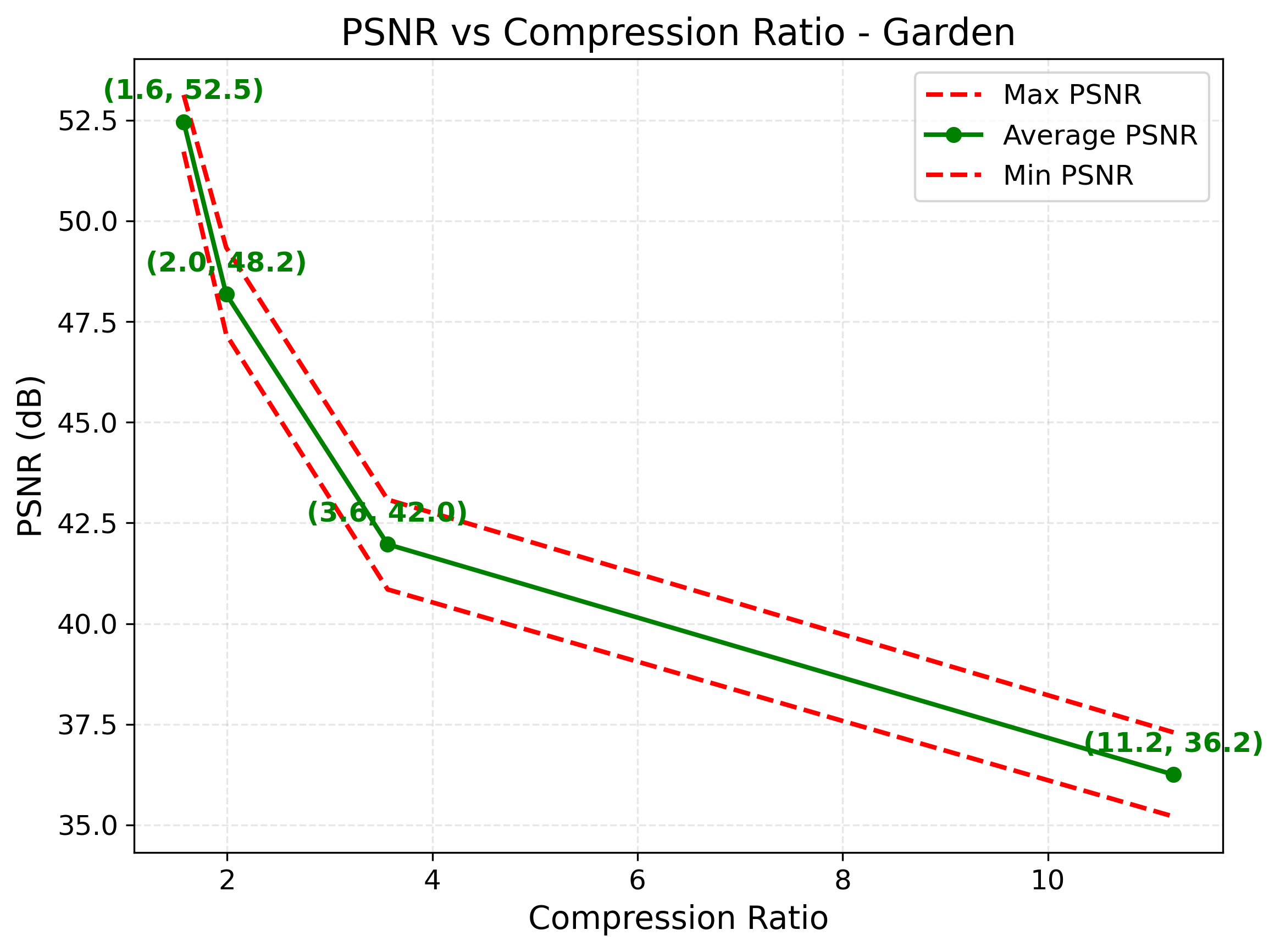}
        \caption{Effect of compression ratio on rendering quality for the \textit{Garden} scene.}
    \end{minipage}
\end{figure}

\begin{figure}[H]
    \centering
    \begin{minipage}{0.44\textwidth}
        \centering
        \includegraphics[width=\linewidth]{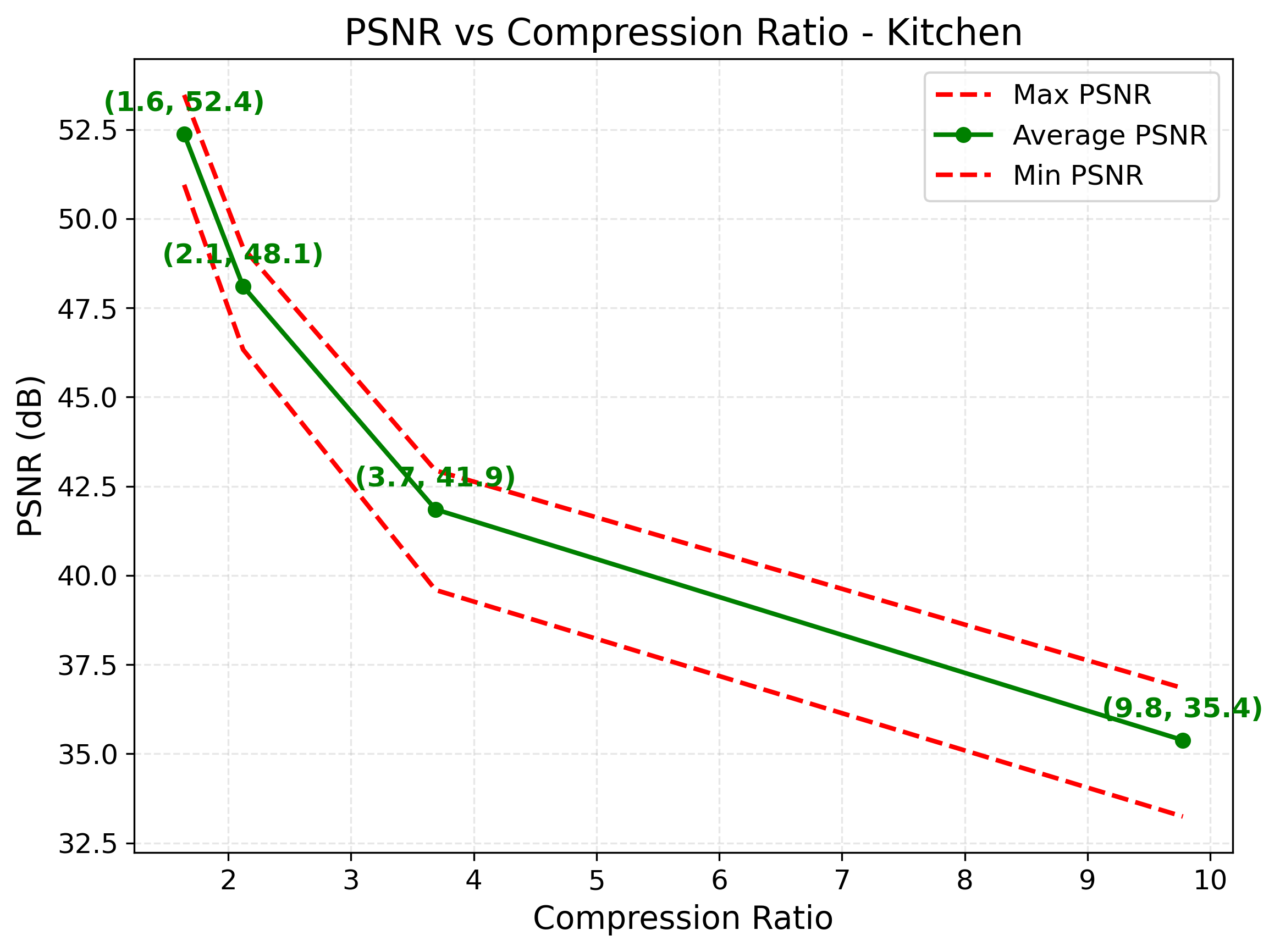}
        \caption{Effect of compression ratio on rendering quality for the \textit{Kitchen} scene.}
    \end{minipage}
    \hfill
    \begin{minipage}{0.44\textwidth}
        \centering
        \includegraphics[width=\linewidth]{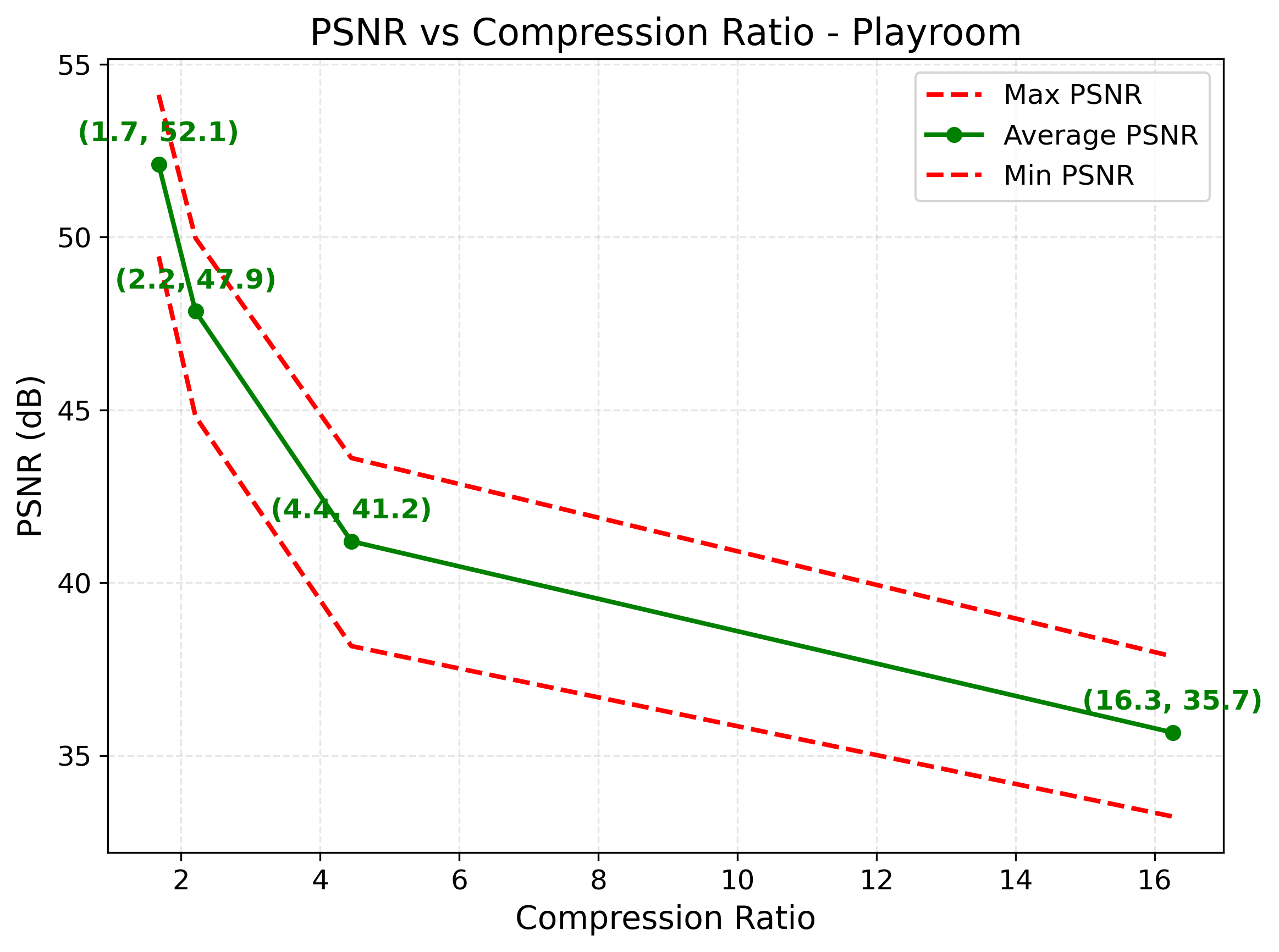}
        \caption{Effect of compression ratio on rendering quality for the \textit{Playroom} scene.}
    \end{minipage}
\end{figure}

\section{Rendering Speed vs.\ Compression Ratio: Cross-Scene and Cross-Method Evaluation}

Here we present, for every test scene and all three methods, the measured rendering throughput (FPS) as a function of compression ratio.

\subsection{3DGS+Comp}

\begin{figure}[H]
\centering
\begin{minipage}[b]{0.44\linewidth}
  \centering
  \includegraphics[width=\linewidth]{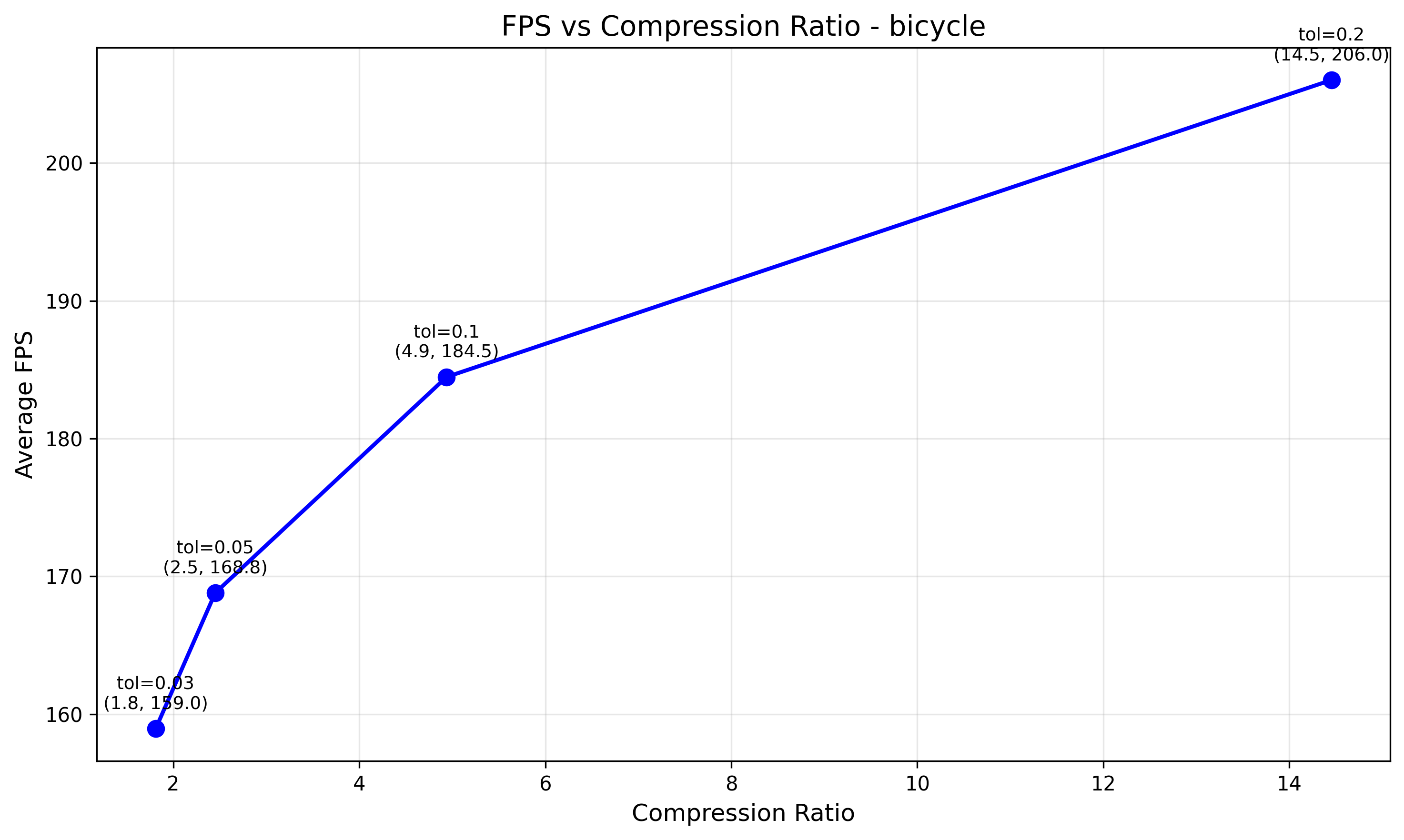}
  \captionsetup{skip=3pt}
  \caption*{Bicycle}
\end{minipage}\hfill
\begin{minipage}[b]{0.44\linewidth}
  \centering
  \includegraphics[width=\linewidth]{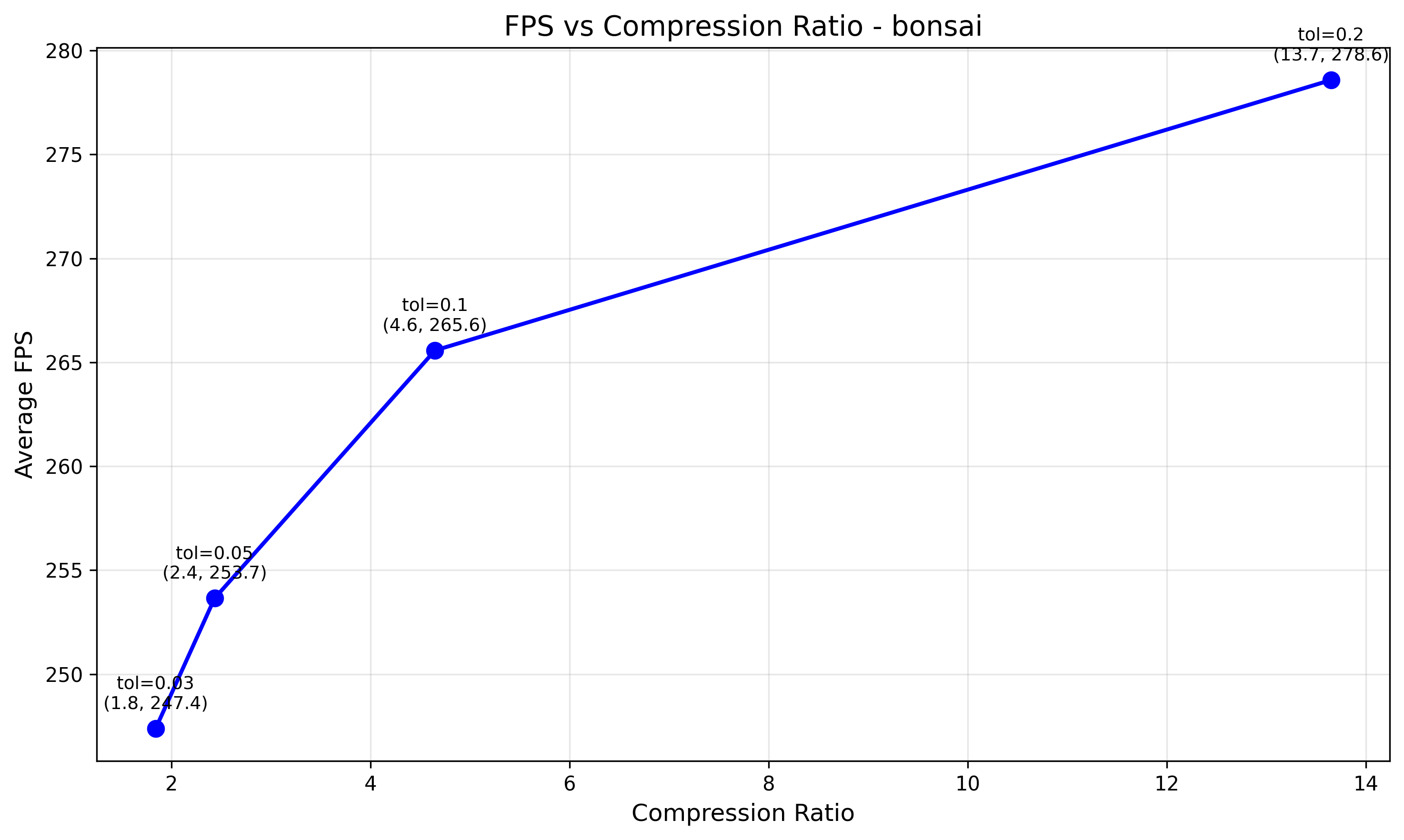}
  \captionsetup{skip=3pt}
  \caption*{Bonsai}
\end{minipage}
\end{figure}

\begin{figure}[H]
\centering
\begin{minipage}[b]{0.44\linewidth}
  \centering
  \includegraphics[width=\linewidth]{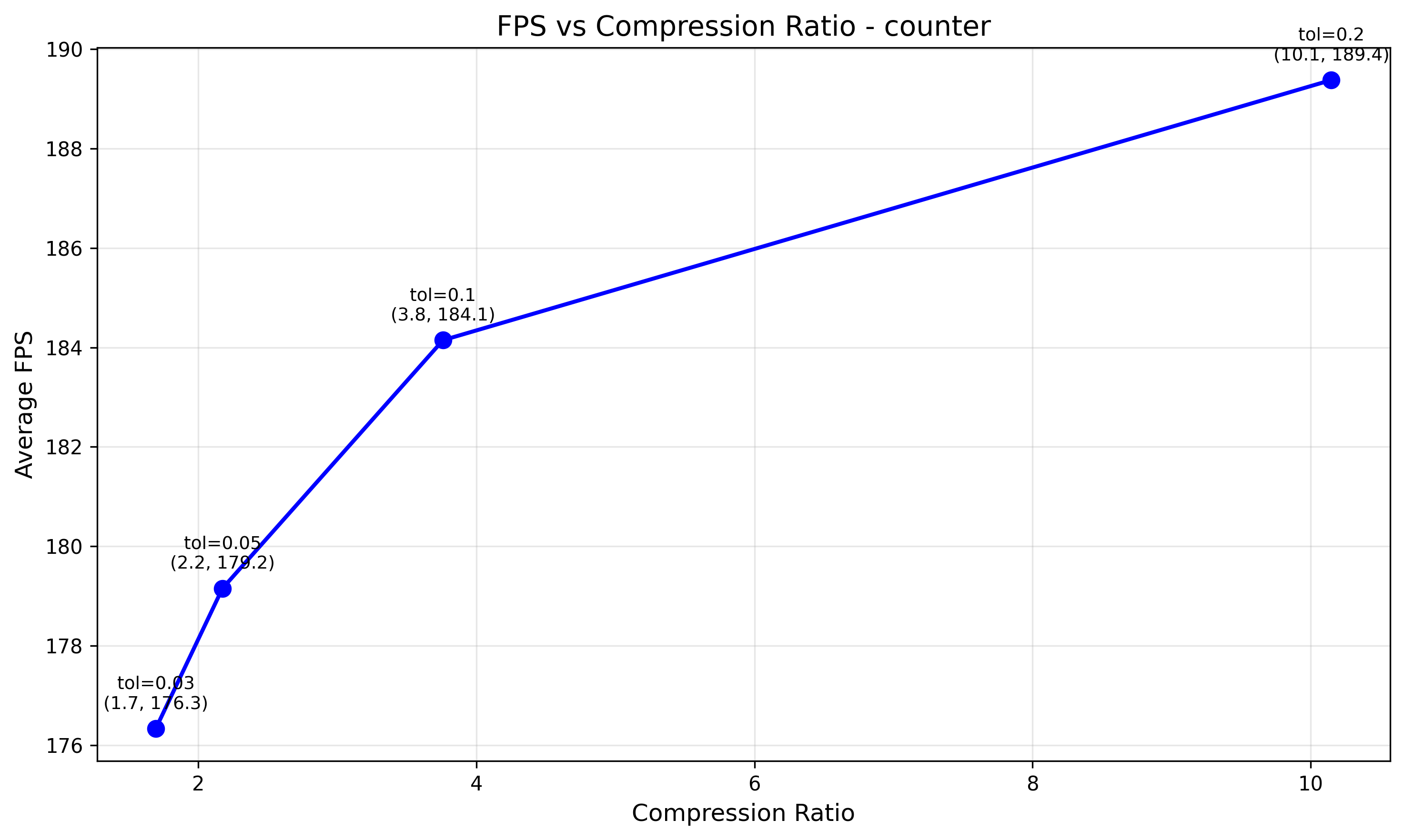}
  \captionsetup{skip=3pt}
  \caption*{Counter}
\end{minipage}\hfill
\begin{minipage}[b]{0.44\linewidth}
  \centering
  \includegraphics[width=\linewidth]{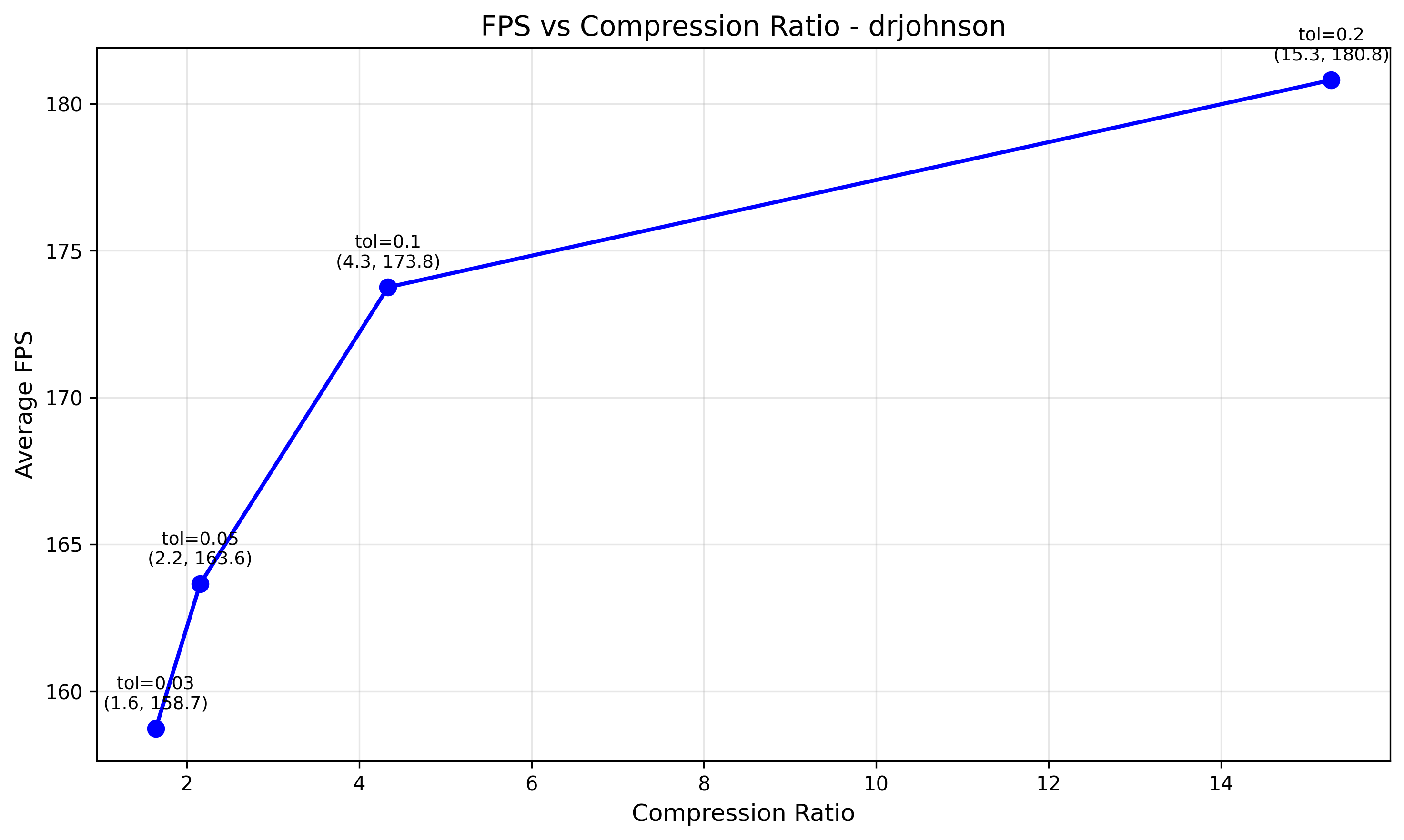}
  \captionsetup{skip=3pt}
  \caption*{Drjohnson}
\end{minipage}
\end{figure}

\begin{figure}[H]
\centering
\begin{minipage}[b]{0.44\linewidth}
  \centering
  \includegraphics[width=\linewidth]{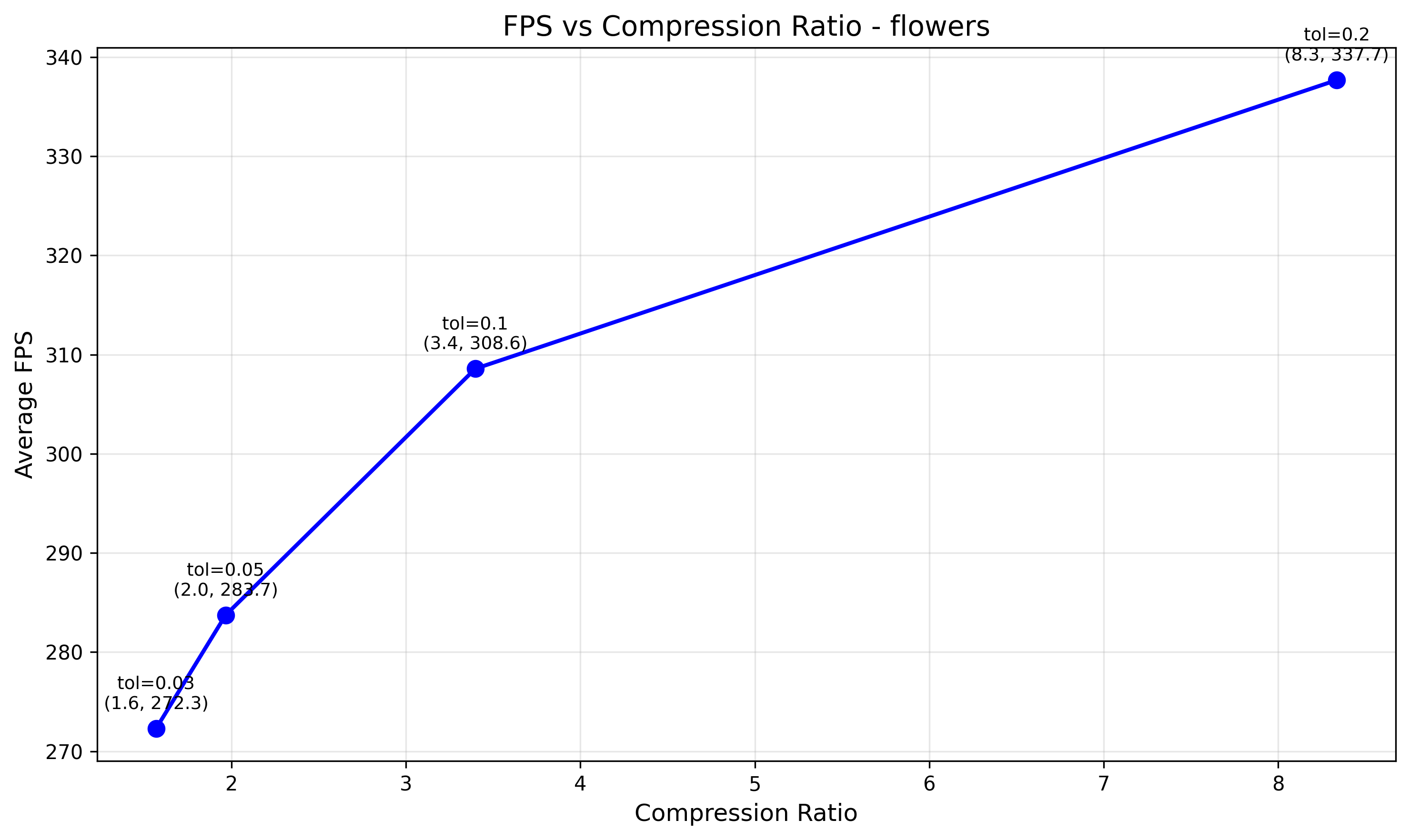}
  \captionsetup{skip=3pt}
  \caption*{Flowers}
\end{minipage}\hfill
\begin{minipage}[b]{0.44\linewidth}
  \centering
  \includegraphics[width=\linewidth]{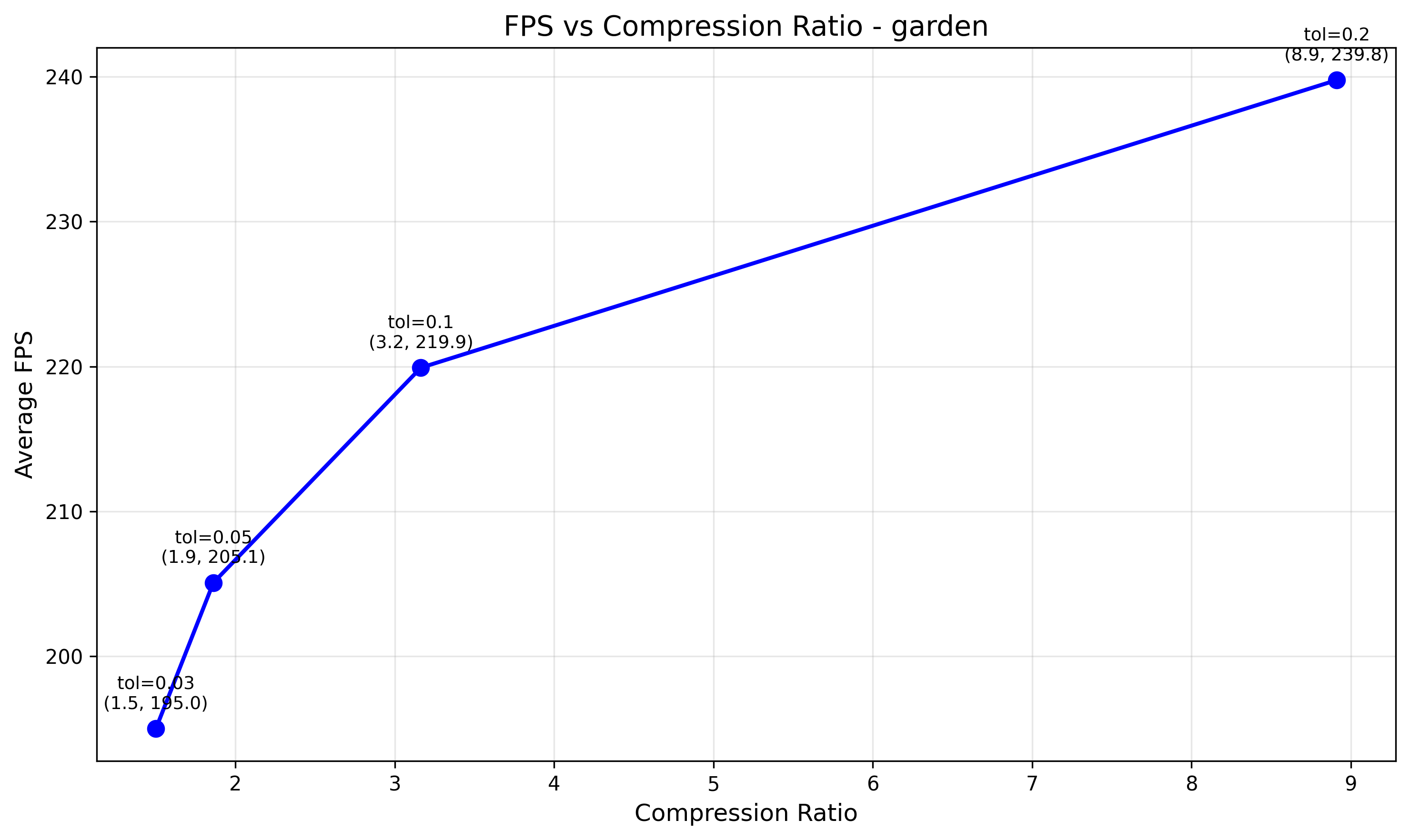}
  \captionsetup{skip=3pt}
  \caption*{Garden}
\end{minipage}
\end{figure}

\begin{figure}[H]
\centering
\begin{minipage}[b]{0.44\linewidth}
  \centering
  \includegraphics[width=\linewidth]{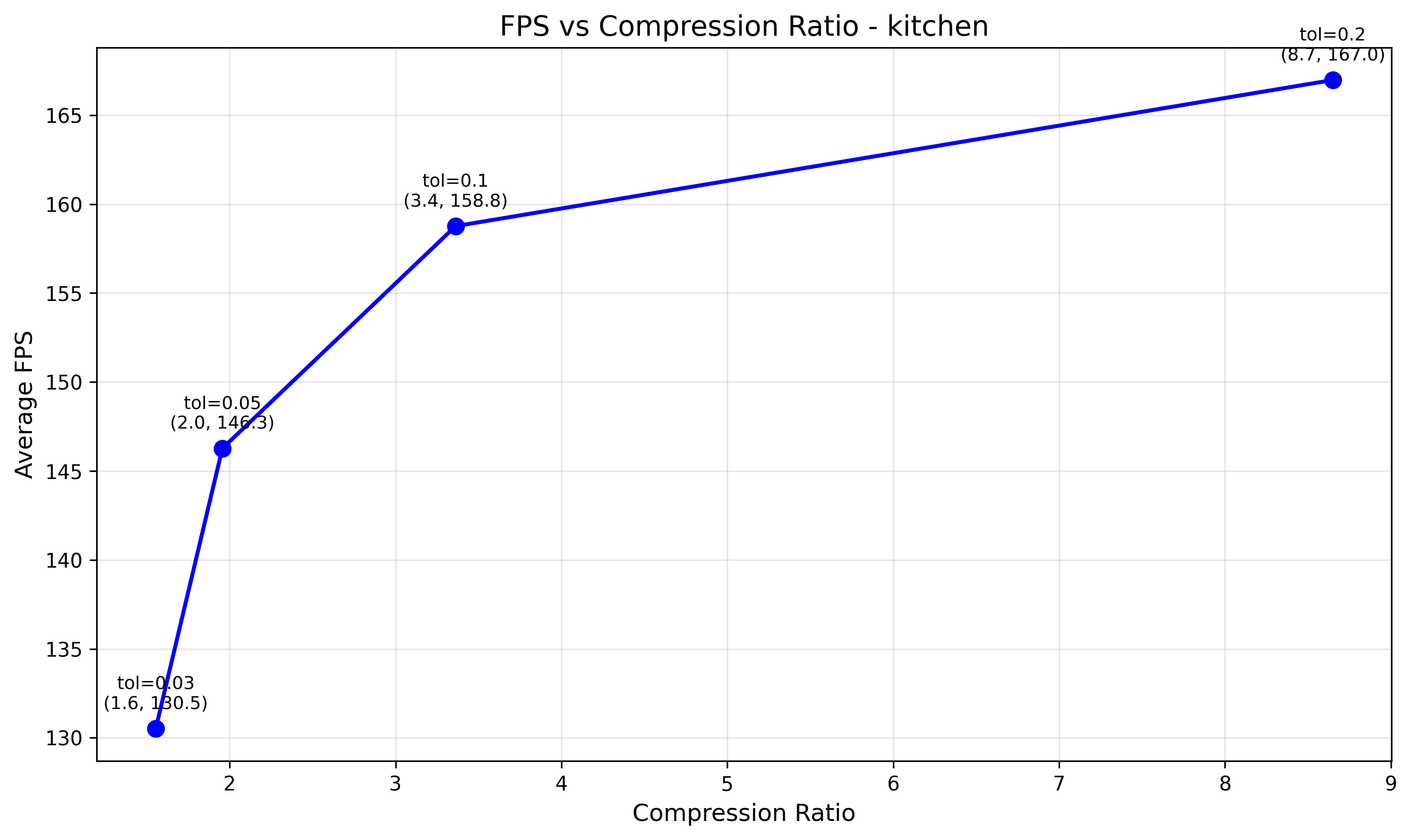}
  \captionsetup{skip=3pt}
  \caption*{Kitchen}
\end{minipage}\hfill
\begin{minipage}[b]{0.44\linewidth}
  \centering
  \includegraphics[width=\linewidth]{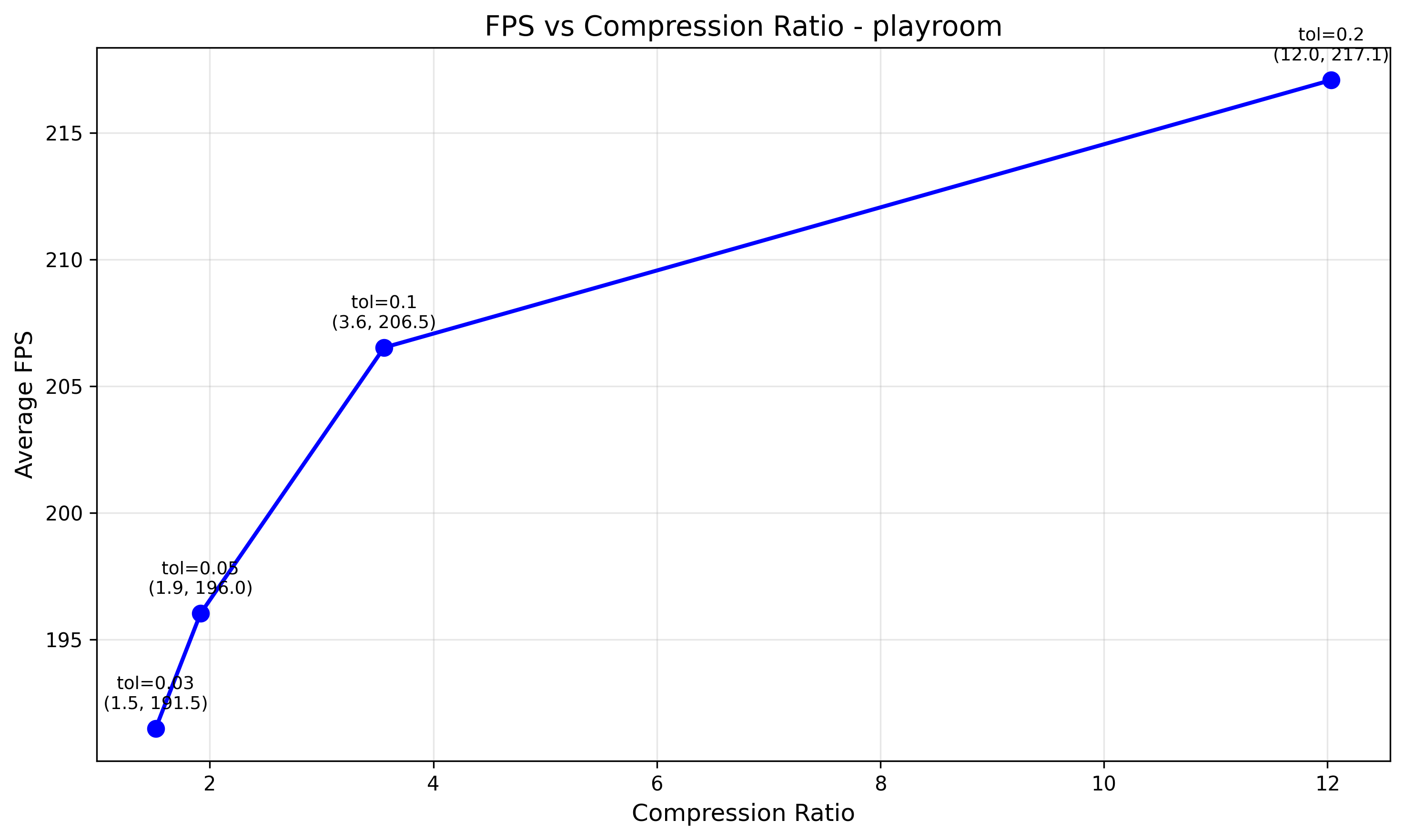}
  \captionsetup{skip=3pt}
  \caption*{Playroom}
\end{minipage}
\end{figure}

\subsection{MCMC+Comp}

\begin{figure}[H]
\centering
\begin{minipage}[b]{0.44\linewidth}
  \centering
  \includegraphics[width=\linewidth]{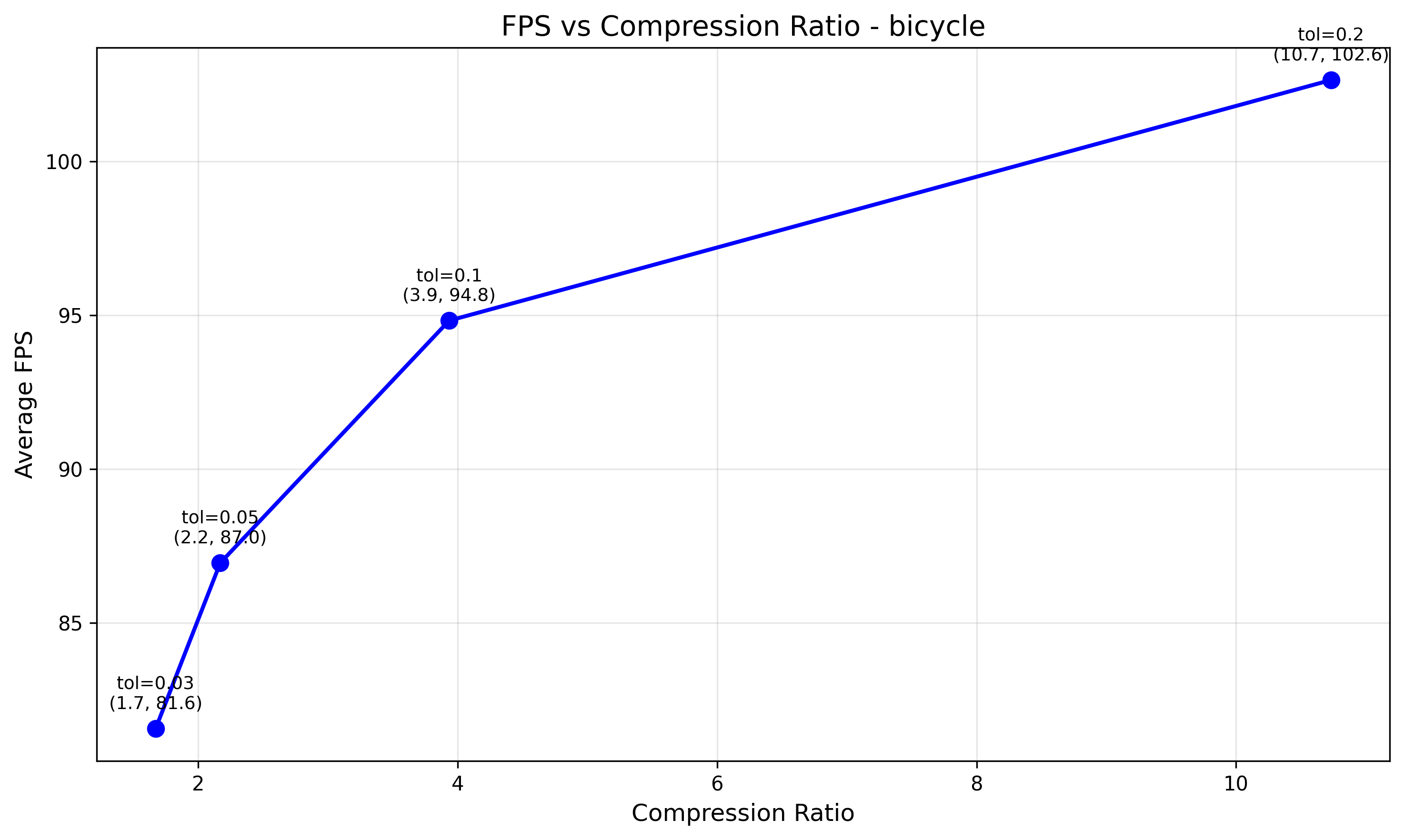}
  \captionsetup{skip=3pt}
  \caption*{Bicycle}
\end{minipage}\hfill
\begin{minipage}[b]{0.44\linewidth}
  \centering
  \includegraphics[width=\linewidth]{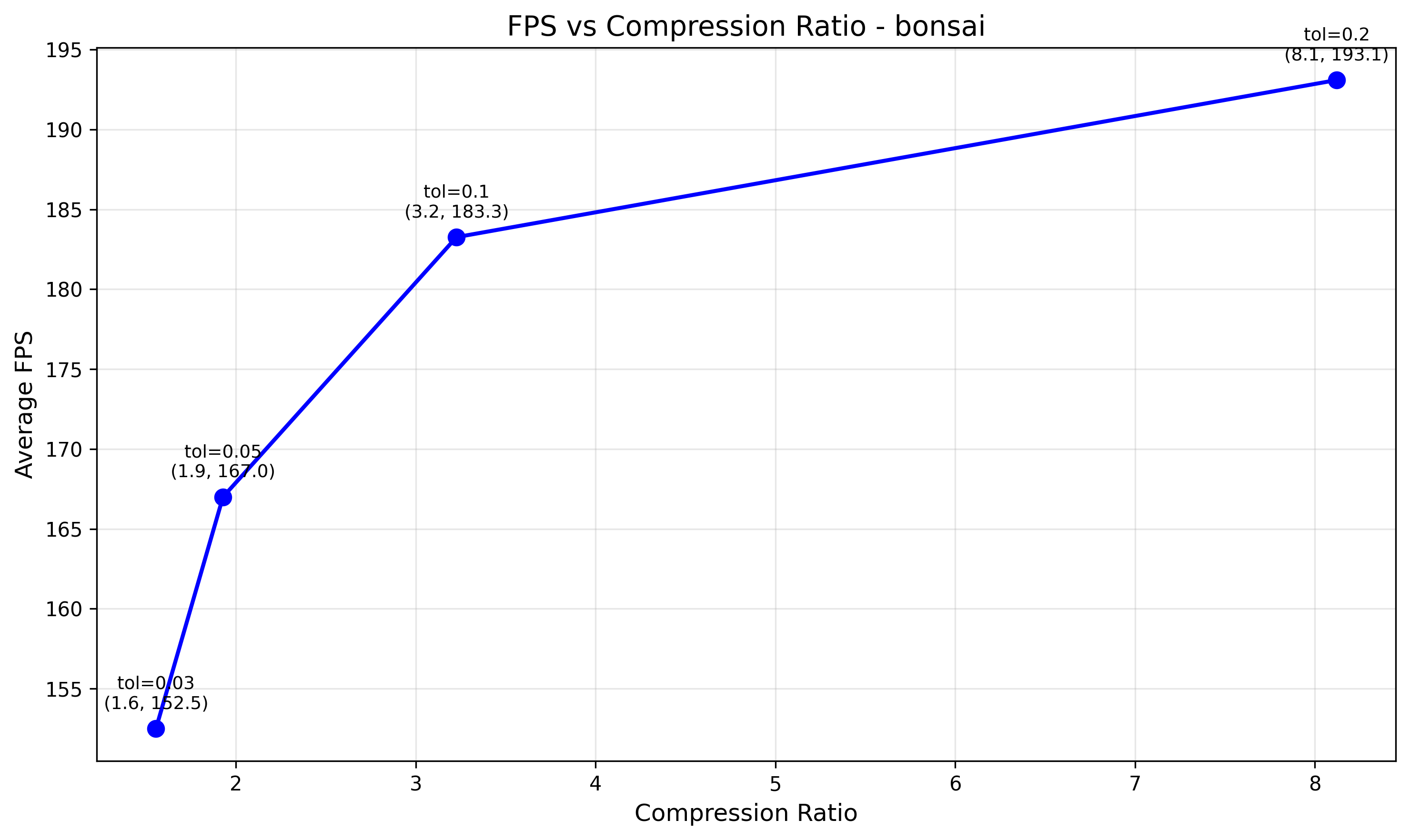}
  \captionsetup{skip=3pt}
  \caption*{Bonsai}
\end{minipage}
\end{figure}

\begin{figure}[H]
\centering
\begin{minipage}[b]{0.44\linewidth}
  \centering
  \includegraphics[width=\linewidth]{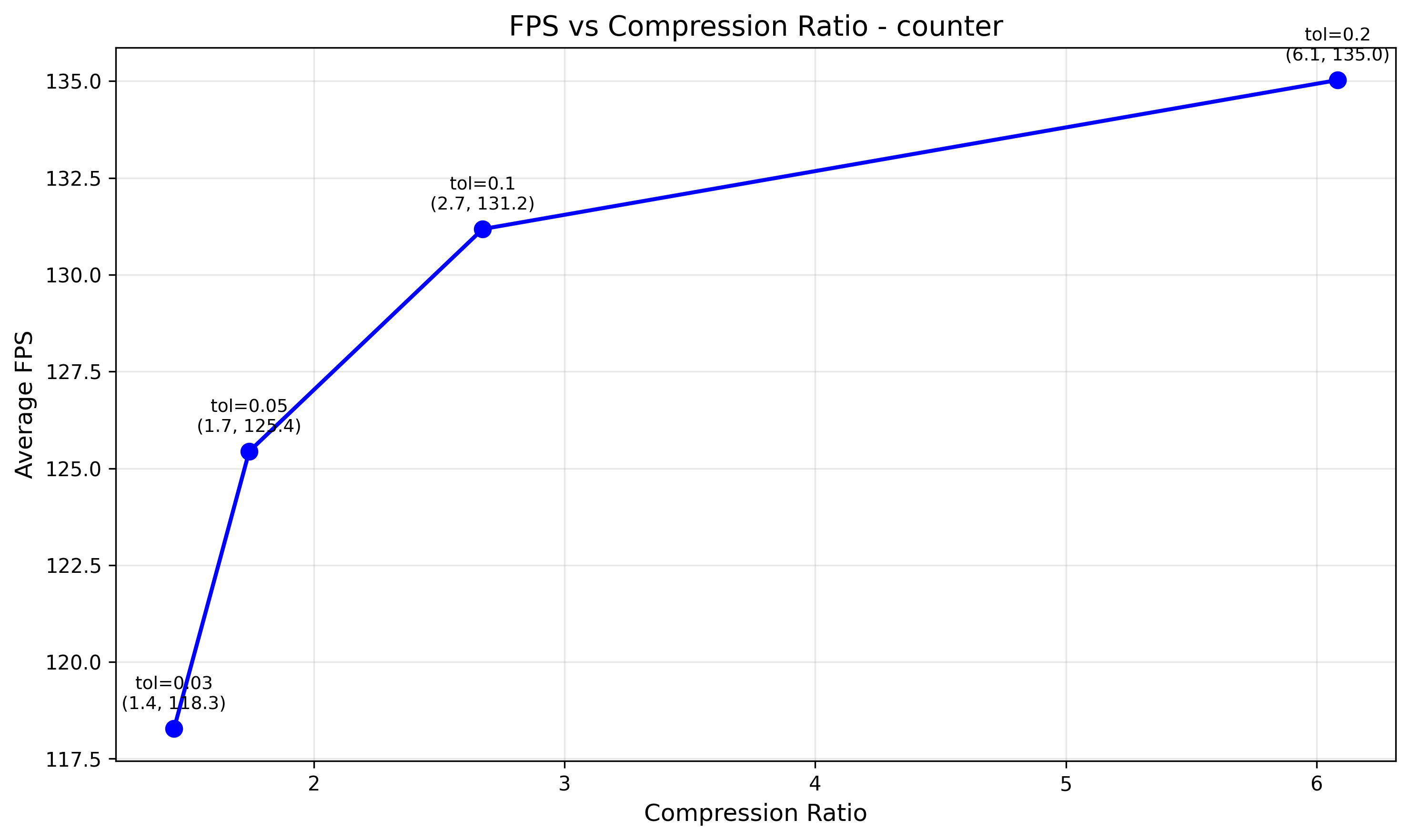}
  \captionsetup{skip=3pt}
  \caption*{Counter}
\end{minipage}\hfill
\begin{minipage}[b]{0.44\linewidth}
  \centering
  \includegraphics[width=\linewidth]{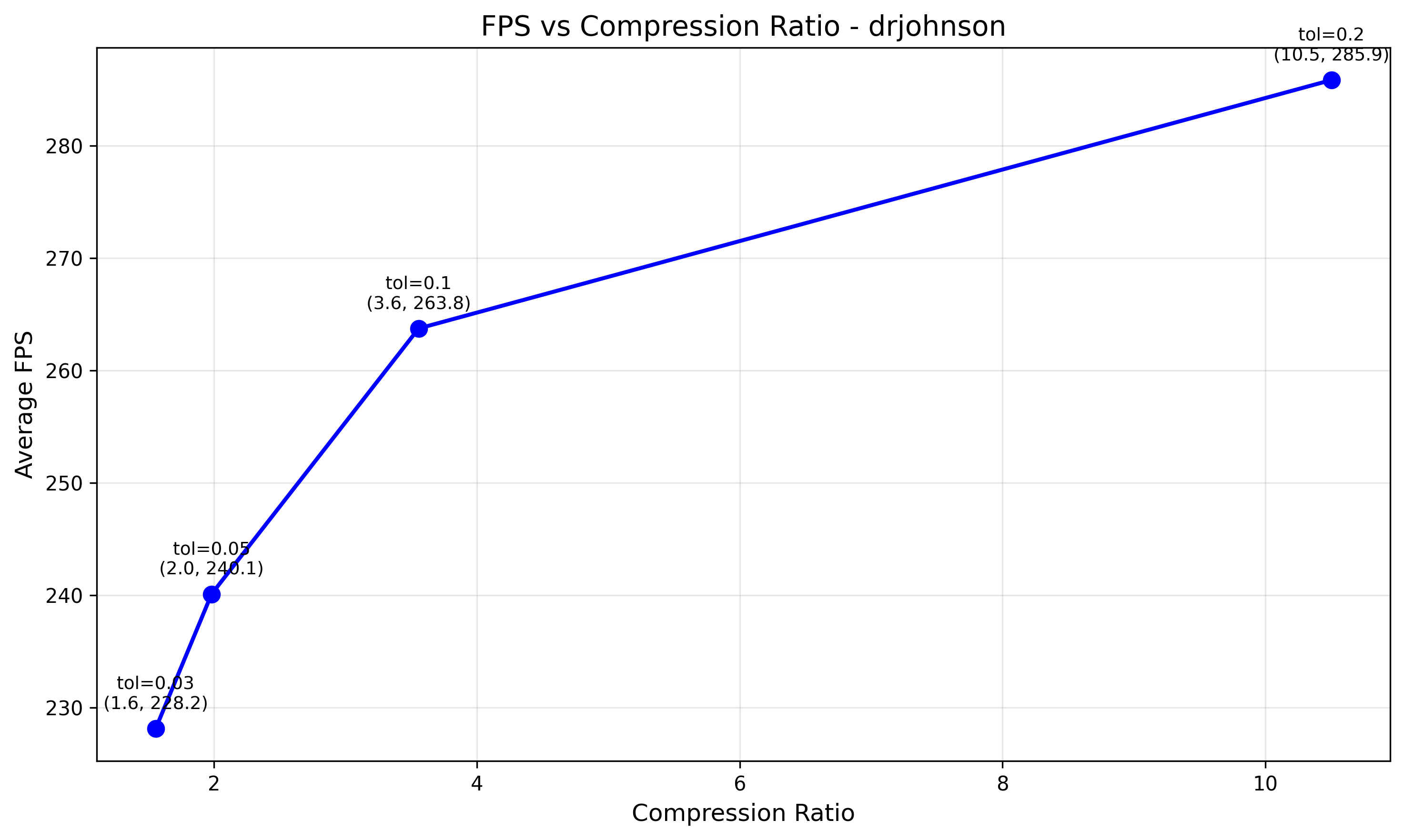}
  \captionsetup{skip=3pt}
  \caption*{Drjohnson}
\end{minipage}
\end{figure}

\begin{figure}[H]
\centering
\begin{minipage}[b]{0.44\linewidth}
  \centering
  \includegraphics[width=\linewidth]{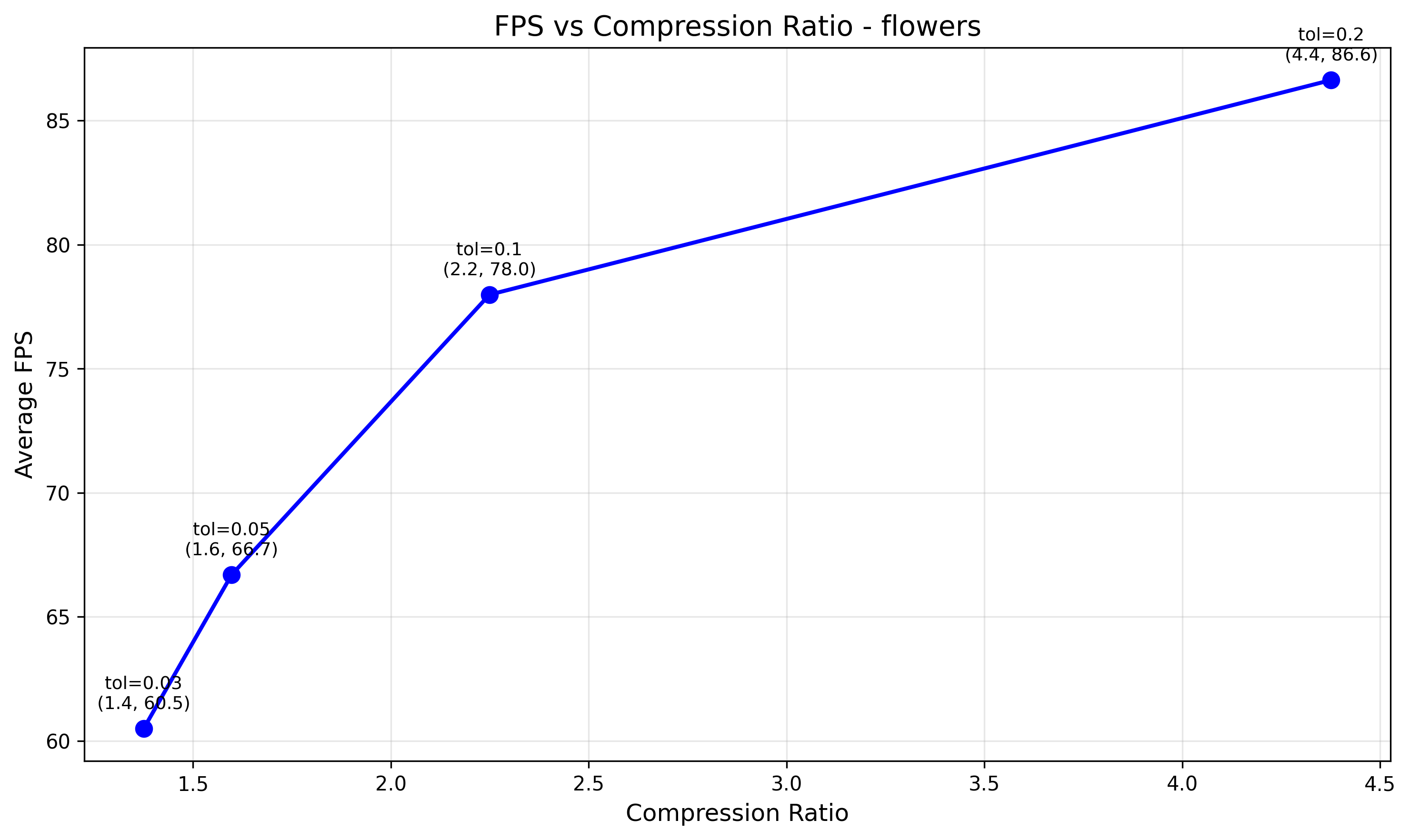}
  \captionsetup{skip=3pt}
  \caption*{Flowers}
\end{minipage}\hfill
\begin{minipage}[b]{0.44\linewidth}
  \centering
  \includegraphics[width=\linewidth]{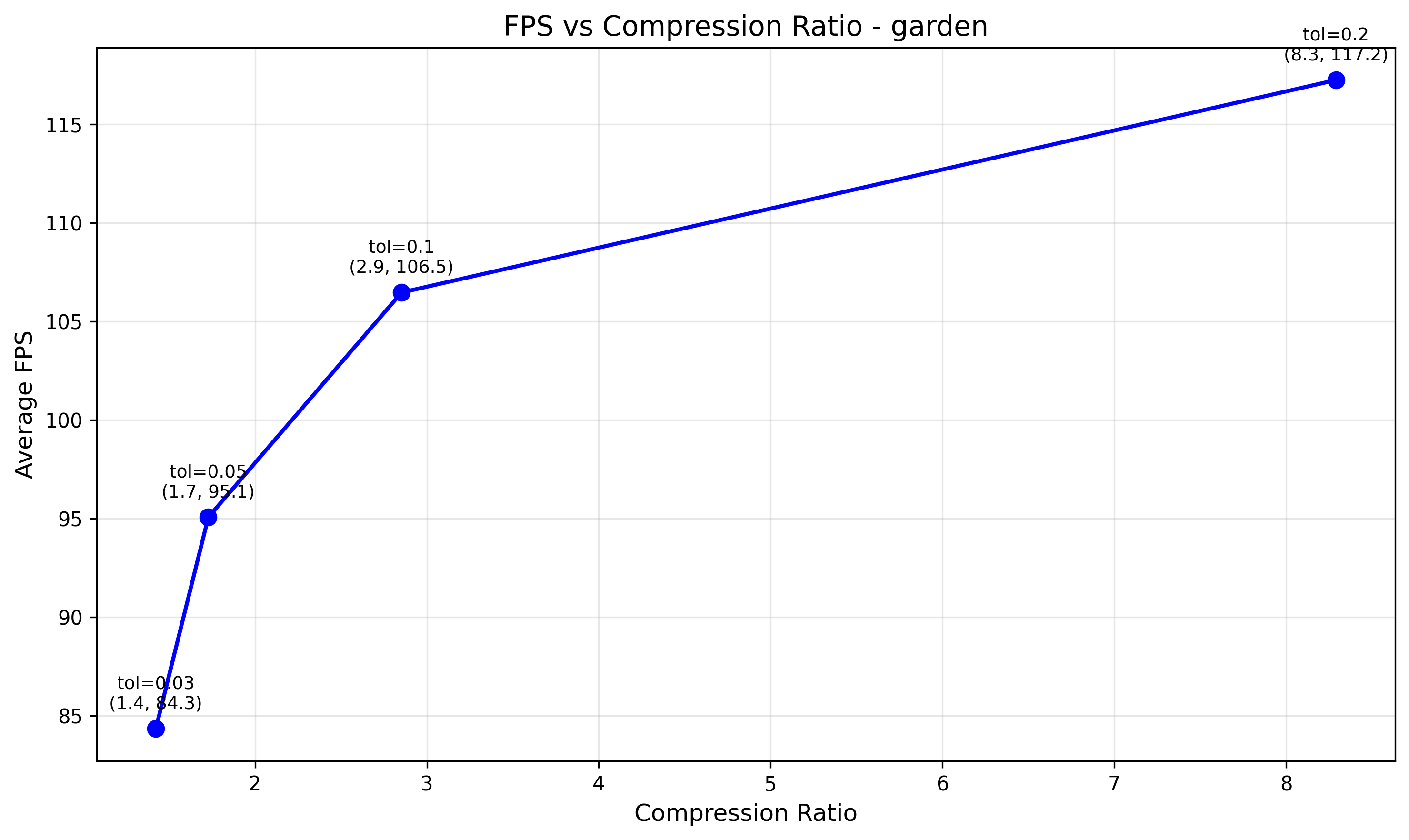}
  \captionsetup{skip=3pt}
  \caption*{Garden}
\end{minipage}
\end{figure}

\begin{figure}[H]
\centering
\begin{minipage}[b]{0.44\linewidth}
  \centering
  \includegraphics[width=\linewidth]{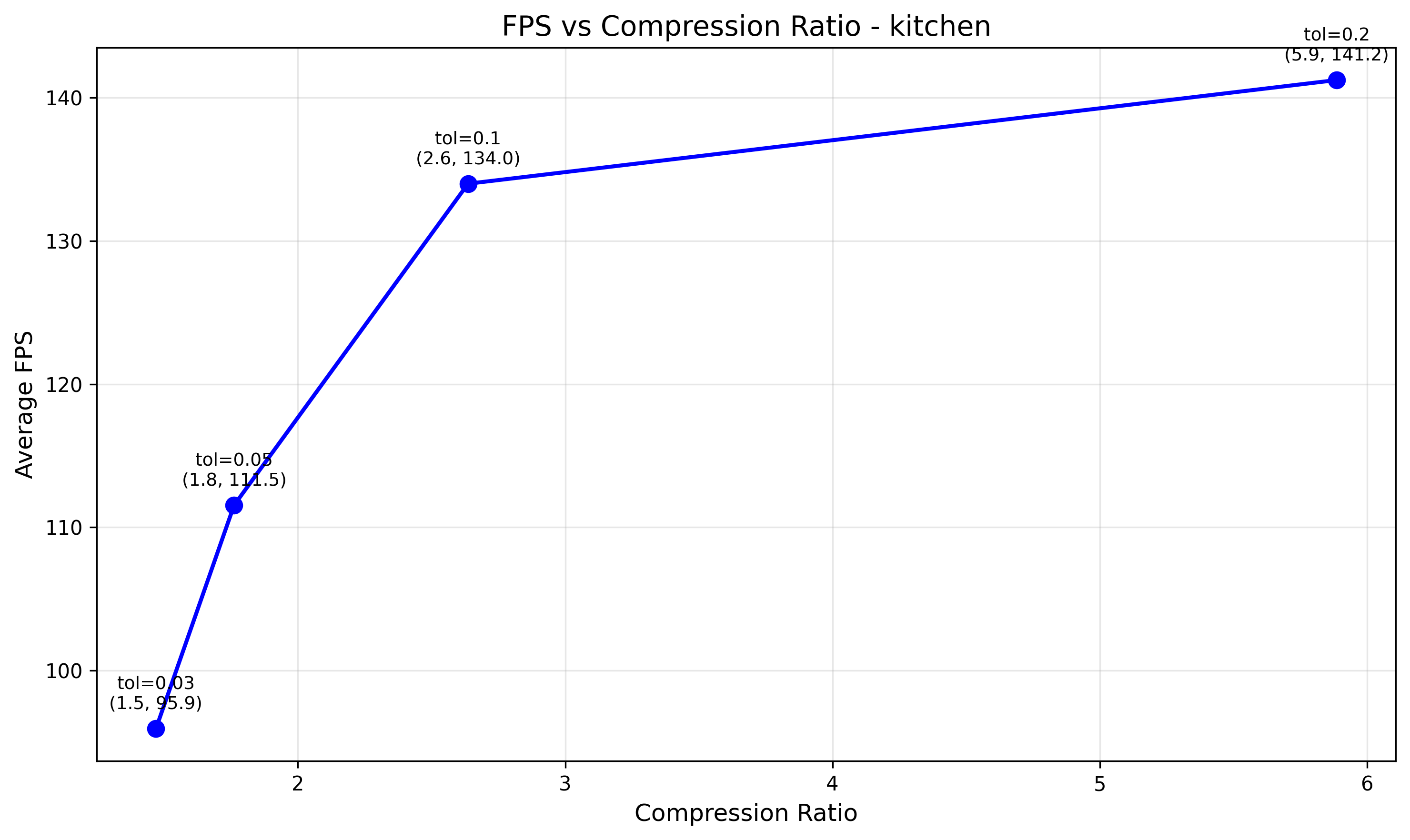}
  \captionsetup{skip=3pt}
  \caption*{Kitchen}
\end{minipage}\hfill
\begin{minipage}[b]{0.44\linewidth}
  \centering
  \includegraphics[width=\linewidth]{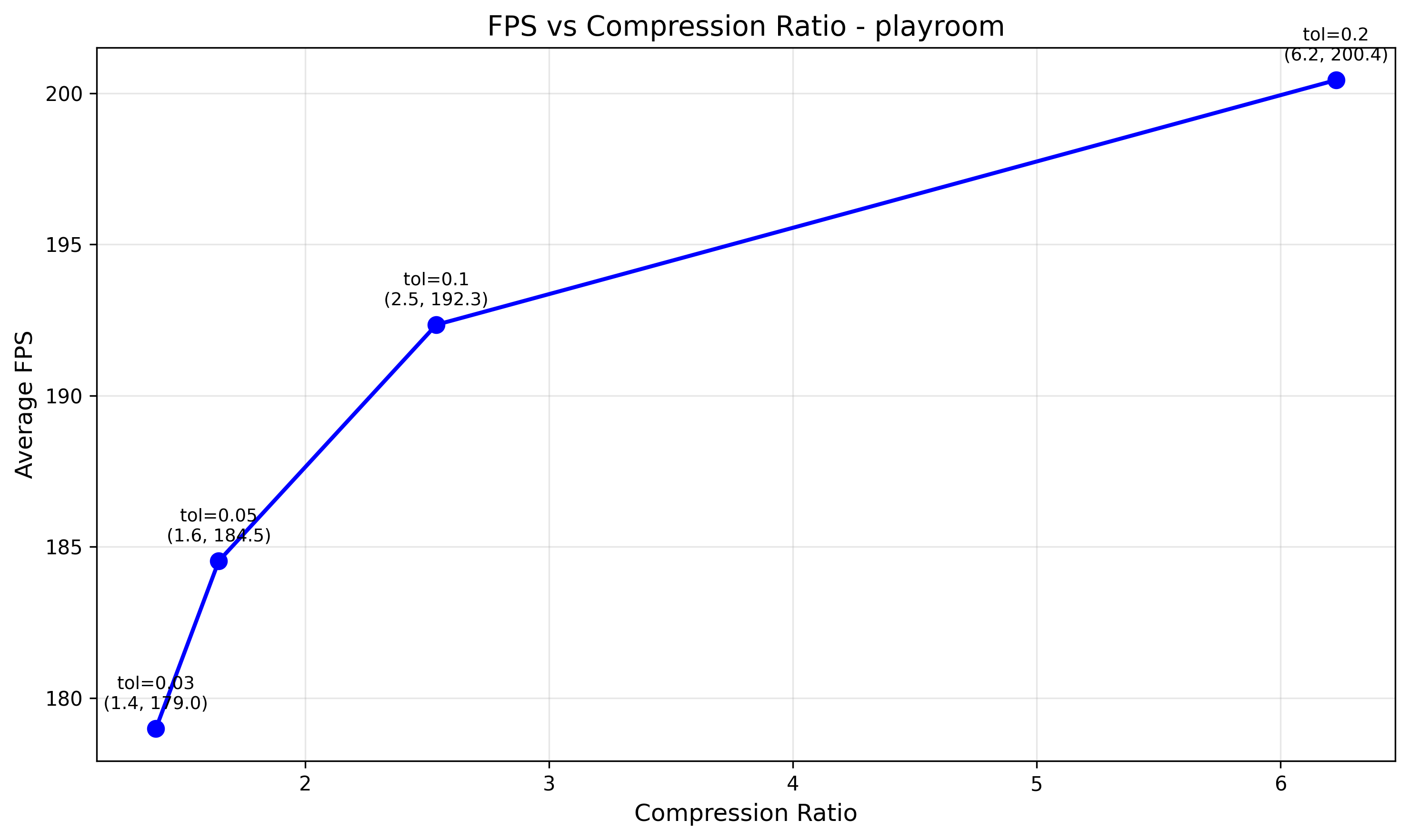}
  \captionsetup{skip=3pt}
  \caption*{Playroom}
\end{minipage}
\end{figure}

\subsection{PixelGS+Comp}

\begin{figure}[H]
\centering
\begin{minipage}[b]{0.44\linewidth}
  \centering
  \includegraphics[width=\linewidth]{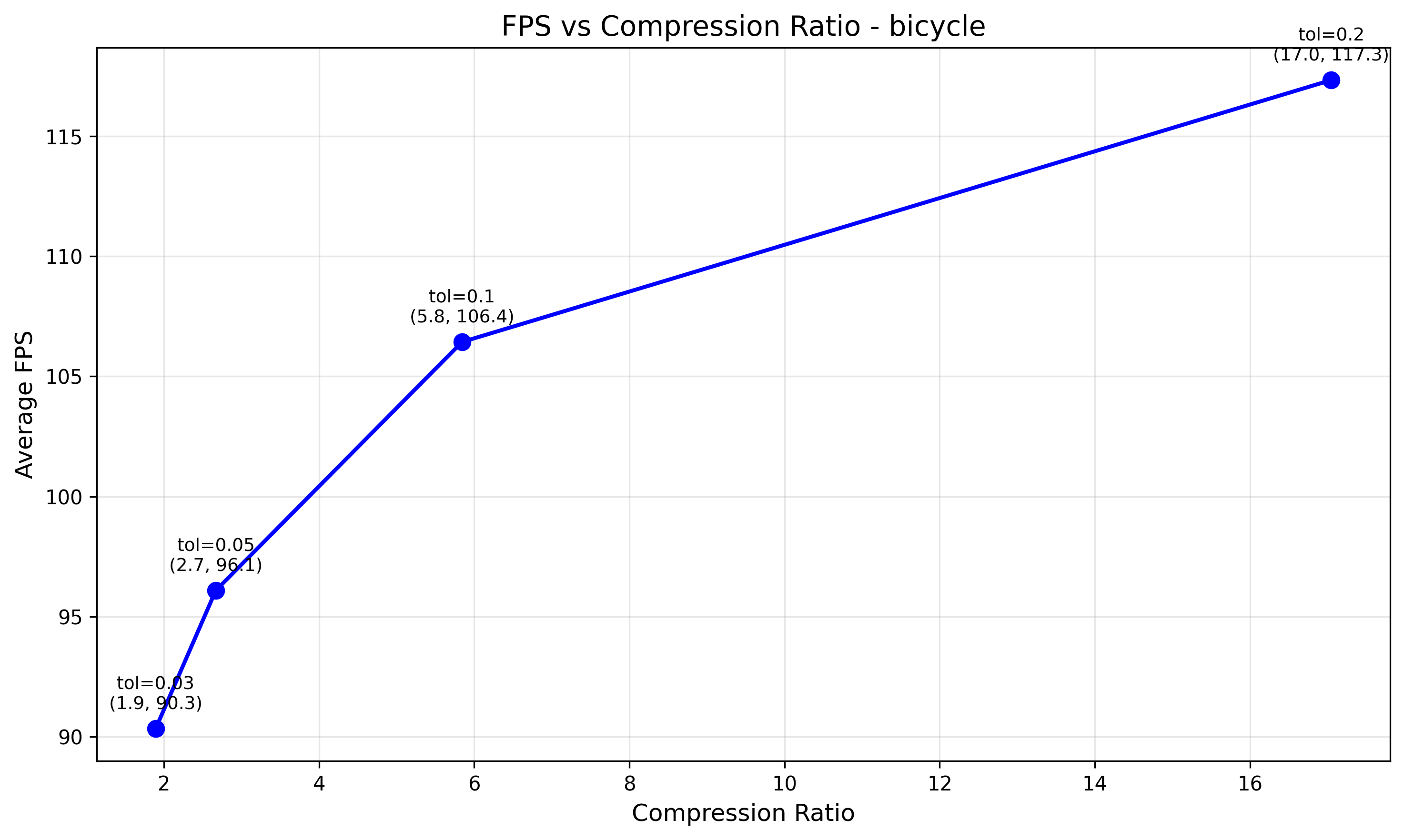}
  \captionsetup{skip=3pt}
  \caption*{Bicycle}
\end{minipage}\hfill
\begin{minipage}[b]{0.44\linewidth}
  \centering
  \includegraphics[width=\linewidth]{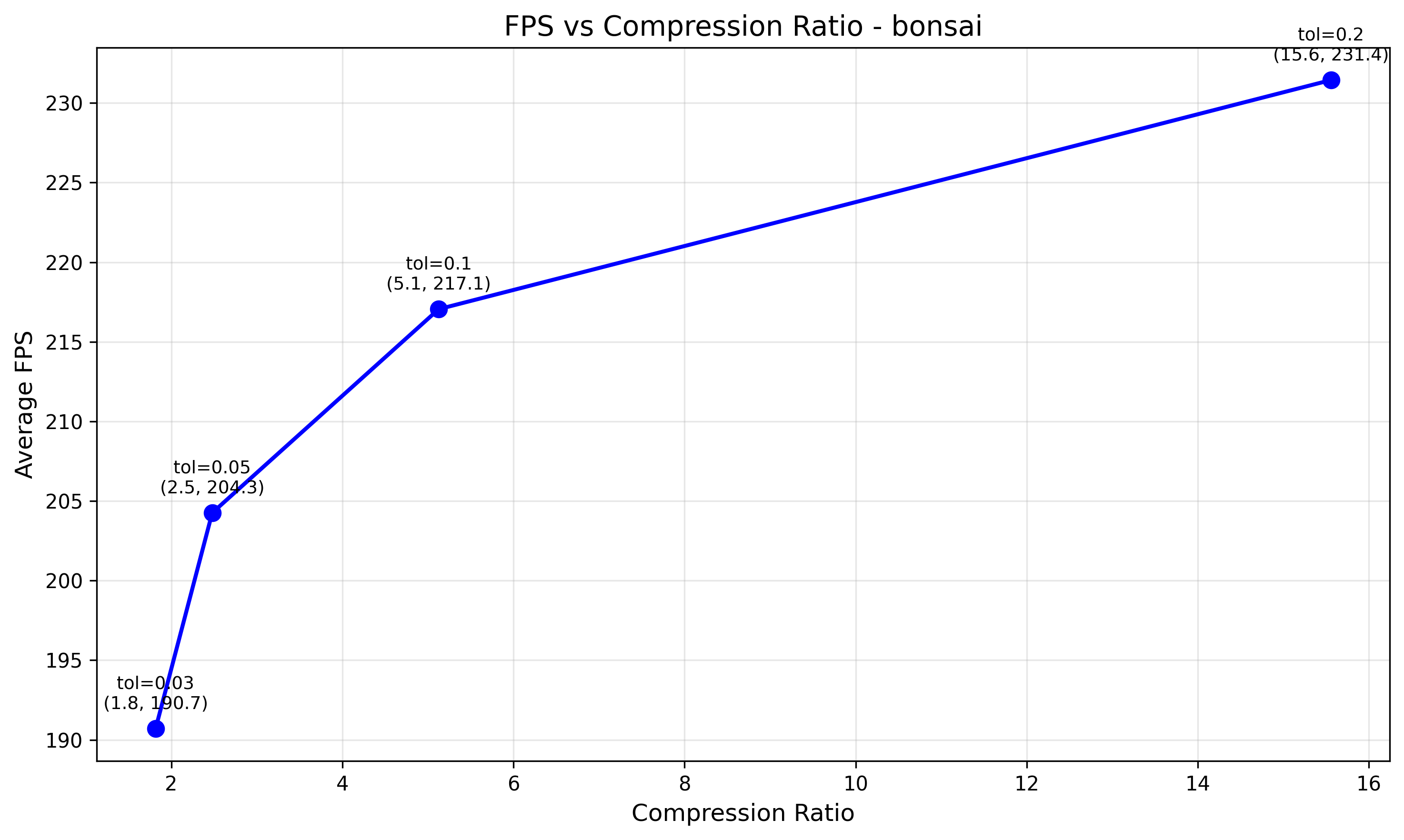}
  \captionsetup{skip=3pt}
  \caption*{Bonsai}
\end{minipage}
\end{figure}

\begin{figure}[H]
\centering
\begin{minipage}[b]{0.44\linewidth}
  \centering
  \includegraphics[width=\linewidth]{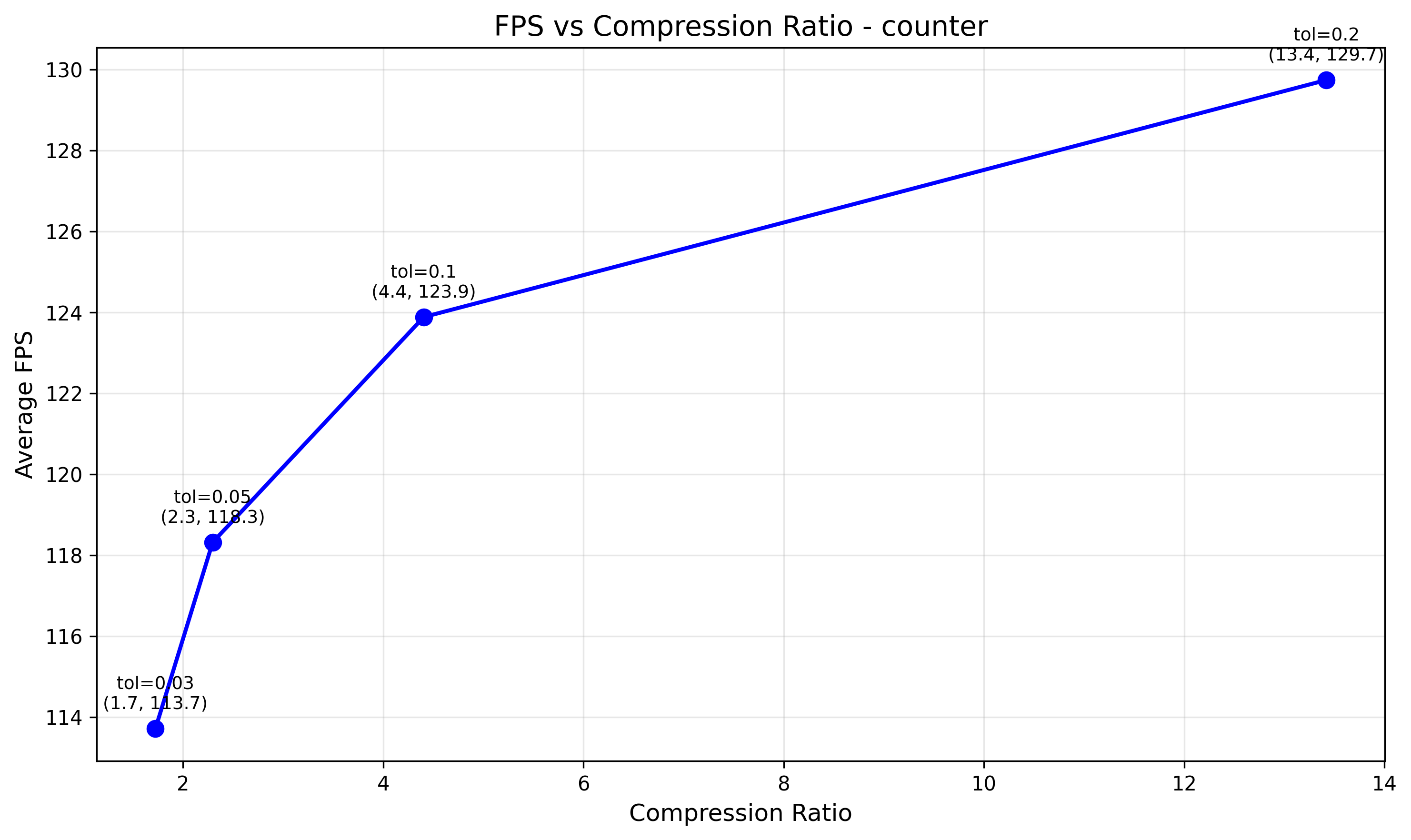}
  \captionsetup{skip=3pt}
  \caption*{Counter}
\end{minipage}\hfill
\begin{minipage}[b]{0.44\linewidth}
  \centering
  \includegraphics[width=\linewidth]{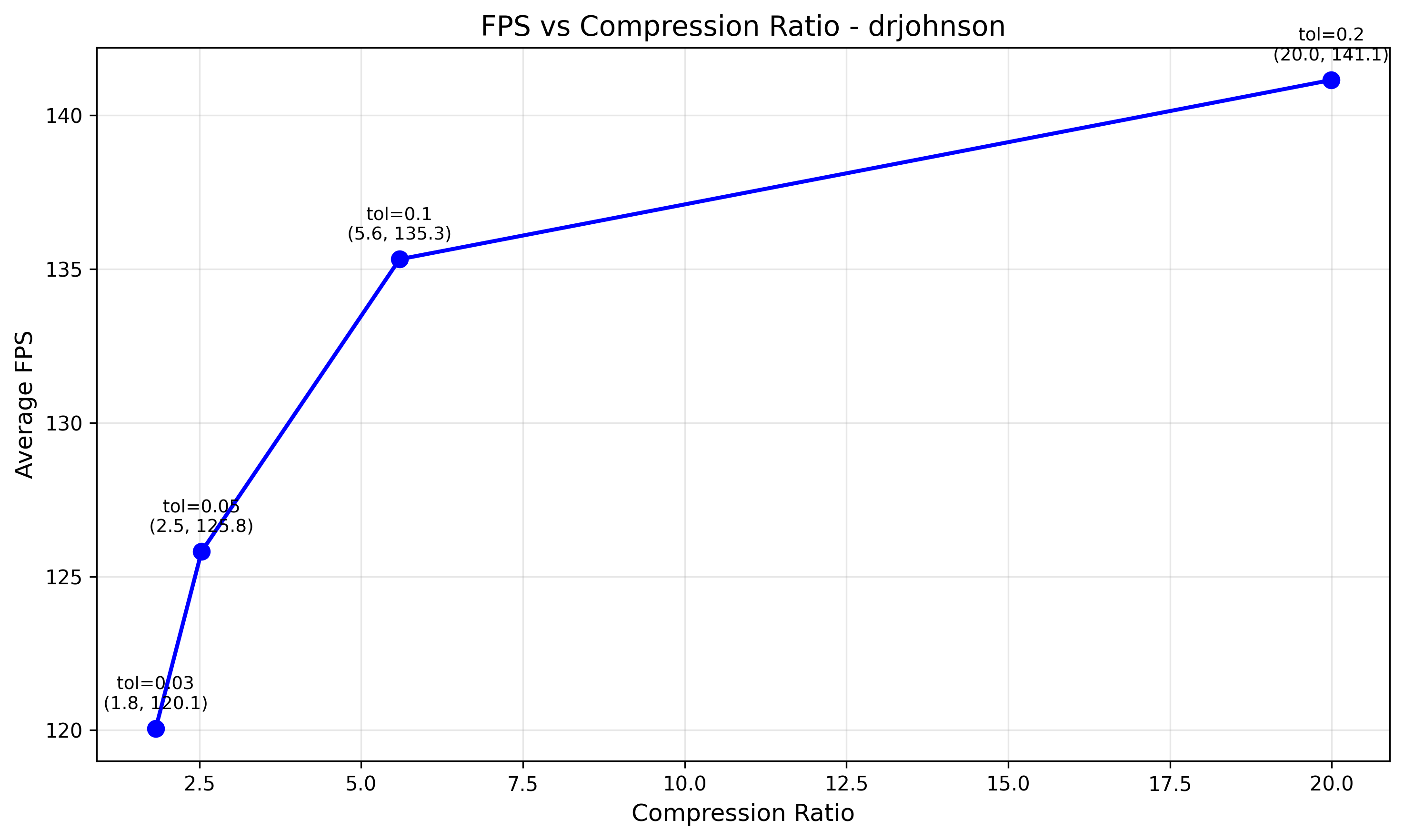}
  \captionsetup{skip=3pt}
  \caption*{Drjohnson}
\end{minipage}
\end{figure}

\begin{figure}[H]
\centering
\begin{minipage}[b]{0.44\linewidth}
  \centering
  \includegraphics[width=\linewidth]{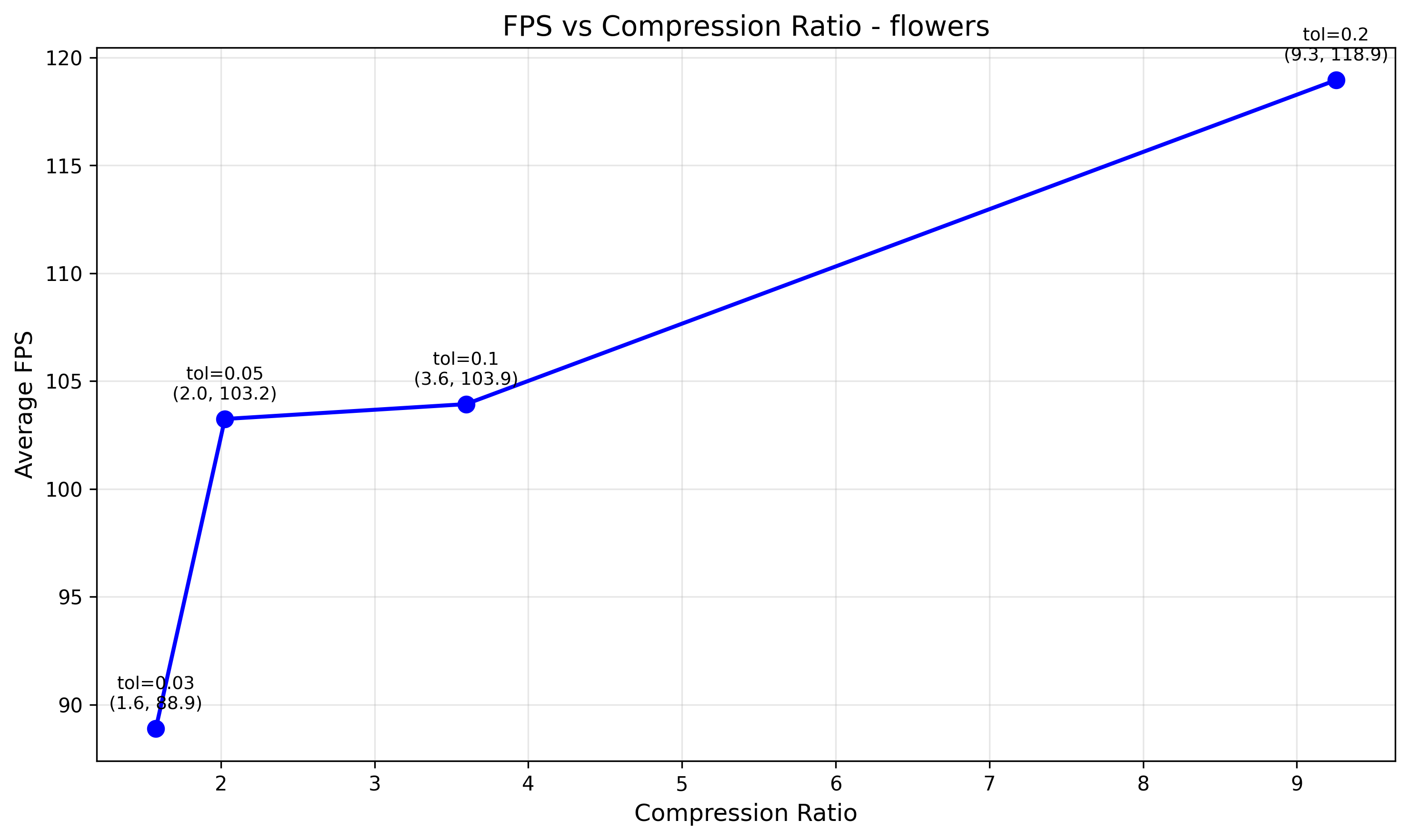}
  \captionsetup{skip=3pt}
  \caption*{Flowers}
\end{minipage}\hfill
\begin{minipage}[b]{0.44\linewidth}
  \centering
  \includegraphics[width=\linewidth]{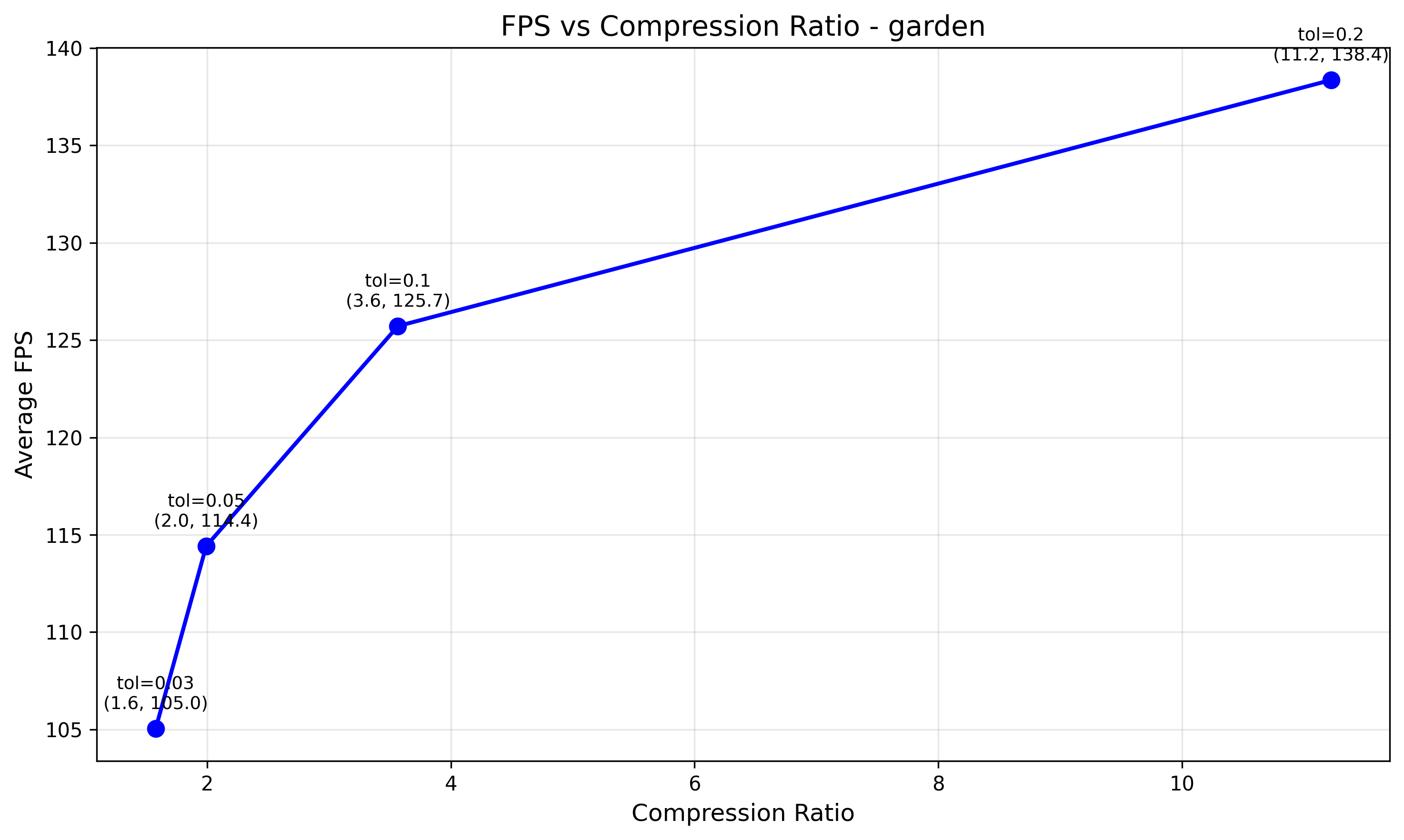}
  \captionsetup{skip=3pt}
  \caption*{Garden}
\end{minipage}
\end{figure}

\begin{figure}[H]
\centering
\begin{minipage}[b]{0.44\linewidth}
  \centering
  \includegraphics[width=\linewidth]{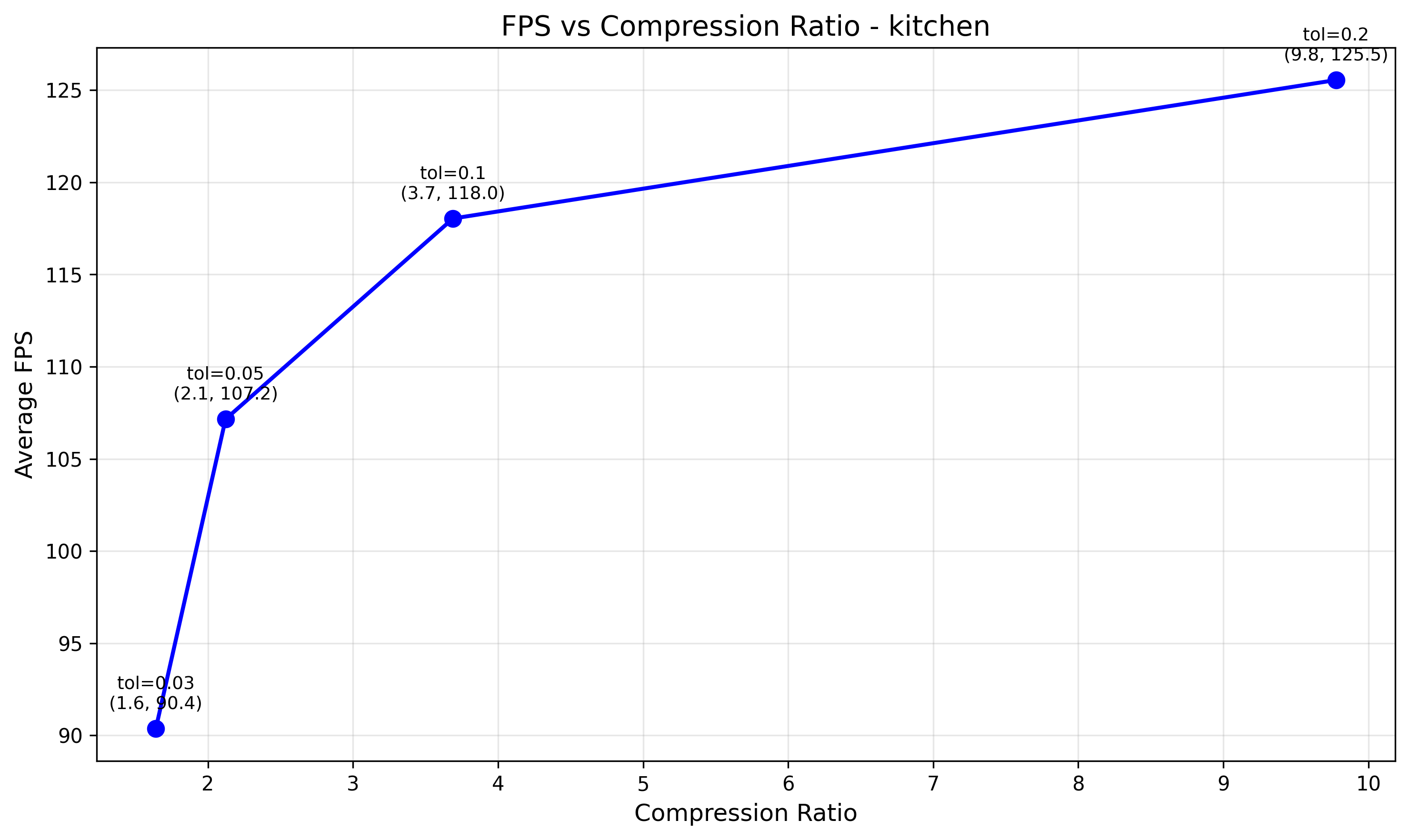}
  \captionsetup{skip=3pt}
  \caption*{Kitchen}
\end{minipage}\hfill
\begin{minipage}[b]{0.44\linewidth}
  \centering
  \includegraphics[width=\linewidth]{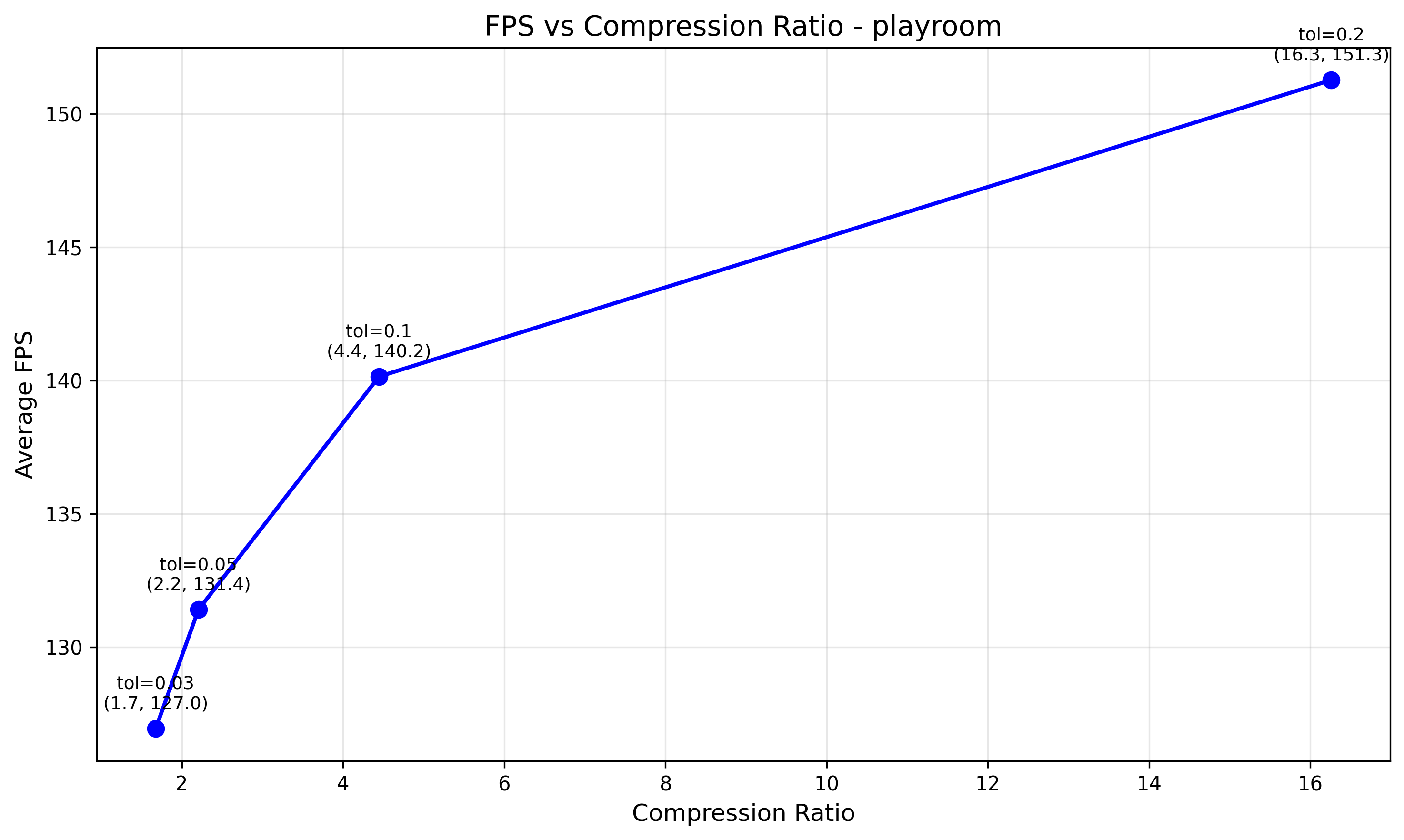}
  \captionsetup{skip=3pt}
  \caption*{Playroom}
\end{minipage}
\end{figure}

\section{Quantitative Evaluation of Rendering Fidelity Across Tolerance Levels}
To quantify the rendering fidelity of our compression method and have a deeper understanding of the effect of the tolerance, Table~\ref{tab:pairwise_comparison1} reports PSNR, SSIM, and LPIPS means over each scene for the compressed renderings of each pipeline \textit{3DGS+comp}, \textit{MCMC+comp}, and \textit{PixelGS+comp} at different tolerance levels (tol = 0.03, 0.05, 0.1, 0.2), against the corresponding original baseline renderings. From Table~\ref{tab:pairwise_comparison1}, we can easily find the same conclusions that as the tolerance value increases (which leads to sparser representations and thus stronger compression), the rendering quality gradually decreases.

\begin{table}[t]
  \small
  \centering
  \caption{Mean per-scene PSNR (dB), SSIM and LPIPS of compressed renderings relative to original renderings at four tolerance levels (tol = 0.03, 0.05, 0.10, 0.20).
    Higher PSNR/SSIM and lower LPIPS indicate closer agreement with the original.}
  \label{tab:pairwise_comparison1}
  \setlength{\tabcolsep}{3pt}
  \resizebox{\textwidth}{!}{
  \begin{tabular}{c c c | ccccccccccccc | c}
    \toprule
    \textbf{Method} & \textbf{tol} & \textbf{Metric}
      & Bicycle & Bonsai & Counter & Garden & Kitchen & Room
      & Stump & Truck & Flowers & Playroom & Train & Treehill & Drjohnson
      & \textbf{Mean} \\
    \midrule

    \multirow{12}{*}{\textbf{3DGS+comp}}
      & \multirow{3}{*}{tol=0.03} & PSNR  & 41.77 & 46.03 & 45.57 & 41.38 & 44.24 & 46.33 & 41.85 & 43.43 & 41.25 & 46.16 & 44.19 & 42.69 & 46.45 & \textbf{43.95} \\
      &  & SSIM  & 0.9898 & 0.9930 & 0.9909 & 0.9893 & 0.9921 & 0.9917 & 0.9906 & 0.9918 & 0.9928 & 0.9912 & 0.9930 & 0.9902 & 0.9931 & \textbf{0.9915} \\
      &  & LPIPS & 0.0176 & 0.0149 & 0.0144 & 0.0156 & 0.0113 & 0.0170 & 0.0190 & 0.0124 & 0.0134 & 0.0148 & 0.0111 & 0.0180 & 0.0171 & \textbf{0.0151} \\
    \cmidrule(lr){2-17}
      & \multirow{3}{*}{tol=0.05} & PSNR  & 40.64 & 44.14 & 43.78 & 40.18 & 42.55 & 44.53 & 40.70 & 41.90 & 40.03 & 44.10 & 42.58 & 41.45 & 44.39 & \textbf{42.38} \\
      &  & SSIM  & 0.9881 & 0.9906 & 0.9878 & 0.9869 & 0.9898 & 0.9890 & 0.9883 & 0.9897 & 0.9910 & 0.9881 & 0.9909 & 0.9883 & 0.9910 & \textbf{0.9892} \\
      &  & LPIPS & 0.0214 & 0.0193 & 0.0188 & 0.0198 & 0.0148 & 0.0211 & 0.0252 & 0.0153 & 0.0171 & 0.0185 & 0.0146 & 0.0219 & 0.0219 & \textbf{0.0192} \\
    \cmidrule(lr){2-17}
      & \multirow{3}{*}{tol=0.10} & PSNR  & 38.54 & 41.03 & 40.69 & 37.97 & 39.59 & 41.20 & 38.51 & 39.34 & 37.80 & 40.03 & 39.64 & 39.18 & 40.84 & \textbf{39.57} \\
      &  & SSIM  & 0.9830 & 0.9863 & 0.9818 & 0.9810 & 0.9851 & 0.9829 & 0.9819 & 0.9859 & 0.9870 & 0.9821 & 0.9873 & 0.9829 & 0.9871 & \textbf{0.9842} \\
      &  & LPIPS & 0.0301 & 0.0264 & 0.0263 & 0.0296 & 0.0217 & 0.0301 & 0.0392 & 0.0210 & 0.0257 & 0.0260 & 0.0210 & 0.0311 & 0.0311 & \textbf{0.0276} \\
    \cmidrule(lr){2-17}
      & \multirow{3}{*}{tol=0.20} & PSNR  & 35.37 & 36.84 & 36.37 & 34.38 & 34.99 & 36.96 & 35.27 & 35.40 & 34.39 & 35.52 & 35.28 & 35.54 & 37.15 & \textbf{35.65} \\
      &  & SSIM  & 0.9656 & 0.9772 & 0.9691 & 0.9629 & 0.9730 & 0.9708 & 0.9638 & 0.9768 & 0.9764 & 0.9725 & 0.9793 & 0.9643 & 0.9799 & \textbf{0.9717} \\
      &  & LPIPS & 0.0510 & 0.0408 & 0.0425 & 0.0506 & 0.0382 & 0.0492 & 0.0647 & 0.0331 & 0.0449 & 0.0421 & 0.0331 & 0.0563 & 0.0472 & \textbf{0.0457} \\
    \midrule

    \multirow{12}{*}{\textbf{MCMC+comp}}
      & \multirow{3}{*}{tol=0.03} & PSNR  & 53.36 & 53.51 & 53.60 & 53.44 & 53.45 & 52.86 & 53.86 & 53.01 & 53.74 & 53.15 & 53.28 & 52.46 & 52.12 & \textbf{53.22} \\
      &  & SSIM  & 0.9993 & 0.9984 & 0.9984 & 0.9993 & 0.9989 & 0.9982 & 0.9994 & 0.9988 & 0.9995 & 0.9982 & 0.9989 & 0.9991 & 0.9980 & \textbf{0.9988} \\
      &  & LPIPS & 0.0012 & 0.0012 & 0.0010 & 0.0008 & 0.0008 & 0.0013 & 0.0013 & 0.0008 & 0.0008 & 0.0010 & 0.0008 & 0.0015 & 0.0017 & \textbf{0.0011} \\
    \cmidrule(lr){2-17}
      & \multirow{3}{*}{tol=0.05} & PSNR  & 49.27 & 49.21 & 49.50 & 49.60 & 49.41 & 49.19 & 50.01 & 48.82 & 49.75 & 49.06 & 48.97 & 48.54 & 47.90 & \textbf{49.17} \\
      &  & SSIM  & 0.9989 & 0.9976 & 0.9975 & 0.9989 & 0.9985 & 0.9975 & 0.9989 & 0.9984 & 0.9991 & 0.9976 & 0.9984 & 0.9986 & 0.9975 & \textbf{0.9983} \\
      &  & LPIPS & 0.0031 & 0.0027 & 0.0022 & 0.0021 & 0.0019 & 0.0023 & 0.0037 & 0.0017 & 0.0022 & 0.0017 & 0.0019 & 0.0035 & 0.0037 & \textbf{0.0025} \\
    \cmidrule(lr){2-17}
      & \multirow{3}{*}{tol=0.10} & PSNR  & 43.20 & 43.09 & 43.56 & 42.85 & 42.64 & 43.64 & 44.21 & 42.51 & 43.41 & 42.17 & 42.29 & 43.22 & 41.57 & \textbf{42.95} \\
      &  & SSIM  & 0.9974 & 0.9951 & 0.9947 & 0.9969 & 0.9964 & 0.9944 & 0.9966 & 0.9969 & 0.9974 & 0.9946 & 0.9970 & 0.9967 & 0.9952 & \textbf{0.9961} \\
      &  & LPIPS & 0.0094 & 0.0067 & 0.0065 & 0.0080 & 0.0060 & 0.0056 & 0.0126 & 0.0051 & 0.0076 & 0.0045 & 0.0062 & 0.0096 & 0.0109 & \textbf{0.0076} \\
    \cmidrule(lr){2-17}
      & \multirow{3}{*}{tol=0.20} & PSNR  & 38.33 & 37.44 & 37.40 & 36.79 & 36.30 & 37.63 & 38.92 & 36.83 & 37.48 & 36.20 & 36.13 & 38.38 & 36.18 & \textbf{37.23} \\
      &  & SSIM  & 0.9920 & 0.9895 & 0.9853 & 0.9895 & 0.9886 & 0.9869 & 0.9894 & 0.9920 & 0.9913 & 0.9878 & 0.9925 & 0.9899 & 0.9882 & \textbf{0.9895} \\
      &  & LPIPS & 0.0197 & 0.0152 & 0.0172 & 0.0206 & 0.0181 & 0.0152 & 0.0285 & 0.0136 & 0.0208 & 0.0138 & 0.0155 & 0.0234 & 0.0278 & \textbf{0.0192} \\
    \midrule

    \multirow{12}{*}{\textbf{PixelGS+comp}}
      & \multirow{3}{*}{tol=0.03} & PSNR  & 51.51 & 52.17 & 52.54 & 52.46 & 52.37 & 51.86 & 52.26 & 51.82 & 52.23 & 52.10 & 52.67 & 50.48 & 51.90 & \textbf{52.03} \\
      &  & SSIM  & 0.9990 & 0.9980 & 0.9981 & 0.9991 & 0.9987 & 0.9977 & 0.9991 & 0.9986 & 0.9993 & 0.9977 & 0.9987 & 0.9988 & 0.9979 & \textbf{0.9985} \\
      &  & LPIPS & 0.0020 & 0.0018 & 0.0015 & 0.0011 & 0.0011 & 0.0021 & 0.0020 & 0.0011 & 0.0012 & 0.0015 & 0.0010 & 0.0023 & 0.0021 & \textbf{0.0016} \\
    \cmidrule(lr){2-17}
      & \multirow{3}{*}{tol=0.05} & PSNR  & 47.56 & 48.03 & 48.67 & 48.18 & 48.10 & 47.94 & 48.39 & 47.82 & 48.18 & 47.85 & 48.14 & 46.88 & 47.60 & \textbf{47.95} \\
      &  & SSIM  & 0.9983 & 0.9970 & 0.9970 & 0.9984 & 0.9980 & 0.9964 & 0.9982 & 0.9980 & 0.9987 & 0.9963 & 0.9982 & 0.9979 & 0.9971 & \textbf{0.9977} \\
      &  & LPIPS & 0.0048 & 0.0039 & 0.0033 & 0.0030 & 0.0027 & 0.0047 & 0.0055 & 0.0023 & 0.0033 & 0.0031 & 0.0025 & 0.0051 & 0.0050 & \textbf{0.0038} \\
    \cmidrule(lr){2-17}
      & \multirow{3}{*}{tol=0.10} & PSNR  & 42.41 & 42.18 & 42.73 & 41.97 & 41.85 & 42.11 & 42.78 & 41.65 & 42.24 & 41.20 & 41.50 & 42.00 & 41.61 & \textbf{42.02} \\
      &  & SSIM  & 0.9954 & 0.9932 & 0.9914 & 0.9950 & 0.9948 & 0.9913 & 0.9943 & 0.9952 & 0.9958 & 0.9915 & 0.9958 & 0.9945 & 0.9940 & \textbf{0.9940} \\
      &  & LPIPS & 0.0123 & 0.0115 & 0.0101 & 0.0104 & 0.0086 & 0.0152 & 0.0164 & 0.0078 & 0.0106 & 0.0114 & 0.0084 & 0.0140 & 0.0164 & \textbf{0.0118} \\
    \cmidrule(lr){2-17}
      & \multirow{3}{*}{tol=0.20} & PSNR  & 37.54 & 36.39 & 36.83 & 36.25 & 35.38 & 36.64 & 37.40 & 35.45 & 36.41 & 35.67 & 34.96 & 36.72 & 36.75 & \textbf{36.34} \\
      &  & SSIM  & 0.9834 & 0.9817 & 0.9788 & 0.9833 & 0.9833 & 0.9781 & 0.9804 & 0.9860 & 0.9853 & 0.9815 & 0.9875 & 0.9804 & 0.9854 & \textbf{0.9827} \\
      &  & LPIPS & 0.0298 & 0.0311 & 0.0292 & 0.0261 & 0.0263 & 0.0407 & 0.0376 & 0.0228 & 0.0284 & 0.0318 & 0.0232 & 0.0365 & 0.0394 & \textbf{0.0310} \\
    \bottomrule
  \end{tabular}
  }
\end{table}

\section{Quantitative Evaluation of Rendering Quality with respect to GT}

To ensure a rigorous and standardized evaluation, we assess the performance of our compression framework by measuring rendering metrics against Ground Truth (GT) images. Table~\ref{tab:per_scene_gt_comparison} presents a comprehensive per-scene comparison across three SOTA baselines (3DGS, 3DGS-MCMC, and PixelGS) and the competitive EAGLES method.

\begin{table}[t]
  \small
  \centering
  \caption{Quantitative comparison across 13 scenes. We report PSNR$\uparrow$, SSIM$\uparrow$, and LPIPS$\downarrow$. ``Base'' refers to the original method, while ``Comp.''\ denotes the application of our compression framework. For instance, ``PixelGS Base'' represents the vanilla PixelGS method, and ``No SQ'' indicates the removal of Scalar Quantization from EAGLES.}
  \label{tab:per_scene_gt_comparison}

  \setlength{\aboverulesep}{0pt}
  \setlength{\belowrulesep}{0pt}
  \setlength{\tabcolsep}{2.8pt}

  \resizebox{\textwidth}{!}{
  \begin{tabular}{l l l ccccccccccccc | c}
    \toprule
    \textbf{Method} & \textbf{Type} & \textbf{Metric}
      & Bicycle & Bonsai & Counter & Drjohnson & Flowers & Garden & Kitchen & Playroom & Room
      & Stump  & Train & Treehill & Truck
      & \textbf{Mean} \\
    \midrule

    \multirow{6}{*}{\textbf{3DGS}}
    & \multirow{3}{*}{Base.}
      & PSNR  & 26.11 & 32.06 & 28.90 & 29.41 & 22.86 & 28.15 & 31.04 & 30.17 & 31.33 & 26.65 & 22.14 & 23.48 & 25.27 & 27.51 \\
    & & SSIM  & 0.8103 & 0.9403 & 0.9007 & 0.9052 & 0.6929 & 0.8812 & 0.9161 & 0.9111 & 0.9133 & 0.7895 & 0.8110 & 0.7026 & 0.8723 & 0.8497 \\
    & & LPIPS & 0.1485 & 0.1240 & 0.1518 & 0.1728 & 0.2533 & 0.0949 & 0.1063 & 0.1540 & 0.1450 & 0.1813 & 0.1775 & 0.2389 & 0.1256 & 0.1595 \\
    \cmidrule(r){2-17}
    & \multirow{3}{*}{Comp.}
      & PSNR  & 26.04 & 31.69 & 28.73 & 29.27 & 22.82 & 28.01 & 30.74 & 30.01 & 31.09 & 26.60 & 22.11 & 23.45 & 25.21 & 27.37 \\
    & & SSIM  & 0.8092 & 0.9378 & 0.8975 & 0.9044 & 0.6917 & 0.8788 & 0.9139 & 0.9107 & 0.9108 & 0.7879 & 0.8099 & 0.7017 & 0.8707 & 0.8481 \\
    & & LPIPS & 0.1525 & 0.1282 & 0.1559 & 0.1761 & 0.2566 & 0.1004 & 0.1111 & 0.1558 & 0.1487 & 0.1878 & 0.1806 & 0.2416 & 0.1286 & 0.1634 \\

    \midrule

    \multirow{6}{*}{\textbf{MCMC}}
    & \multirow{3}{*}{Base.} & PSNR  & 27.21 & 32.65 & 29.34 & 29.35 & 23.28 & 29.76 & 32.06 & 29.94 & 32.13 & 27.84 & 22.71 & 24.29 & 26.43 & 28.23 \\
    & & SSIM  & 0.8654 & 0.9476 & 0.9162 & 0.9024 & 0.7335 & 0.9269 & 0.9333 & 0.9112 & 0.9277 & 0.8443 & 0.8415 & 0.7379 & 0.9007 & 0.8761 \\
    & & LPIPS & 0.1054 & 0.1905 & 0.1854 & 0.2344 & 0.1835 & 0.0524 & 0.1208 & 0.2377 & 0.1988 & 0.1297 & 0.1816 & 0.1850 & 0.1084 & 0.1626 \\
    \cmidrule(r){2-17}
    & \multirow{3}{*}{Comp.} & PSNR  & 27.11 & 32.21 & 29.17 & 29.14 & 23.25 & 29.55 & 31.64 & 29.67 & 31.82 & 27.77 & 22.64 & 24.25 & 26.33 & 28.04 \\
    & & SSIM  & 0.8643 & 0.9443 & 0.9133 & 0.9013 & 0.7324 & 0.9250 & 0.9316 & 0.9107 & 0.9246 & 0.8429 & 0.8403 & 0.7369 & 0.8993 & 0.8744 \\
    & & LPIPS & 0.1110 & 0.1945 & 0.1890 & 0.2379 & 0.1880 & 0.0585 & 0.1246 & 0.2392 & 0.2009 & 0.1373 & 0.1852 & 0.1898 & 0.1117 & 0.1677 \\

    \midrule

    \multirow{6}{*}{\textbf{PixelGS}}
    & \multirow{3}{*}{Base.} & PSNR  & 26.74 & 32.34 & 29.17 & 28.10 & 22.91 & 29.32 & 31.54 & 29.91 & 31.23 & 27.19 & 22.26 & 23.35 & 25.47 & 27.66 \\
    & & SSIM  & 0.8483 & 0.9453 & 0.9139 & 0.8876 & 0.7234 & 0.9215 & 0.9302 & 0.9044 & 0.9179 & 0.8197 & 0.8289 & 0.7095 & 0.8866 & 0.8644 \\
    & & LPIPS & 0.1131 & 0.1916 & 0.1825 & 0.2548 & 0.1787 & 0.0557 & 0.1194 & 0.2403 & 0.2104 & 0.1424 & 0.1783 & 0.1986 & 0.1208 & 0.1682 \\
    \cmidrule(r){2-17}
    & \multirow{3}{*}{Comp.} & PSNR  & 26.63 & 31.82 & 28.97 & 27.99 & 22.89 & 29.09 & 31.14 & 29.74 & 30.89 & 27.12 & 22.22 & 23.32 & 25.38 & 27.48 \\
    & & SSIM  & 0.8459 & 0.9415 & 0.9086 & 0.8867 & 0.7219 & 0.9182 & 0.9275 & 0.9040 & 0.9134 & 0.8173 & 0.8272 & 0.7074 & 0.8843 & 0.8619 \\
    & & LPIPS & 0.1198 & 0.1970 & 0.1874 & 0.2591 & 0.1835 & 0.0632 & 0.1245 & 0.2430 & 0.2153 & 0.1508 & 0.1824 & 0.2054 & 0.1252 & 0.1736 \\

    \midrule

    \multirow{6}{*}{\textbf{EAGLES}}
    & \multirow{3}{*}{Base.} & PSNR  & 26.31 & 31.24 & 28.32 & 29.32 & 22.74 & 28.53 & 30.50 & 30.22 & 31.43 & 26.97 & 21.43 & 23.80 & 25.01 & 27.37 \\
    & & SSIM  & 0.8200 & 0.9368 & 0.8998 & 0.9083 & 0.6856 & 0.9001 & 0.9224 & 0.9128 & 0.9191 & 0.8032 & 0.7988 & 0.7134 & 0.8755 & 0.8689 \\
    & & LPIPS & 0.1754 & 0.2152 & 0.2148 & 0.2413 & 0.2793 & 0.0889 & 0.1373 & 0.2513 & 0.2235 & 0.1825 & 0.2369 & 0.2729 & 0.1647 & 0.2065 \\
    \cmidrule(r){2-17}
    & \multirow{3}{*}{No SQ} & PSNR  & 26.54 & 32.00 & 28.94 & 29.03 & 23.01 & 28.88 & 31.11 & 29.96 & 31.35 & 27.09 & 21.62 & 23.77 & 25.29 & 27.58 \\
    & & SSIM  & 0.8303 & 0.9410 & 0.9083 & 0.9031 & 0.7013 & 0.9086 & 0.9281 & 0.9085 & 0.9199 & 0.8102 & 0.8018 & 0.7127 & 0.8779 & 0.8578 \\
    & & LPIPS & 0.1562 & 0.2096 & 0.2024 & 0.2503 & 0.2618 & 0.0742 & 0.1289 & 0.2519 & 0.2205 & 0.1646 & 0.2325 & 0.2663 & 0.1624 & 0.1986 \\

    \bottomrule
  \end{tabular}
  }
\end{table}

Experimental results in Table~\ref{tab:per_scene_gt_comparison} demonstrate that our framework maintains high rendering fidelity across diverse SOTA methods. Notably, our framework is method-agnostic and can be seamlessly integrated as a plug-and-play module. On average, the PSNR degradation remains within a marginal range of $0.14$ to $0.19$ dB, with negligible impacts on SSIM and LPIPS, illustrating a near-lossless preservation of visual quality post-compression.

Furthermore, our framework exhibits superior stability across various base architectures. While compression often introduces artifacts in complex geometries, our scheme incurs only marginal fidelity loss even in the most challenging scenarios. For instance, when integrated with 3DGS, the maximum per-scene PSNR drop is strictly capped at $0.37$ dB; across all evaluated methods, the worst-case degradation remains below $0.52$ dB. In stark contrast, EAGLES' SQ exhibits significant instability. Its PSNR plunges by as much as $0.76$ dB in the ``Bonsai'' scene and exceeds $0.6$ dB in ``Counter'' and ``Kitchen.'' These results highlight the robustness of our framework in preserving scene-specific details without the unpredictable volatility inherent in existing quantization baselines.

Beyond rendering quality, our framework offers significant advantages in simplicity and efficiency. Unlike EAGLES, which relies on Quantization-Aware Training (QAT) and necessitates the Straight-Through Estimator (STE) to bypass non-differentiability, our method is a strictly post-training compression scheme. Specifically, EAGLES requires modifying gradient propagation, imposing heavy training overhead and implementation complexity. In contrast, our approach can be directly applied to any pre-trained model without further tuning. This nature makes our framework a more scalable solution for large-scale deployment where rapid compression of existing assets is paramount.

\end{document}